\documentclass[11pt]{article}

\usepackage{fullpage}

\usepackage{latexsym}
\usepackage{amsmath}
\usepackage{amssymb}
\usepackage{mathtools}
\usepackage{url}
\usepackage{xspace}
\usepackage{color}

\usepackage{booktabs}
\usepackage{listings}

\usepackage{tikz}
\usepackage{url}

\newcommand{\leaveout}[1]{}

\newcommand{\eg}{e.g.\@\xspace}

\newcommand{\ceiling}[1]{\lceil #1\rceil}
\newcommand{\floor}[1]{\lfloor #1\rfloor}

\newcommand{\piperepetition}{50}
\newcommand{\repetition}{100}

\newcommand{\openmpiversion}{Open\,MPI 4.0.1\xspace}
\newcommand{\intelmpiversion}{Intel\,MPI 2018\xspace}
\newcommand{\mpichversion}{\texttt{mpich} 3.3\xspace}
\newcommand{\mvapichversion}{\texttt{mvapich} 2-2-3.1\xspace}
\newcommand{\gccversion}{\texttt{gcc 8.3.0}\xspace}

\newcommand{\mpisendrecv}{\texttt{MPI\_\-Sendrecv}\xspace}

\newcommand{\mpibarrier}{\texttt{MPI\_\-Barrier}\xspace}
\newcommand{\mpibcast}{\texttt{MPI\_\-Bcast}\xspace}
\newcommand{\mpigather}{\texttt{MPI\_\-Gather}\xspace}
\newcommand{\mpigatherv}{\texttt{MPI\_\-Gatherv}\xspace}
\newcommand{\mpiscatter}{\texttt{MPI\_\-Scatter}\xspace}
\newcommand{\mpiscatterv}{\texttt{MPI\_\-Scatterv}\xspace}
\newcommand{\mpiallgather}{\texttt{MPI\_\-Allgather}\xspace}
\newcommand{\mpiallgatherv}{\texttt{MPI\_\-Allgatherv}\xspace}
\newcommand{\mpialltoall}{\texttt{MPI\_\-Alltoall}\xspace}

\newcommand{\mpireduce}{\texttt{MPI\_\-Reduce}\xspace}
\newcommand{\mpiallreduce}{\texttt{MPI\_\-Allreduce}\xspace}
\newcommand{\mpireducescatter}{\texttt{MPI\_\-Reduce\_\-scatter}\xspace}
\newcommand{\mpireducescatterblock}{\texttt{MPI\_\-Reduce\_\-scatter\_\-block}\xspace}
\newcommand{\mpiscan}{\texttt{MPI\_\-Scan}\xspace}
\newcommand{\mpiexscan}{\texttt{MPI\_\-Exscan}\xspace}

\newcommand{\mpicommsplit}{\texttt{MPI\_\-Comm\_\-split}\xspace}
\newcommand{\mpicommsplittype}{\texttt{MPI\_\-Comm\_\-split\_\-type}\xspace}
\newcommand{\mpicommcreate}{\texttt{MPI\_\-Comm\_\-create}\xspace}

\newcommand{\mpiinplace}{\texttt{MPI\_\-IN\_\-PLACE}\xspace}

\newcommand{\mpiint}{\texttt{MPI\_\-INT}\xspace}

\newcommand{\mpicommworld}{\texttt{MPI\_\-COMM\_\-WORLD}\xspace}

\usepackage{tikz}

\newtheorem{proposition}{Proposition}

\title{Decomposing Collectives for Exploiting Multi-lane Communication}
\author{Jesper Larsson Tr\"aff\\
TU Wien, Faculty of Informatics\\
Favoritenstrasse 16/191-4, 1040 Vienna, Austria}
\date{October 31, 2019\\
Revised and expanded December 2019, January 2020}

\begin{document}
\maketitle

\begin{abstract}
  Many modern, high-performance systems increase the cumulated
  node-bandwidth by offering more than a single communication network
  and/or by having multiple connections to the network. Efficient
  algorithms and implementations for collective operations as found
  in, \eg, MPI must be explicitly designed for such multi-lane
  capabilities. We discuss a model for the design of multi-lane
  algorithms, and in particular give a recipe for converting any
  standard, one-ported, (pipelined) communication tree algorithm into
  a multi-lane algorithm that can effectively use $k$ lanes
  simultaneously.

  We first examine the problem from the perspective of
  \emph{self-consistent performance guidelines}, and give simple,
  \emph{full-lane, mock-up implementations} of the MPI broadcast,
  reduction, scan, gather, scatter, allgather, and alltoall operations
  using only similar operations of the given MPI library itself in
  such a way that multi-lane capabilities can be exploited.  These
  implementations which rely on a decomposition of the communication
  domain into communicators for nodes and lanes are full-fledged and
  readily usable implementations of the MPI collectives. The mock-up
  implementations, contrary to expectation, in many cases show
  surprising performance improvements with different MPI libraries on
  a small 36-node dual-socket, dual-lane Intel OmniPath cluster,
  indicating severe problems with the native MPI library
  implementations. Our full-lane implementations are in many cases
  considerably more than a factor of two faster than the corresponding
  MPI collectives. We see similar results on the larger Vienna
  Scientific Cluster, VSC-3. These experiments indicate considerable
  room for improvement of the MPI collectives in current libraries
  including more efficient use of multi-lane communication.

  We complement our findings by comparing also to traditional,
  \emph{hierarchical implementations} of the MPI collectives where
  only a single process per node is communicating with other
  nodes. Also these implementations follow the decomposition of the
  communication domain, and in many case improve significantly over
  native MPI library collectives, although less than the full-lane
  implementations.
\end{abstract}

\section{Introduction}

Almost all current distributed memory (HPC) parallel computers are
clusters with a marked, hierarchical structure, \eg, islands
consisting of racks consisting of nodes consisting of sockets of
multi-core processors, or whatever the specific (vendor) terminology
may be. Hierarchy matters, with different levels having different
communication capabilities.  Communication interfaces must take the
hierarchy into account, and for any sets of processes, communicate at
the most efficient hierarchy levels.  Collective operations, as
eminently found in MPI~\cite{MPI-3.1}, can be given algorithms that
take the communication hierarchy into account, and implementations of
the MPI collectives in high-quality MPI libraries should utilize the
best, such hierarchical algorithms. Part of the motivation of this
report is to investigate in a portable manner whether this might be the
case.

Traditional hierarchical collective algorithms for clustered,
high-performance systems have often focussed on minimizing contention
on a single communication network through which the compute nodes are
connected by letting a single, node-local \emph{root process} on each
node be responsible for the communication with other nodes, see, \eg,
early work on clustered, wide area
systems~\cite{KielmannBalGorlatchVerstoepHofman01}. The collective
operations of MPI can readily be decomposed into hierarchical
applications of similar collective operations as described and
implemented in, \eg,
\cite{GrahamVenkataLaddShamisRabinovitzFilipovShainer11,KaronisSupinski00,Traff06:mpisxcoll,ThakurGroppRabenseifner05,Traff14:alltoall,Traff14:gather}
and many, many other works.

Many modern high-performance systems are equipped with multiple,
high-bandwidth communication networks and/or multiple connections to
the network(s) from the compute nodes, with bandwidth in some cases so
large that it cannot be saturated by a single processor-core.  We will
refer to systems with such capabilities as \emph{multi-lane}
(alternatively, \emph{multi-rail}) systems with $k$ being both the
number of physical lanes and the number of processor-cores required to
fully saturate the network(s). Our assumption is that processes (or
threads) that are close in some sense to the respective \emph{lanes}
can communicate independently, such that the cumulated bandwidth of
the compute nodes can be increased proportionally to the number of
available lanes. As an example, the cluster that we primarily use in
this study has compute nodes with two sockets, each of which is
connected to an own (Intel OmniPath) network, and MPI processes on
either socket can communicate independently of processes on the other
socket. Note that these assumptions are quite different from the
standard $k$-ported assumption where each processor can exploit $k$
communication ports simultaneously,
\eg,~\cite{BarNoyHo99,Bruck97}. Algorithm design under the $k$-lane
assumptions will be different from algorithm design under traditional,
$k$-ported assumptions.

An approach to exploit potential multi-lane communication capabilities
of modern cluster systems is to let several processes, each of which
are close to a network lane, communicate concurrently. For this
approach to be beneficial, data to be communicated across the nodes
must be effectively distributed across the communicating processes,
such that the total amount of data in and out of the cluster nodes
does not exceed that of a traditional, hierarchical algorithm. This
approach seems to have been pioneered by Panda et al.\ over a number
of
papers~\cite{BayatpourChakrabortySubramoniLuPanda17,KandallaSubramoniSanthanaramanKoopPanda09,KumarMamidalePanda08}
addressing different collectives of the MPI standard (allgather,
alltoall, allreduce). The idea of these \emph{multi-root} algorithms
is to divide the collective communication over several virtual roots
per compute node, with proportionally smaller parts of the total data
per virtual root.  Similar decompositions were proposed and evaluated
by K\"uhnemann et al.~\cite{KuhnemannRauberRunger06}, but without
always distributing the amount of data (for operations like broadcast
and allreduce). Improvememnts alomng these lines for \mpiallreduce
were recently explored in~\cite{BienzOlsonGropp19}. All these papers
demonstrated improvements in applications by improved implementations
of the MPI collectives considered, and we will not repeat such
application studies here.

In this report we explore algorithms and implementations of the MPI
collectives that can possibly exploit multi-lane capabilities. We do
so similarly to the papers by Panda et al.; but consider
\emph{full-lane implementations} that spread the data to be
communicated evenly across all MPI processes on the nodes.  In the best
case, such implementations could possibly have a $k$ fold speed-up of
the inter-node communication over implementations where only a single
MPI process per node communicates (as in traditional, hierarchical
implementations) on systems with $k$ communication lanes.  Whether
this can actually be accomplished, depends on specific system
(software) characteristics, and on the algorithms used. We discuss
algorithm design and modeling in the last section of the report.

We give full-lane implementations for \emph{all} the regular
(non-vector) MPI collectives \mpibcast, \mpigather, \mpiscatter,
\mpigather, \mpialltoall, \mpireduce, \mpiallreduce,
\mpireducescatterblock, \mpiscan and \mpiexscan that can possibly
exploit multiple lanes. These implementations are intended for MPI
communicators populating the compute nodes with the same number of MPI
processes, ranked consecutively. We call such communicators
\emph{regular}, but our concrete implementations actually work for any
communicator. Regular communicators are a common case on clustered
systems, since \mpicommworld is usually regular. Our complete code,
including our benchmark program is available via
\url{www.par.tuwien.ac.at/Downloads/TUWMPI/tuw_lanecoll.zip}.

Our implementations can be viewed as performance
guidelines~\cite{Traff16:autoguide,Traff10:selfcons} that formalize
expectations on the performance of the MPI collectives. A performance
guideline is typically an implementation of some MPI functionality,
\eg, \mpibcast, in terms of other, similar MPI functionality, \eg,
\mpiscatter followed by \mpiallgather. A good MPI library
implementation of the native \mpibcast could reasonably be expected to
perform at least as well as any such guideline implementation: If not,
\mpibcast could readily be replaced with the guideline
implementation. We give such guideline, \emph{mock-up
  implementations}~\cite{HunoldCarpenAmarie18} for regular,
consecutively ranked communicators that can exploit multi-lane
communication capabilities for all regular collectives. It is
important to note that these mock-ups are full-fledged, correct
implementations for the corresponding collectives, and can thus
readily be used to (auto)tune an MPI library that exhibits performance
defects by using the mock-up in the cases where it performs better
than a library native implementation.  In our implementations, we
furthermore use MPI derived datatypes to perform the necessary
reordering of data, thus our mock-ups are in almost all cases
\emph{zero-copy}~\cite{HoeflerGottlieb10,Traff14:alltoall,Traff14:gather}
in the sense that there are no explicit data movement operations
before or after the collective operations. The decomposition of the
regular communicator can be used also to give more traditional,
\emph{hierarchical} implementations of the collectives, and we do so
for all collectives except \mpialltoall and compare the performance of
these implementations against the full-lane mock-ups.

The remainder of the report is structured as follows. In
Section~\ref{sec:multilane}, two simple benchmarks are used to explore
whether multiple lanes can be exploited with MPI point-to-point and
collective communication, and whether multiple, physical lanes can
give the expected, proportional increase in cumulated communication
bandwidth. Section~\ref{sec:mockups} presents the MPI performance
guideline, zero-copy, full-lane implementations, contrast these
against hierarchical implementations, and analyze the implementations
under the most optimistic, best known assumptions on the component
collective operations. This analysis reveals bottlenecks and
limitations to the full-lane
implementations. Section~\ref{sec:evaluation} gives an experimental
comparison of the performance guideline implementations to the library
native collectives, in many cases showing substantial and unexpected
improvements. Section~\ref{sec:modelalgorithms} discusses a model for
multi-lane systems that can possible be used to guide the design of
better $k$-lane algorithms. As an example, we give a pipelined,
$k$-lane broadcast algorithm that is directly derived from a simple,
linear pipeline broadcast designed under the single-ported model. We
also discuss $k$-lane algorithms for non-pipelined tree algorithms.

We use the following notation throughout. The number of MPI processes
is $p$, the number of compute nodes $N$, and the number of MPI
processes per node $n$, such that $p=nN$. The number of (physical)
lanes is $k$. All benchmarks communicate data as integers,
\mpiint. The amount of data \emph{per process} is given as a count $c$
of such integers (following the conventions of MPI).

\section{Communication performance with multiple lanes}
\label{sec:multilane}

\begin{table}
  \caption{Systems (hardware and software) used for the experimental
    evaluation.  For more information on the VSC-3 system, see
    \protect\url{www.vsc3.vsc.ac.at}.}
  \label{tab:systems}
  \begin{center}
    \begin{tabular}{crrrccc}
      Name & $n$ & $N$ & $p$ & Processor & Interconnect & MPI library \\
      \toprule
      Hydra & 32 & 36 & 1152 & Intel Xeon Gold 6130, 2.1 GHz & OmniPath & \openmpiversion \\
      & & & & & & with \gccversion \\
      & & & & Dual socket & Dual (2-lane) & \intelmpiversion \\
      & & & & & & \mpichversion \\
      & & & & & & with \gccversion \\
      \midrule
      VSC-3 & 16 & 2020 & 32320 & Intel Xeon E5-2650v2, 2.6 GHz & InfiniBand & \intelmpiversion \\
      & & & & Dual socket & Intel QDR-80 & \\
      & & & & & Dual (2-lane) & \\
      \bottomrule
    \end{tabular}
  \end{center}
\end{table}

\begin{table}
  \caption{Lane pattern benchmark results on ``Hydra'' for increasing
    number of virtual lanes $k$ used for communicating the data (count
    $c$ \mpiint) per compute node. The MPI library used is \openmpiversion.}
  \label{tab:laneopenmpi}
  \begin{center}
    \begin{tabular}{rrrrrrr}
      $k$ & $n$ & $N$ & $p$ & $c$ & avg ($\mu s$) & min ($\mu s$) \\
\toprule
1 & 32 & 36 & 1152 & 1152 & 141.22 & 133.77 \\
2 & 32 & 36 & 1152 & 1152 & 110.63 & 104.37 \\
4 & 32 & 36 & 1152 & 1152 & 95.48 & 90.72 \\
8 & 32 & 36 & 1152 & 1152 & 89.66 & 83.32 \\
16 & 32 & 36 & 1152 & 1152 & 94.96 & 80.86 \\
32 & 32 & 36 & 1152 & 1152 & 97.89 & 85.18 \\
\midrule
1 & 32 & 36 & 1152 & 11520 & 697.84 & 644.32 \\
2 & 32 & 36 & 1152 & 11520 & 387.72 & 351.86 \\
4 & 32 & 36 & 1152 & 11520 & 324.32 & 304.06 \\
8 & 32 & 36 & 1152 & 11520 & 188.89 & 175.76 \\
16 & 32 & 36 & 1152 & 11520 & 165.29 & 150.06 \\
32 & 32 & 36 & 1152 & 11520 & 157.93 & 144.66 \\
\midrule
1 & 32 & 36 & 1152 & 115200 & 3377.29 & 3260.22 \\
2 & 32 & 36 & 1152 & 115200 & 2040.01 & 1923.35 \\
4 & 32 & 36 & 1152 & 115200 & 1842.07 & 1685.66 \\
8 & 32 & 36 & 1152 & 115200 & 1384.20 & 1283.65 \\
16 & 32 & 36 & 1152 & 115200 & 1441.00 & 1412.02 \\
32 & 32 & 36 & 1152 & 115200 & 1373.89 & 1331.35 \\
\midrule
1 & 32 & 36 & 1152 & 1152000 & 26025.29 & 24584.80 \\
2 & 32 & 36 & 1152 & 1152000 & 13760.84 & 13249.18 \\
4 & 32 & 36 & 1152 & 1152000 & 11191.25 & 10961.43 \\
8 & 32 & 36 & 1152 & 1152000 & 10518.30 & 10464.52 \\
16 & 32 & 36 & 1152 & 1152000 & 10272.04 & 10202.46 \\
32 & 32 & 36 & 1152 & 1152000 & 10646.15 & 10492.62 \\
\midrule
1 & 32 & 36 & 1152 & 11520000 & 203868.07 & 202095.87 \\
2 & 32 & 36 & 1152 & 11520000 & 105835.37 & 104168.32 \\
4 & 32 & 36 & 1152 & 11520000 & 101349.99 & 100660.73 \\
8 & 32 & 36 & 1152 & 11520000 & 101106.96 & 100766.08 \\
16 & 32 & 36 & 1152 & 11520000 & 100646.86 & 100264.96 \\
32 & 32 & 36 & 1152 & 11520000 & 101030.17 & 100591.41 \\
\bottomrule
    \end{tabular}
  \end{center}
\end{table}

\begin{table}
  \caption{Lane pattern benchmark results on VSC-3 benchmark for increasing
    number of virtual lanes $k$ used for communicating the data (count
    $c$ \mpiint) per compute node. The MPI library used is \intelmpiversion.}
  \label{tab:vsclane}
\begin{center}
\begin{tabular}{rrrrrrr}
  $k$ & $n$ & $N$ & $p$ & $c$ & avg ($\mu s$) & min ($\mu s$) \\
  \toprule
1 & 16 & 100 & 1600 & 16 & 191.13 & 124.93 \\
2 & 16 & 100 & 1600 & 16 & 162.71 & 126.84 \\
4 & 16 & 100 & 1600 & 16 & 199.29 & 133.04 \\
8 & 16 & 100 & 1600 & 16 & 181.17 & 119.92 \\
16 & 16 & 100 & 1600 & 16 & 188.98 & 123.98 \\
\midrule
1 & 16 & 100 & 1600 & 160 & 226.37 & 153.06 \\
2 & 16 & 100 & 1600 & 160 & 189.77 & 138.04 \\
4 & 16 & 100 & 1600 & 160 & 204.43 & 133.99 \\
8 & 16 & 100 & 1600 & 160 & 204.65 & 132.08 \\
16 & 16 & 100 & 1600 & 160 & 335.04 & 142.10 \\
\midrule
1 & 16 & 100 & 1600 & 1600 & 866.72 & 449.18 \\
2 & 16 & 100 & 1600 & 1600 & 691.91 & 370.98 \\
4 & 16 & 100 & 1600 & 1600 & 300.03 & 216.96 \\
8 & 16 & 100 & 1600 & 1600 & 240.02 & 190.02 \\
16 & 16 & 100 & 1600 & 1600 & 255.62 & 177.15 \\
\midrule
1 & 16 & 100 & 1600 & 16000 & 3509.08 & 2555.13 \\
2 & 16 & 100 & 1600 & 16000 & 2230.83 & 1792.91 \\
4 & 16 & 100 & 1600 & 16000 & 2278.26 & 1677.99 \\
8 & 16 & 100 & 1600 & 16000 & 1980.68 & 1499.89 \\
16 & 16 & 100 & 1600 & 16000 & 1864.47 & 1277.92 \\
\midrule
1 & 16 & 100 & 1600 & 160000 & 25752.80 & 18226.15 \\
2 & 16 & 100 & 1600 & 160000 & 17478.72 & 14142.99 \\
4 & 16 & 100 & 1600 & 160000 & 17273.14 & 15020.13 \\
8 & 16 & 100 & 1600 & 160000 & 15398.98 & 12808.08 \\
16 & 16 & 100 & 1600 & 160000 & 27259.58 & 12650.97 \\
\midrule
1 & 16 & 100 & 1600 & 1600000 & 240925.21 & 209810.02 \\
2 & 16 & 100 & 1600 & 1600000 & 132452.19 & 117342.00 \\
4 & 16 & 100 & 1600 & 1600000 & 151751.88 & 150255.92 \\
8 & 16 & 100 & 1600 & 1600000 & 132714.86 & 127789.97 \\
16 & 16 & 100 & 1600 & 1600000 & 134652.29 & 128103.97 \\
  \bottomrule
\end{tabular}
\end{center}
\end{table}

\begin{table}
  \caption{Multi-collective pattern benchmark results on ``Hydra'' for
    increasing number of virtual lanes $k$ used for communicating the
    data (count $c$ \mpiint) per lane. The collective function is
    \mpialltoall. The MPI library used is \openmpiversion.}
  \label{tab:multicollopenmpi}
  \begin{center}
    \begin{tabular}{rrrrrrr}
      $k$ & $n$ & $N$ & $p$ & $c$ & avg ($\mu s$) & min ($\mu s$) \\
\toprule
1 & 32 & 36 & 1152 & 1152 & 24.38 & 21.92 \\
2 & 32 & 36 & 1152 & 1152 & 24.37 & 21.80 \\
4 & 32 & 36 & 1152 & 1152 & 24.87 & 22.35 \\
8 & 32 & 36 & 1152 & 1152 & 26.60 & 23.11 \\
16 & 32 & 36 & 1152 & 1152 & 30.07 & 25.93 \\
32 & 32 & 36 & 1152 & 1152 & 40.13 & 34.59 \\
\midrule
1 & 32 & 36 & 1152 & 11520 & 40.66 & 36.41 \\
2 & 32 & 36 & 1152 & 11520 & 39.73 & 37.02 \\
4 & 32 & 36 & 1152 & 11520 & 42.38 & 38.62 \\
8 & 32 & 36 & 1152 & 11520 & 49.86 & 44.96 \\
16 & 32 & 36 & 1152 & 11520 & 74.92 & 63.81 \\
32 & 32 & 36 & 1152 & 11520 & 112.87 & 101.47 \\
\midrule
1 & 32 & 36 & 1152 & 115200 & 257.20 & 194.04 \\
2 & 32 & 36 & 1152 & 115200 & 261.65 & 236.02 \\
4 & 32 & 36 & 1152 & 115200 & 301.16 & 273.93 \\
8 & 32 & 36 & 1152 & 115200 & 381.70 & 356.11 \\
16 & 32 & 36 & 1152 & 115200 & 629.15 & 589.88 \\
32 & 32 & 36 & 1152 & 115200 & 938.57 & 897.69 \\
\midrule
1 & 32 & 36 & 1152 & 1152000 & 1060.70 & 1041.76 \\
2 & 32 & 36 & 1152 & 1152000 & 1074.68 & 1048.08 \\
4 & 32 & 36 & 1152 & 1152000 & 1364.49 & 1325.78 \\
8 & 32 & 36 & 1152 & 1152000 & 1989.84 & 1952.72 \\
16 & 32 & 36 & 1152 & 1152000 & 3574.45 & 3510.26 \\
32 & 32 & 36 & 1152 & 1152000 & 6875.56 & 6825.13 \\
\midrule
1 & 32 & 36 & 1152 & 11520000 & 6203.28 & 6102.96 \\
2 & 32 & 36 & 1152 & 11520000 & 6250.37 & 6164.13 \\
4 & 32 & 36 & 1152 & 11520000 & 9412.57 & 9209.53 \\
8 & 32 & 36 & 1152 & 11520000 & 16750.09 & 16561.37 \\
16 & 32 & 36 & 1152 & 11520000 & 32254.22 & 31886.48 \\
32 & 32 & 36 & 1152 & 11520000 & 63994.62 & 63532.38 \\
\bottomrule
    \end{tabular}
  \end{center}
\end{table}

\begin{table}
  \caption{Multi-collective pattern benchmark results on the VSC-3 system for
    increasing number of virtual lanes $k$ used for communicating the
    data (count $c$ \mpiint) per lane. The collective function is
    \mpialltoall. The MPI library used is \intelmpiversion.}
  \label{tab:multicollintel}
  \begin{center}
    \begin{tabular}{rrrrrrr}
      $k$ & $n$ & $N$ & $p$ & $c$ & avg ($\mu s$) & min ($\mu s$) \\
      \toprule
1 & 16 & 100 & 1600 & 16 & 69.31 & 48.88 \\
2 & 16 & 100 & 1600 & 16 & 60.99 & 48.88 \\
4 & 16 & 100 & 1600 & 16 & 68.27 & 50.07 \\
8 & 16 & 100 & 1600 & 16 & 68.08 & 51.98 \\
16 & 16 & 100 & 1600 & 16 & 68.14 & 51.98 \\
\midrule
1 & 16 & 100 & 1600 & 160 & 75.33 & 51.98 \\
2 & 16 & 100 & 1600 & 160 & 65.46 & 52.21 \\
4 & 16 & 100 & 1600 & 160 & 64.45 & 51.98 \\
8 & 16 & 100 & 1600 & 160 & 64.13 & 53.17 \\
16 & 16 & 100 & 1600 & 160 & 67.00 & 54.84 \\
\midrule
1 & 16 & 100 & 1600 & 1600 & 105.92 & 82.02 \\
2 & 16 & 100 & 1600 & 1600 & 106.94 & 82.97 \\
4 & 16 & 100 & 1600 & 1600 & 106.17 & 90.12 \\
8 & 16 & 100 & 1600 & 1600 & 138.32 & 119.92 \\
16 & 16 & 100 & 1600 & 1600 & 330.42 & 279.90 \\
\midrule
1 & 16 & 100 & 1600 & 16000 & 165.61 & 152.83 \\
2 & 16 & 100 & 1600 & 16000 & 172.73 & 155.93 \\
4 & 16 & 100 & 1600 & 16000 & 201.66 & 190.02 \\
8 & 16 & 100 & 1600 & 16000 & 374.63 & 320.20 \\
16 & 16 & 100 & 1600 & 16000 & 803.39 & 776.05 \\
\midrule
1 & 16 & 100 & 1600 & 160000 & 1212.14 & 1029.01 \\
2 & 16 & 100 & 1600 & 160000 & 1436.99 & 1271.96 \\
4 & 16 & 100 & 1600 & 160000 & 2160.90 & 2024.17 \\
8 & 16 & 100 & 1600 & 160000 & 3680.81 & 3510.95 \\
16 & 16 & 100 & 1600 & 160000 & 8654.56 & 8312.94 \\
\midrule
1 & 16 & 100 & 1600 & 1600000 & 7545.93 & 7040.02 \\
2 & 16 & 100 & 1600 & 1600000 & 12080.37 & 11605.02 \\
4 & 16 & 100 & 1600 & 1600000 & 20710.37 & 19948.96 \\
8 & 16 & 100 & 1600 & 1600000 & 37365.79 & 36098.00 \\
16 & 16 & 100 & 1600 & 1600000 & 89516.15 & 73323.01 \\
\bottomrule
    \end{tabular}
  \end{center}
\end{table}

To explore the potential performance benefits of communicating over
multiple lanes simultaneously, we use two different MPI benchmarks for
two different use-cases. We primarily rely on our small, Intel Skylake
dual-socket, dual-rail OmniPath ``Hydra'' cluster described in
Table~\ref{tab:systems}. The nodes of this cluster consists of two
16-core sockets, each with a direct OmniPath connection to a separate
network. We also experiment on the much larger Vienna Scientific
Cluster VSC-3, another Intel-InfiniBand, dual-rail system,
see~\url{vsc.ac.at}. Our hypothesis is that MPI processes residing on
different sockets can communicate independently and effectively use
the two independent OmniPath/InfiniBand networks to achieve twice as
high bandwidth as when only one process (or processes on assigned to a
single socket) is communicating. Communication latency per process
should stay the same.

The \emph{lane pattern benchmark} divides the calling processes into
sets of processes for each compute node, assuming that the number of
MPI processes $p$ is a multiple of the number of compute nodes $N$,
and each compute nodes hosts $n$ processes with $n=p/N$. The
communicator used in the call to the benchmark routine is not
explicitly split. Each compute node sends and receives a count of $c$
data elements (of type \mpiint). Sending and receiving is repeated
(without any barriers) a certain number of times (here
$\piperepetition$). This is done by the assumption that multiple lanes
will be used in pipelined algorithms, see
Section~\ref{sec:modelalgorithms}. A parameter $k$ for the number of
\emph{virtual lanes} determines how the $c$ data elements are sent and
received from the compute nodes: The count $c$ is divided evenly over
the $k$ first processes on each node, which then independently
communicate these $\floor{c/k}$ elements (with $c\bmod k$ extra
elements for the first process). The process with rank $i$ sends to
process $(i+n)\bmod p$ and receives from process $(i-n)\bmod p$ using
a blocking \mpisendrecv operation (we have also experimented with
other communication patterns over the nodes). We repeat this
experiment $\repetition$ times (disposing of the first few warmup
repetitions), each repetition separated by an \mpibarrier. The
completion time of an experiment is the completion time of the slowest
process, and we report both the average over the completion times for
all repetitions, as well as the minimum completion time seen.

The question here is how many times faster the $c$ elements per node can
be communicated when sent and received over $k$ virtual lanes per
node. Our assumption is that the processes are assigned to the nodes
in such a way that the available, physical lanes can be
active simultaneously. For this, the MPI processes are assigned
consecutively to the compute nodes, and are pinned alternatingly over
the two sockets.  In case there are $k'$ physical lanes, we expect a
$k'$ fold speed-up, when our parameter $k$ is chosen with $k\geq k'$,
and the MPI processes properly spread evenly over the lanes.

A possible drawback for interpreting the results of this benchmark, is
that the send-receive operations work on counts of $c=n/k$
elements. The decrease with increasing $k$ may for some element counts
lead to a change of send-receive protocol with possibly better
running times, out of proportion with the factor of $k$ virtual
lanes. The next benchmark is used to alleviate such effects.

The \emph{multi-collective benchmark} measures how many executions of
the same collective over the lanes (here \mpialltoall) can be
sustained concurrently at no extra cost in running time compared to
only one execution. The benchmark splits the calling communicator into
$n$ communicators, each one spanning $N$ compute nodes (see
Section~\ref{sec:mockups}). With $k$ virtual lanes, the $k$ first of
these \emph{lane communicators}, \texttt{lanecomm}, execute the
\mpialltoall collective with $c$ being the total number of elements
per MPI process. The measurement is repeated $\repetition$ times, and
running times collected as explained above for the lane pattern
benchmark. Our hypothesis for this benchmark is that a system with
$k'$ physical lanes can, for $k\geq k'$, sustain $k'$ concurrent
executions of the collective, that is, that the running time for $k$
concurrent executions is about $k/k'$ times the time for one execution
(preferably already with $k=k'$).

The lane pattern benchmark results for the ``Hydra'' cluster, see
Table~\ref{tab:systems}, for $N=36$ nodes with $n=1152$ processes with
varying $k$ and different $c$ (chosen such that all active processes
on the nodes communicate exactly the same count $c/k$) with
\openmpiversion are shown in Table~\ref{tab:laneopenmpi}. A result on
the larger VSC-3 system, see again Table~\ref{tab:systems}, with
$N=100$ and $n=16$ and \intelmpiversion is shown in
Table~\ref{tab:vsclane}. Additional results with the other MPI
libraries and with other count values can be found in
Appendix~\ref{app:openmpi}.

The results on the two systems are somewhat similar. For very small
data, there is almost no benefit from communication over multiple,
virtual lanes $k$, but no large latency degradation either. For $k=16$ and
$k=32$, the ``Hydra'' system does give a notable improvement, though. For larger
counts beyond $c=11520$, the small systems exhibits an improvement by
(almost) the expected factor of two when $k$ goes up from 1 to 2 (and all
the way up to 32). For $c=115200$ there is a local minimum with $k=8$ where
the improvement over $k=1$ is significantly more than a factor of 2.

The improvements on the VSC-3 system are generally less conspicuous,
and often require $k=16$ to reach the maximum factor, see for instance
the cases $c=160000, c=1600000$ , which is then in many cases more than the
expected factor of 2.

Together, the observations on both systems showing no extreme
penalty by letting all $n$ processes on the node communicate their
share $c/n$ of the data, motivate our \emph{full-lane
  implementations} of Section~\ref{sec:mockups} where all $n$
processes on the nodes are partaking in independent collective
operations.

Results with the multi-collective benchmark can be found in
Table~\ref{tab:multicollopenmpi} and Table~\ref{tab:multicollintel}
for the two systems. On the ``Hydra'' system, it is noteworthy that
for small counts $c=1152$, even $k=n$ concurrent executions of
\mpialltoall each with the same total message size $c$ can be
sustained at the same running time as only one execution. As the count
grow large(r), it seems clear that (more than) two concurrent
executions can be sustained: The running time is less than $k/k'$
times larger than the running time for one execution ($k=1$) even with
$k=k'$.  Only for the case with $c=115200$, the running time increases
from $k=1$ to $k=2$, but after $k=4$ is even less than $k/k'$ times
the running time with $k=1$. On the VSC-3 system, for the small counts
$c=16, c=160$, the system can sustain $k=16$ concurrent \mpialltoall
operations. As $c$ increases, the behavior is not as clear-cut as on
the ``Hydra'' system. There is an increase in running time from $k=1$
to $k=2$, but overall the running time with $k=16$ is less than a
factor of $16/2=8$ times the running time with only $k=1$ \mpialltoall
operation, except for the very large count $c=1600000$ where the
increase roughly matches the expected factor of $8$.

Also, these observations justifies the \emph{full-lane
  implementations} which employ $k=n$ concurrent collective operations
each on $c/n$ of the input data.

As a sanity check, we have also run these (and all the following)
benchmarks on a system with only a single Infiniband network. For this
system, there is little to no advantage shown in using multiple,
virtual lanes. These results can be found in
Appendix~\ref{app:singlelane}.

\section{Performance guideline implementations}
\label{sec:mockups}

In this section, we give simple, multi-lane implementations of the MPI
collectives in terms of other, similar collective operations. Our
assumption, which seems justified by the lane pattern and
multi-collective benchmark results of Section~\ref{sec:multilane}, is
that these implementations by construction will be able to exploit
possible multi-lane capabilities of the hardware, similarly to the way
the benchmark patterns seem to do.

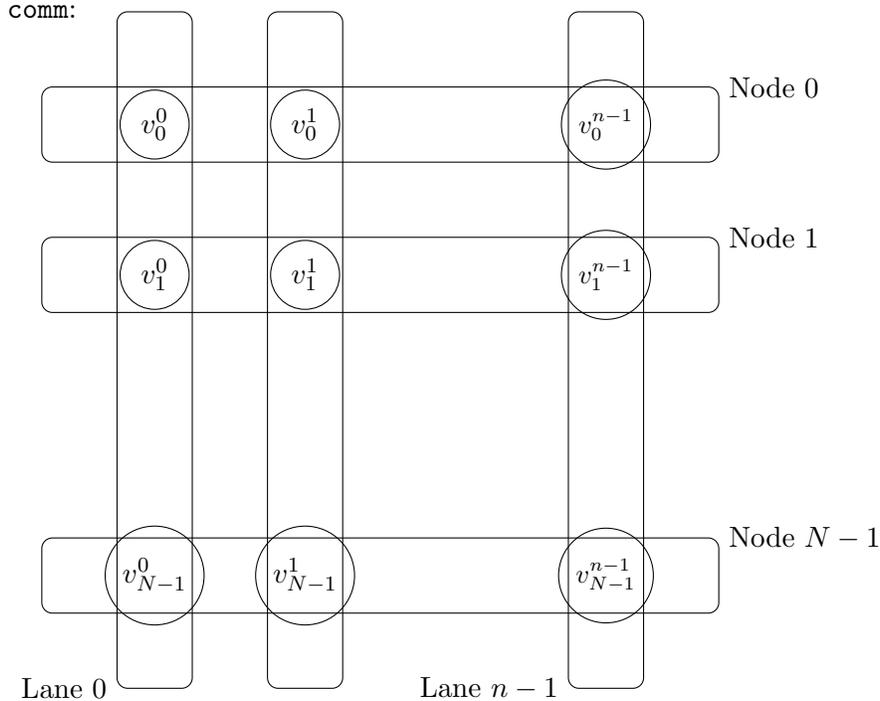
\begin{figure}
  \begin{center}
    \begin{tikzpicture}
      \draw node at (0,9) {\texttt{comm}:};
  \draw[rounded corners] (0,1) rectangle (9,2) node[right] {Node $N-1$};
  \draw[rounded corners] (0,5) rectangle (9,6) node[right] {Node $1$};
  \draw[rounded corners] (0,7) rectangle (9,8) node[right] {Node $0$};

  \draw[rounded corners] (1,0) node[left] {Lane $0$} rectangle (2,9);
  \draw[rounded corners] (3,0) rectangle (4,9);
  \draw[rounded corners] (7,0) node[left] {Lane $n-1$} rectangle (8,9);

  \node[shape=circle,draw=black] (v00) at (1.5,7.5) {$v_0^0$};
  \node[shape=circle,draw=black] (v01) at (3.5,7.5) {$v_0^1$};
  \node[shape=circle,draw=black] (v0n-1) at (7.5,7.5) {\small $v_0^{n-1}$};

  \node[shape=circle,draw=black] (v10) at (1.5,5.5) {$v_1^0$};
  \node[shape=circle,draw=black] (v11) at (3.5,5.5) {$v_1^1$};
  \node[shape=circle,draw=black] (v1n-1) at (7.5,5.5) {\small $v_1^{n-1}$};

  \node[shape=circle,draw=black] (vN-10) at (1.5,1.5) {$v_{N-1}^0$};
  \node[shape=circle,draw=black] (vN-11) at (3.5,1.5) {$v_{N-1}^1$};
  \node[shape=circle,draw=black] (vN-1n-1) at (7.5,1.5) {\small $v_{N-1}^{n-1}$};
\end{tikzpicture}
\end{center}
  \caption{The node and lane communicator decomposition of
    \texttt{comm} into disjoint \texttt{nodecomm} and
    \texttt{lanecomm} communicators as used in the full-lane
    collectives. Each MPI process $v^i_j$ belongs to one of each such
    communicator, and has rank $i$ in its \texttt{nodecomm} and rank
    $j$ in its \texttt{lanecomm}.}
\label{fig:comms}
\end{figure}

Our implementations assume \emph{regular communicators} where all
compute nodes host the same number of MPI processes. Furthermore, we
assume that the processes in the communicators are consecutively
ranked over the nodes. Let \texttt{comm} be such a regular,
consecutively ranked communicator.  The MPI functionality
\mpicommsplittype can be used to partition \texttt{comm} into disjoint
\emph{node communicators}, \texttt{nodecomm}, and we will assume the
given MPI library to split communicators non-trivially in this way,
see~\cite[Chapter 6]{MPI-3.1}.  Using \mpicommsplit (or
\mpicommcreate) it is likewise easy to partition \texttt{comm} into as
many disjoint \emph{lane communicators} \texttt{lanecomm} as there are
processes per node in \texttt{comm}. That is, we assume $n$ lanes
(number of processes in \texttt{nodecomm}) and $N$ processes in each
\texttt{lanecomm} communicator. In the code that follows
\texttt{noderank}, \texttt{nodesize}, and \texttt{lanerank},
\texttt{lanesize} will be the ranks and the number of processes in the
\texttt{nodecomm} and \texttt{lanecomm} communicators,
respectively. Note that we can check with a few allreduce operations
whether \texttt{comm} is actually regular; if not, we let
\texttt{lanecomm} be a duplicate of \texttt{comm} and
\texttt{nodecomm} just a self-communicator with one process. This way,
our implementations will work on any communicator \texttt{comm}.  The
communicator decomposition is shown in Figure~\ref{fig:comms}.  The
processes in the communicator \texttt{comm} are shown as nodes with
labels $v^i_j$, and the rank of each such process in \texttt{comm} is
by the regularity requirement $jn+i$. On the other hand, process
$v^i_j$ has rank $i$ in its \texttt{nodecomm} and rank $j$ in its
\texttt{lanecomm}.

Communicator splitting is done only once. The first time any of the
full-lane collectives is called on some communicator \texttt{comm},
the splitting into \texttt{lanecomm} and \texttt{nodecomm}
communicators is performed, and the two new communicators are cached
as MPI attributes with \texttt{comm}. Also, the communicators are
stored as static variables with each mock-up collective, so that even
attribute lookup time can be saved. This means that all the full-lane
collective implementations shown in the following can be used just as
the MPI collectives without the need for any special
initialization. At most the first call to a collective may take more
time than subsequent calls. We take this into account in the benchmark
results by not timing the first few invocations whenever we benchmark
a collective function.

The key idea of all the mock-up implementations is to divide the data
with element count $c$ evenly over the virtual lanes using a suitable
collective operation on the node(s), perform the collective operation
concurrently over the lanes, and finally put the pieces together using
again a suitable collective on the nodes. For all collectives
considered here, these decompositions are very similar to performance
guidelines often stated for collective
operations~\cite{Traff16:autoguide,Traff10:selfcons}.  We use MPI
user-defined datatypes to avoid explicit copying of data between
intermediate buffers.

Different mock-ups for verifying performance of traditionally
hierarchy sensitive collective implementations were discussed
in~\cite{Traff14:alltoall,Traff14:gather}, and have also been
implemented. For each collective operation, all data for the
collective operation are collected at only one process per node, which
then perform a collective operations over the nodes with only one
process per node and thus eliminate conflicting communication from
several processes per node. Finally, a collective operation on the
nodes may be needed.

\subsection{Broadcast}

\begin{lstlisting}[float=*,caption={The full-lane broadcast guideline implementation.},label=lst:lanebcast]
int Bcast_lane(void *buffer, int count, MPI_Datatype datatype, int root,
               MPI_Comm comm)
{
  rootnode = root/nodesize;
  noderoot = root%nodesize;
    
  block = count/nodesize;
  for (i=0; i<nodesize; i++) counts[i] = block;
  counts[nodesize-1] += count%nodesize;
  displs[0] = 0;
  for (i=1; i<nodesize; i++) displs[i] = displs[i-1]+counts[i-1];
  blockcount = counts[noderank];

  if (lanerank==rootnode) {
    void *recbuf =
      (noderank==noderoot) ? MPI_IN_PLACE : (char*)buffer+noderank*block*extent;
      
    MPI_Scatterv(buffer,counts,displs,datatype,
                 recbuf,blockcount,datatype,noderoot,nodecomm);
  }
  MPI_Bcast((char*)buffer+noderank*block*extent,blockcount,datatype,
            rootnode,lanecomm);
  MPI_Allgatherv(MPI_IN_PLACE,blockcount,datatype,
                 buffer,counts,displs,datatype,nodecomm);
  
  return MPI_SUCCESS;
}
\end{lstlisting}

Our full-lane broadcast implementation first divides the data $c$ to
be broadcast from the root process evenly over the processes on the
compute node hosting the root by an \mpiscatterv operation. Each
process on the root node, now responsible for $c/n$ data elements,
broadcasts its data on its lane communicator. Finally, all processes
perform an \mpiallgatherv operation on the node communicator to
assemble the full $c$ data elements. The mock-up is thus similar to
the \mpiscatter followed by \mpiallgather performance guideline for
\mpibcast, with just an \mpibcast on proportionally smaller data
inbetween.  Under the assumption that processes are consecutively
ranked over the nodes, the node hosting the broadcast root $r$ can be
found easily as $\floor{r/n}$ where $n$ is the number of processes per
node. Likewise, the node rank of the root in its \texttt{nodecomm} is
$r\bmod n$. The mock-up implementation is shown in
Listing~\ref{lst:lanebcast} with obvious declarations and
initializations left out. The irregular \mpiscatterv and
\mpiallgatherv operations are used to cater for the case where $c$ is
not divisible by $n$. If $n$ divides $c$, regular (non-vector)
\mpiscatter and \mpiallgather collectives can be used instead and
might (or might not) perform better. We benchmark both variants in
Section~\ref{sec:evaluation}.

The best possible performance of this \mpibcast mock-up can be
estimated as
follows~\cite{Bruck97,ChanHeimlichPurkayasthavandeGeijn07}. An optimal
\mpiscatter algorithm on \texttt{nodecomm}, assuming fully connected,
bidirectional send-receive communication
capabilities~\cite{BarNoyKipnisSchieber00}, takes $\ceiling{\log n}$
communication rounds, and communicates $\frac{n-1}{n}c$
data. Broadcast of the $c/n$ data on a \texttt{lanecomm} takes another
$\ceiling{\log N}$ communication rounds (again, assuming fully
connected communication), and the $c/n$ data are sent once. The final
\mpiallgather in the best case takes $\ceiling{\log n}$ communication
rounds, and sends and receives $\frac{n-1}{n}c$ data elements.  Thus
the total number of communication rounds is at most $2\ceiling{\log
  n}+\ceiling{\log N}\leq \ceiling{\log p}+1+\ceiling{\log n}$, which
is $1+\ceiling{\log n}$ rounds more than optimal. The total volume of
data sent or received by a process is
$2\frac{n-1}{n}c+c/n=2c-c/n$. The latter is almost a factor of two off
from what an optimal broadcast algorithm could do. However, the total
amount of data broadcast from or into a node is $n (c/n)=c$: The $c$
data elements are sent from the broadcast root node once in chunks
over the $n$ lane communicators. With $k$ physical lanes, the $n$
concurrent broadcast operations on the $n$ lane communicators could be
sped up by a factor of $k$ (as seen with the lane pattern
benchmark). The scatter and allgather operations on the node
communicators thus turn out a bottleneck with increasing $n$. The
\mpiallgatherv on the root node could be replaced by a simpler
collective operation, since the root process of course does not have
to gather data back, but such a restricted allgather-like collective
is not part of MPI.
  
Worth noticing is also that our implementations all make heavy use of
\mpiinplace. We assume that the MPI library is such that the block of
data of size $c/n$ that does not have to be scattered and allgathered
on the \texttt{nodecomm} communicators is actually not copied. The
implementations tacitly assume that all processes calling the
full-lane broadcast provide the same count $c$, otherwise the
\mpiscatterv might be incorrect. This condition is not checked, but is
the users responsibility. This is a limitation the might affect some
uses of broadcast. The full-lane implementations of reduce and
allreduce make the same assumption.

\begin{lstlisting}[float=*,caption={The hierarchical broadcast guideline implementation.},label=lst:hierbcast]
int Bcast_hier(void *buffer, int count, MPI_Datatype datatype, int root,
               MPI_Comm comm)
{
  rootnode = root/nodesize;
  noderoot = root%nodesize;
    
  if (noderank==noderoot) {
    MPI_Bcast(buffer,count,datatype,rootnode,lanecomm);
  }
  MPI_Bcast(buffer,count,datatype,noderoot,nodecomm);
  
  return MPI_SUCCESS;
}
\end{lstlisting}

A hierarchical broadcast using the partition of \texttt{comm} into
\texttt{nodecomm} and \texttt{lanecomm} is shown in
Listing~\ref{lst:hierbcast}. The lane corresponding to the rank of the
root on the nodes is chosen for a broadcast operation over the
nodes. Subsequently, all nodes complete the broadcast operation by
a node local broadcast on \texttt{nodecomm}. This implementation has
the important property that the $c$ data elements are broadcast once
over the nodes, with a best possible number of communication rounds of
$\ceiling{\log N}$, followed by the node internal broadcast of at most
$\ceiling{\log n}$ communication rounds. In total, this is at most one
round more than the best possible number of rounds $\ceiling{\log p}$.

\subsection{Gather and Scatter}

\begin{lstlisting}[float=*,caption={The full-lane gather guideline implementation.},label=lst:lanegather]
int Gather_lane(void *sendbuf, int sendcount, MPI_Datatype sendtype,
                void *recvbuf, int recvcount, MPI_Datatype recvtype, int root,
                MPI_Comm comm)
{
  rootnode = root/nodesize;
  noderoot = root%nodesize;

  if (lanerank==rootnode) {
    if (noderank==noderoot) {
      MPI_Datatype nt, nodetype, lt, lanetype;
      
      MPI_Type_get_extent(recvtype,&lb,&extent);

      MPI_Type_contiguous(recvcount,recvtype,&nt);
      MPI_Type_create_resized(nt,0,nodesize*recvcount*extent,&nodetype);
      MPI_Type_commit(&nodetype);

      MPI_Type_vector(lanesize,recvcount,nodesize*recvcount,recvtype,&lt);
      MPI_Type_create_resized(lt,0,recvcount*extent,&lanetype);
      MPI_Type_commit(&lanetype);

      MPI_Gather(sendbuf,sendcount,sendtype,
                (char*)recvbuf+noderank*recvcount*extent,1,nodetype,
                rootnode,lanecomm);
      MPI_Gather(MPI_IN_PLACE,lanesize*sendcount,sendtype,
                 recvbuf,1,lanetype,noderoot,nodecomm);
    } else {
      MPI_Gather(sendbuf,sendcount,sendtype,tempbuf,sendcount,sendtype,
                 rootnode,lanecomm);
      MPI_Gather(tempbuf,lanesize*sendcount,sendtype,
                 recvbuf,lanesize*recvcount,recvtype,noderoot,nodecomm);
    }
  } else {
    MPI_Gather(sendbuf,sendcount,sendtype,recvbuf,recvcount,recvtype,rootnode,
	       lanecomm);
  }
  
  return MPI_SUCCESS;
}
\end{lstlisting}

The gather and scatter full-lane implementations also work by dividing
the $p c$ data elements to be gathered at or scattered from the root
process into blocks for each \texttt{lanecomm} of $N c= p c/n$ data
elements, similarly to the decompositions proposed
in~\cite{KuhnemannRauberRunger06}. For the \mpiscatter mock-up
implementation, first an \mpiscatter of blocks of size $Nc$ is
performed on the root node. Second, each lane communicator scatters
$N-1$ blocks of size $c$ from the lane root at the root node.  The
total number of communication rounds under best-case assumptions is
$\ceiling{\log n}+\ceiling{\log N}\leq \ceiling{\log p}+1$ which is at
most one round off from optimal. The amount of data scattered is
$(n-1)Nc+(N-1)c=(p-1)c$ which is the same as a best possible scatter
algorithm~\cite{ChanHeimlichPurkayasthavandeGeijn07}. The total amount
of data leaving the root node is $n(N-1)c=(p-n)c$ which is also
optimal. The implementation and analysis of \mpigather is
similar. Thus for \mpigather and \mpiscatter, our mock-up
implementations can perform very close to optimal in terms of
communication rounds and data volume, provided that the operations on
\texttt{nodecomm} and \texttt{lanecomm} are optimally implemented.
However, the \mpiscatter/\mpigather on the root node are a bottleneck
to achieving a $k$-fold speed-up, since almost all data are
scattered/gathered on the root node.

The code for the gather mock-up is shown in
Listing~\ref{lst:lanegather}.  MPI derived datatypes are used at the
root process to receive blocks directly into the supplied receive
buffer in the process rank order of \texttt{comm}. The blocks gathered
on the lane communicators are from processes $i, i+n, i+2n, i+3n,
\ldots$ where $i$ is the process rank on the \texttt{nodecom}, and
must thus be placed with strides of $n$ in the receive buffer. The
non-root processes do all communication on consecutive buffers.

\begin{lstlisting}[float=*,caption={The hierarchical gather guideline implementation.},label=lst:hiergather]
int Gather_hier(void *sendbuf, int sendcount, MPI_Datatype sendtype,
                void *recvbuf, int recvcount, MPI_Datatype recvtype, int root,
                MPI_Comm comm)
{
  rootnode = root/nodesize;
  noderoot = root%nodesize;

  if (lanerank==rootnode) {
    MPI_Gather(sendbuf,sendcount,sendtype,
              (char*)recvbuf+lanerank*nodesize*recvcount*extent,
              recvcount,recvtype,noderoot,nodecomm);
  } else {
    MPI_Gather(sendbuf,sendcount,sendtype,
               tempbuf,sendcount,sendtype,noderoot,nodecomm);
  }
  
  if (noderank==noderoot) {
    if (lanerank==rootnode) {
      MPI_Gather(MPI_IN_PLACE,nodesize*sendcount,sendtype,
                 recvbuf,nodesize*recvcount,recvtype,
                 rootnode,lanecomm);
    } else {
      MPI_Gather(tempbuf,nodesize*sendcount,sendtype,
                 recvbuf,nodesize*recvtype,recvtype,
                 rootnode,lanecomm);
    }
  }
  
  return MPI_SUCCESS;
}
\end{lstlisting}

A hierarchical implementation for the gather operation is shown in
Listing~\ref{lst:hiergather}. All processes node locally gather data,
for processes that are not on the node of the gather root into a
temporary buffer.  After that, the lane of the node local roots is
chosen for a gather operation over the nodes.

\subsection{Allgather}

\begin{lstlisting}[float=*,caption={The full-lane allgather guideline implementation.},label=lst:laneallgather]
int Allgather_lane(void *sendbuf, int sendcount, MPI_Datatype sendtype,
                   void *recvbuf, int recvcount, MPI_Datatype recvtype,
                   MPI_Comm comm)
{
  MPI_Type_contiguous(recvcount,recvtype,&lt);
  MPI_Type_create_resized(lt,0,nodesize*recvcount*extent,&lanetype);
  MPI_Type_commit(&lanetype);
  
  MPI_Type_vector(lanesize,recvcount,nodesize*recvcount,recvtype,&nt);
  MPI_Type_create_resized(nt,0,recvcount*extent,&nodetype);
  MPI_Type_commit(&nodetype);

  if (sendbuf!=MPI_IN_PLACE) {
    MPI_Sendrecv(sendbuf,sendcount,sendtype,0,ALLGATHER,
                 (char*)recvbuf+rank*recvcount*extent,recvcount,recvtype,
                 0,ALLGATHER,MPI_COMM_SELF,MPI_STATUS_IGNORE);
  }
  
  MPI_Allgather(MPI_IN_PLACE,sendcount,sendtype,
                (char*)recvbuf+noderank*recvcount*extent,1,lanetype,
                lanecomm);
  MPI_Allgather(MPI_IN_PLACE,sendcount,sendtype,
                recvbuf,1,nodetype,nodecomm);
  
  return MPI_SUCCESS;
}
\end{lstlisting}

In the full-lane allgather implementation, all processes first
perform an \mpiallgather on their \texttt{lanecomm}, resulting in $Nc$
elements gathered per process. Then all processes on each node perform
an \mpiallgather over their \texttt{nodecomm}, resulting in $nNc=pc$
data elements gathered per process. With best possible implementations
of the component allgather operations, the number of communication
rounds is again at most $\ceiling{\log p}+1$ (at most one round off
from optimal~\cite{Bruck97}), and the number of data elements sent and
received by each process exactly $(N-1)c+(n-1)Nc=(p-1)c$ which is
optimal (all data, except the process' own block sent and received
once). The total amount of data communicated from and to a node is
$n(N-1)c=(p-n)c$, thus with $k$ physical lanes, a speed up of a factor
of $k$ is possible for the simultaneous \mpiallgather on
\texttt{lanecomm}.  Unfortunately, again the \mpiallgather on
\texttt{nodecomm} sends and receives $(n-1)Nc$ data elements, which
prevents $k$ fold speed-up with increasing $n$.

The implementation shown in Listing~\ref{lst:laneallgather} is
completely zero-copy, meaning no explicit data movements, and also
uses no intermediate buffer space. This is possible by the
regularity assumption for the communicator \texttt{comm}, by which the
blocks to be gathered on the lane communicators are spaced $nc$ elements
apart in the final receive buffer. Such strided data block layouts can
easily be expressed with MPI derived vector data types taking care to
set the datatype extents correctly such that the \mpiallgather
operations can tile the received data blocks.  This use of MPI derived
datatypes was also discussed in~\cite{Traff14:alltoall,Traff14:gather}
where it was noted that this implementation strategy cannot always be
used when the communicator is not regular and consecutively ranked.
Whether this is the best performing implementation will depend on the
way derived datatypes are handled by the MPI implementation (a
straightforward \mpiallgather implementation on datatypes may pack and
unpack the same data many times). We provide some discussion on this
in Section~\ref{sec:evaluation}.

\begin{lstlisting}[float=*,caption={The hierarchical allgather guideline implementation.},label=lst:hierallgather]
int Allgather_hier(void *sendbuf, int sendcount, MPI_Datatype sendtype,
		   void *recvbuf, int recvcount, MPI_Datatype recvtype,
		   MPI_Comm comm)
{
  if (sendbuf==MPI_IN_PLACE&&noderank!=0) {
    takebuf   = (char*)recvbuf+rank*recvcount*extent;
    takecount = recvcount;
    taketype  = recvtype;
  } else {
    takebuf   = sendbuf;
    takecount = sendcount;
    taketype  = sendtype;
  }
  
  MPI_Gather(takebuf,takecount,taketype,
	     (char*)recvbuf+lanerank*nodesize*recvcount*extent,
	     recvcount,recvtype,0,nodecomm);
  
  if (noderank==0) {
    MPI_Allgather(MPI_IN_PLACE,nodesize*recvcount,recvtype,
		  recvbuf,nodesize*recvcount,recvtype,lanecomm);
  }

  MPI_Bcast(recvbuf,size*recvcount,recvtype,0,nodecomm);
	     
  return MPI_SUCCESS;
}
\end{lstlisting}

A hierarchical allgather implementation is shown in
Listing~\ref{lst:hierallgather}, and consists in node local gather
operations to some chosen local root (here the process with rank 0 in
\texttt{nodecomm}), an allgather over the nodes using the
\texttt{lanecomm} of the node local roots, and finally node local
broadcast operations of the full gathered result. Also this
implementation is zero-copy, and actually does not need derived
datatypes (or any reorderings of data) to achieve this.

\subsection{Reduction}

\begin{lstlisting}[float=*,caption={The full-lane allreduce guideline implementation.},label=lst:laneallreduce]
int Allreduce_lane(void *sendbuf,
                   void *recvbuf, int count, MPI_Datatype datatype,
                   MPI_Op op, MPI_Comm comm)
{
  block = count/nodesize;
  
  for (i=0; i<nodesize; i++) counts[i] = block;
  counts[nodesize-1] += count%nodesize;
  displs[0] = 0;
  for (i=1; i<nodesize; i++) displs[i] = displs[i-1]+counts[i-1];
    
  MPI_Reduce_scatter(sendbuf,
                     (char*)recvbuf+noderank*block*extent,
                     counts,datatype,op,nodecomm);
  MPI_Allreduce(MPI_IN_PLACE,(char*)recvbuf+noderank*block*extent,
                counts[noderank],datatype,op,lanecomm);    
  MPI_Allgatherv(MPI_IN_PLACE,counts[noderank],datatype,
                 recvbuf,counts,displs,datatype,nodecomm);
  
  return MPI_SUCCESS;
}
\end{lstlisting}

The full-lane reduction implementations rely on the observation that
reduction can be performed as a reduce-scatter followed by a(n)
(all)gather operation (which can be phrased as yet another performance
guideline). The full-lane \mpiallreduce first performs an
\mpireducescatter on the \texttt{nodecomm} communicator followed by an
\mpiallreduce on the \texttt{lanecomm} over $c/n$ of the data
elements, finally followed by an \mpiallgatherv to get the final
result together. Since the reductions are not performed in rank order
(the lane communicator does not have the processes of \texttt{comm}
consecutively ranked), this works for commutative operators only. A
mock-up implementation for \mpiallreduce is shown in
Listing~\ref{lst:laneallreduce}. As was the case for the broadcast
mock-up, the irregular \mpireducescatter and \mpiallgatherv operations
can be replaced by their regular counterparts \mpireducescatterblock
and \mpiallgather when $c$ is divisible by $n$ which might perform
better.

Under best-case assumptions the implementation takes at most
$2(\ceiling{\log p}+1)$ communication rounds, with $2\frac{p-1}{p}c$
data elements being exchanged ($\frac{n-1}{n}c$ elements for
\mpireducescatterblock and \mpiallgather on \texttt{nodecomm}, and
$2\frac{N-1}{N}c/n$ for \mpiallreduce on \texttt{lanecomm}). This is
the same as best known allreduce algorithms. The bottleneck for
achieving a $k$ fold speed-up is again the collective operations on
\texttt{nodecomm}.

\begin{lstlisting}[float=*,caption={The full-lane, regular reduce-scatter guideline implementation.},label=lst:lanereducescatter]
int Reduce_scatter_block_lane(void *sendbuf,
                             void *recvbuf, int count, MPI_Datatype datatype,
                             MPI_Op op, MPI_Comm comm)
{
  MPI_Type_get_extent(datatype,&lb,&extent);

  MPI_Type_vector(lanesize,count,nodesize*count,datatype,&lt);
  MPI_Type_create_resized(lt,0,count*extent,&bt);
  MPI_Type_contiguous(nodesize,bt,&permtype);
  MPI_Type_commit(&permtype);
  
  if (sendbuf!=MPI_IN_PLACE) {
    MPI_Sendrecv(sendbuf,1,permtype,0,SELFCOPY,
                 permbuf,nodesize*lanesize*count,datatype,0,SELFCOPY,
                 MPI_COMM_SELF,MPI_STATUS_IGNORE);
  } else {
    MPI_Sendrecv(recvbuf,1,permtype,0,SELFCOPY,
                 permbuf,nodesize*lanesize*count,datatype,0,SELFCOPY,
                 MPI_COMM_SELF,MPI_STATUS_IGNORE);
  }

  MPI_Reduce_scatter_block(permbuf,tempbuf,lanesize*count,datatype,op,nodecomm);
  MPI_Reduce_scatter_block(tempbuf,recvbuf,count,datatype,op,lanecomm);

  return MPI_SUCCESS;
}
\end{lstlisting}

For \mpireduce, the \mpiallreduce on \texttt{lanecomm} is replaced by
an \mpireduce operation, and the final \mpiallgatherv by an
\mpigatherv operation. This could possibly be further improved by
replacing the \mpireducescatter on the root node by a final \mpigather
and local reductions on the root process at the root node; but we have
not tried this.  The regular \mpireducescatterblock collective, shown
in Listing~\ref{lst:lanereducescatter}, is particularly simple, and
consists of just an \mpireducescatterblock on the \texttt{nodecomm}
followed by an \mpireducescatterblock on the \texttt{lanecomm} on
smaller data. There is one subtlety to handle, though;
\mpireducescatterblock delivers consecutive blocks of $c$ elements
consecutively to the processes, but the processes for the
subsequent \mpireducescatterblock on the \texttt{lanecomm}
communicator are not consecutive in the original communicator. To
still be able to use \mpireducescatterblock as is, the blocks can be
permuted into the process order on \texttt{lanecomm} in advance,
similarly to the reordering trick in~\cite{Traff05:redscat}.

\begin{lstlisting}[float=*,caption={The hierarchical allreduce guideline implementation.},label=lst:hierallreduce]
int Allreduce_hier(void *sendbuf, 
                   void *recvbuf, int count, MPI_Datatype datatype,
                   MPI_Op op, MPI_Comm comm)
{
  takebuf = (sendbuf==MPI_IN_PLACE&&noderank!=0) ? recvbuf : sendbuf;
  
  MPI_Reduce(takebuf,recvbuf,count,datatype,op,0,nodecomm);
  
  if (noderank==0) {
    MPI_Allreduce(MPI_IN_PLACE,recvbuf,count,datatype,op,lanecomm);
  }

  MPI_Bcast(recvbuf,count,datatype,0,nodecomm);

  return MPI_SUCCESS;
}
\end{lstlisting}

\begin{lstlisting}[float=*,caption={The hierarchical, regular reduce-scatter guideline implementation.},label=lst:hierreducescatter]
int Reduce_scatter_block_hier(void *sendbuf,
                              void *recvbuf, int count, MPI_Datatype datatype,
                              MPI_Op op, MPI_Comm comm)
{
  if (noderank==0) {
    tempbuf = mpitalloc(size*count,datatype,&lb,&extent);
  }
  
  if (sendbuf==MPI_IN_PLACE) {
    MPI_Reduce(recvbuf,tempbuf,size*count,datatype,op,0,nodecomm);
  } else {
    MPI_Reduce(sendbuf,tempbuf,size*count,datatype,op,0,nodecomm);
  }

  if (noderank==0) {
    MPI_Reduce_scatter_block(MPI_IN_PLACE,tempbuf,nodesize*count,datatype,op,
                            lanecomm);
  }
  
  MPI_Scatter(tempbuf,count,datatype,
              recvbuf,count,datatype,0,nodecomm);

  return MPI_SUCCESS;
}
\end{lstlisting}

Hierarchical allreduce and reduce-scatter implementations are shown in
Listing~\ref{lst:hierallreduce} and in
Listing~\ref{lst:hierreducescatter}, and should be immediately
understandable. In particular, reduce-scatter is implemented as a
reduce operation (on the nodes), followed by a reduce-scatter (over
the nodes), followed by a scatter operation (on the nodes), similarly
to the implementation of reduce-scatter as a reduce operation followed
by scatter.

\subsection{Scan}

\begin{lstlisting}[float=*,caption={The full-lane scan guideline implementation.},label=lst:lanescan]
int Scan_lane(void *sendbuf, void *recvbuf, int count, MPI_Datatype datatype,
              MPI_Op op, MPI_Comm comm)
{
  block = count/nodesize;
  
  for (i=0; i<nodesize; i++) counts[i] = block;
  counts[nodesize-1] += count%nodesize;
  displs[0] = 0;
  for (i=1; i<nodesize; i++) displs[i] = displs[i-1]+counts[i-1];

  takebuf = (sendbuf==MPI_IN_PLACE) ? recvbuf : sendbuf;
  MPI_Reduce_scatter(takebuf,
                     (char*)tempbuf+noderank*block*extent,counts,datatype,
                     op,nodecomm);
  
  MPI_Exscan(MPI_IN_PLACE,
	     (char*)tempbuf+noderank*block*extent,counts[noderank],datatype,
	     op,lanecomm);

  MPI_Scan(sendbuf,recvbuf,count,datatype,op,nodecomm); // could be overlapped
  if (lanerank>0) {
    MPI_Allgatherv(MPI_IN_PLACE,counts[noderank],datatype,
                   tempbuf,counts,displs,datatype,nodecomm);
    
    MPI_Reduce_local(tempbuf,recvbuf,count,datatype,op);
  }

  return MPI_SUCCESS;
}
\end{lstlisting}

The \mpiscan and \mpiexscan operations can also be given full-lane
implementations as shown (for \mpiscan) in
Listing~\ref{lst:lanescan}. The input data of $c$ elements contributed
by the processes are reduced and split into partial results of roughly
$c/n$ elements per process. The processes on each \texttt{lanecomm}
then perform an exclusive scan over the locally reduced elements, and
collect the full $c$ element scan over the node results by an
allgather operation on \texttt{nodecomm}. To complete the scan per
process, a node local scan on the $c$ element vectors is performed, to
which each process has to add the result of the exclusive scan.

\begin{lstlisting}[float=*,caption={The hierarchical scan guideline implementation.},label=lst:hierscan]
int Scan_hier(void *sendbuf,
	      void *recvbuf, int count, MPI_Datatype datatype,
	      MPI_Op op, MPI_Comm comm)
{
  MPI_Scan(sendbuf,recvbuf,count,datatype,op,nodecomm);

  if (noderank==nodesize-1) {
    MPI_Exscan(recvbuf,tempbuf,count,datatype,op,lanecomm);
  }
  if (lanerank>0) {
    MPI_Bcast(tempbuf,count,datatype,nodesize-1,nodecomm);
    MPI_Reduce_local(tempbuf,recvbuf,count,datatype,op);
  }

  return MPI_SUCCESS;
}
\end{lstlisting}

The hierarchical Scan in Listing~\ref{lst:hierscan} performs a node
local scan, and uses the result on the last process on the nodes for
the exclusive scan operation over the nodes. The node global exclusive
scan result is broadcast locally on each node, and added to the local
scan for each process to give the result of the global scan. To see
that this is correct, observe that the global scan for process $i$ on
node $j$ with inputs $x_k$ for process $k=jn+i$ is computed as
\begin{displaymath}
  \sum_{j'=0}^{j-1}(\sum_{i'=0}^{n-1} x_{j'n+i'}) + \sum_{i'=0}^{i} x_{jn+i'}.
\end{displaymath}
where the first double sum term is the result of the exclusive scan
over the nodes.

\subsection{Alltoall}

\begin{lstlisting}[float=*,caption={The full-lane alltoall guideline implementation.},label=lst:lanealltoall]
int Alltoall_lane(void *sendbuf, int sendcount, MPI_Datatype sendtype,
                  void *recvbuf, int recvcount, MPI_Datatype recvtype,
                  MPI_Comm comm)
		  
{
  MPI_Type_vector(lanesize,recvcount,nodesize*recvcount,recvtype,&nt);
  MPI_Type_create_resized(nt,0,recvcount*extent,&nodetype);
  MPI_Type_commit(&nodetype);

  MPI_Alltoall(sendbuf,nodesize*sendcount,sendtype,
               tempbuf,nodesize*recvcount,recvtype,lanecomm);
  MPI_Alltoall(tempbuf,1,nodetype,recvbuf,1,nodetype,nodecomm);

  return MPI_SUCCESS;
}
\end{lstlisting}

Finally, also \mpialltoall can be given a full-lane, zero-copy
implementation using MPI derived datatypes by following the same idea
as the allgather mock-up.  The implementation performs first an
alltoall on \texttt{lanecomm}, and then an alltoall on
\texttt{nodecomm} (or the other way round, both variations would
work).  An implementation is shown in
Listing~\ref{lst:lanealltoall}. Note that here an intermediate buffer
seems to be necessary.

If we assume that linear round \mpialltoall implementations are used
over the \texttt{nodecomm} and \texttt{lanecomm} communicators, the
total amount of data elements sent and received per process is
$(N-1)nc+(n-1)Nc=2pc-(N+n)c$ and thus higher than for a direct
algorithm with $(p-1)c$ elements sent and received per
process. However, no indirect algorithm for alltoall can reach this
communication volume~\cite{Bruck97}. The advantage of this
decomposition of \mpialltoall is that the concurrent communication
over the lane communicators of $(N-1)nc$ data per lane can possibly be
sped up by a factor of $k$, but the \mpialltoall on \texttt{nodecomm}
remains a bottleneck. The same would hold if if the order
\mpiallgather on \texttt{nodecomm} and \texttt{lanecomm} is
interchanged.

We do not give a hierarchical implementation for the alltoall
collective, since this would require the selected local root processes
to first gather an excessive amount of data, namely $n$ times the
total data from each process on the node to all other processes. Such
an implementation was discussed in~\cite{Traff14:alltoall}.

\section{Experimental results}
\label{sec:evaluation}


We have benchmarked the mock-up implementations of the collectives
described in Section~\ref{sec:mockups}, on the small ``Hydra'' cluster
and the larger VSC-3 cluster, see Table~\ref{tab:systems}.  We have
benchmarked with different MPI libraries available on the cluster
(Intel MPI, mpich, OpenMPI), but report only for \openmpiversion here;
the results for the other MPI libraries are in the appendix. For all
libraries we have used the defult settings which should enable the
most efficient collective implementations for the system at hand. For
the \intelmpiversion we have tried setting \texttt{I\_MPI\_ADJUST}
environment variables, also the \texttt{I\_MPI\_COLL\_EXTERNAL}
variable, but without observing positive performance effects.

Our full-lane and hierarchical implementations are benchmarked as
performance guidelines against the native MPI implementations of the
corresponding collectives~\cite{Traff16:autoguide}. This reveals
differences between two different implementations of the same
functionality, the native and closed, and the open mock-ups of
Section~\ref{sec:mockups} that rely entirely on similar MPI
functionality on the smaller \texttt{nodecomm} and \texttt{lanecomm}
communicators. Results do not show whether differences are due to
different ways of exploiting the multi-lane capabilities of the
system, but points to defects in the native implementations in cases
where the mock-ups perform significantly better. And in cases where
the mock-ups do not perform as well as might be expected from the lane
pattern benchmark in Section~\ref{sec:multilane}, reasons could be
that component functionality (collectives and derived datatypes) does
not perform as well as could be expected. Measurements are done as
explained in Section~\ref{sec:multilane}, in particular we report the
best seen completion times over 100 measurements synchronized with
\mpibarrier, but only for one \texttt{mpirun}. Results can vary
somewhat over different \texttt{mpirun}s, as observed
in~\cite{HunoldCarpenamarie16}, but differences between full-lane
mock-up and native MPI library function are in most cases so large
that we disregard this experimental factor. On the VSC-3 system, the
variations between different runs with different allocations were
often very significant (sometimes by a factor of two), but in all
cases, the relative differences between mock-up and
full-lane/hierarchical implementation seem stable. Reported here are
the best results seen over a sample of (about five) runs.

\subsection{Broadcast}

\begin{table}
  \caption{Results for the native \mpibcast compared against the
    mock-up guideline implementations on the ``Hydra'' system. The MPI library
    used is \openmpiversion.}
  \label{tab:bcastopenmpi}
  \begin{center}
    \begin{tabular}{rrrrrr}
      \toprule
      \multicolumn{6}{c}{BcastLane} \\
      $n$ & $N$ & $p$ & $c$ & avg ($\mu s$) & min ($\mu s$) \\
      \midrule
      32 & 36 & 1152 & 1152 & 52.27 & 29.61 \\
      32 & 36 & 1152 & 11520 & 111.81 & 83.99 \\
      32 & 36 & 1152 & 115200 & 354.83 & 314.01 \\
      32 & 36 & 1152 & 1152000 & 3934.19 & 3798.20 \\
      32 & 36 & 1152 & 11520000 & 38805.02 & 38629.53 \\
      \midrule
      \multicolumn{6}{c}{BcastLane (irregular \texttt{nodecomm} collectives)} \\
      $n$ & $N$ & $p$ & $c$ & avg ($\mu s$) & min ($\mu s$) \\
      \midrule
      32 & 36 & 1152 & 1152 & 89.24 & 44.62 \\
      32 & 36 & 1152 & 11520 & 88.00 & 82.88 \\
      32 & 36 & 1152 & 115200 & 337.42 & 301.16 \\
      32 & 36 & 1152 & 1152000 & 3826.27 & 3670.76 \\
      32 & 36 & 1152 & 11520000 & 37757.87 & 37482.68 \\
      \midrule
      \multicolumn{6}{c}{BcastHier} \\
      $n$ & $N$ & $p$ & $c$ & avg ($\mu s$) & min ($\mu s$) \\
      \midrule
      32 & 36 & 1152 & 1152 & 45.78 & 30.44 \\
      32 & 36 & 1152 & 11520 & 140.03 & 132.81 \\
      32 & 36 & 1152 & 115200 & 1465.79 & 1364.15 \\
      32 & 36 & 1152 & 1152000 & 16339.26 & 13290.90 \\
      32 & 36 & 1152 & 11520000 & 53409.95 & 47903.34 \\
      \midrule
      \multicolumn{6}{c}{\mpibcast} \\
      $n$ & $N$ & $p$ & $c$ & avg ($\mu s$) & min ($\mu s$) \\
      \midrule
      32 & 36 & 1152 & 1152 & 41.98 & 34.32 \\
      32 & 36 & 1152 & 11520 & 145.00 & 130.11 \\
      32 & 36 & 1152 & 115200 & 8418.56 & 8093.52 \\
      32 & 36 & 1152 & 1152000 & 19178.54 & 18414.16 \\
      32 & 36 & 1152 & 11520000 & 120751.21 & 104535.94 \\
      \bottomrule
    \end{tabular}
  \end{center}
\end{table}

\begin{table}
  \caption{Results for native \mpibcast compared against the mock-up
    guideline implementations on the VSC-3 system. The MPI library used is
    \intelmpiversion.}
  \label{tab:bcastintelmpi}
  \begin{center}
    \begin{tabular}{rrrrrrr}
      \toprule
    \multicolumn{6}{c}{BcastLane} \\
    $n$ & $N$ & $p$ & $c$ & avg ($\mu s$) & min ($\mu s$) \\
    \midrule
16 & 100 & 1600 & 16 & 28.88 & 18.12 \\
16 & 100 & 1600 & 160 & 32.27 & 23.13 \\
16 & 100 & 1600 & 1600 & 48.69 & 41.01 \\
16 & 100 & 1600 & 16000 & 172.65 & 152.11 \\
16 & 100 & 1600 & 160000 & 1330.24 & 1236.92 \\
16 & 100 & 1600 & 1600000 & 15648.27 & 15357.02 \\
    \midrule
    \multicolumn{6}{c}{BcastHier} \\
    $n$ & $N$ & $p$ & $c$ & avg & min \\
    \midrule
16 & 100 & 1600 & 16 & 23.67 & 15.97 \\
16 & 100 & 1600 & 160 & 27.10 & 20.98 \\
16 & 100 & 1600 & 1600 & 70.52 & 61.04 \\
16 & 100 & 1600 & 16000 & 592.39 & 564.10 \\
16 & 100 & 1600 & 160000 & 2854.70 & 2669.10 \\
16 & 100 & 1600 & 1600000 & 20800.17 & 19978.05 \\
    \midrule
    \multicolumn{6}{c}{\mpibcast} \\
    $n$ & $N$ & $p$ & $c$ & avg & min \\
    \midrule
16 & 100 & 1600 & 16 & 21.36 & 15.02 \\
16 & 100 & 1600 & 160 & 24.40 & 19.07 \\
16 & 100 & 1600 & 1600 & 62.49 & 56.03 \\
16 & 100 & 1600 & 16000 & 535.58 & 522.85 \\
16 & 100 & 1600 & 160000 & 9867.32 & 9790.90 \\
16 & 100 & 1600 & 1600000 & 26829.20 & 22873.88 \\
    \bottomrule
    \end{tabular}
    \end{center}
\end{table}

We have benchmarked with $c$ divisible by $n$ and $N$ and ranging from
$1152$ to $11520000$ \mpiint on the ``Hydra'' system. The results with
the \openmpiversion library are shown in
Table~\ref{tab:bcastopenmpi}. We have tried two versions of the
full-lane broadcast mock-up, one as shown directly in
Listing~\ref{lst:lanebcast}, and a version that uses regular
\mpiallgather and \mpiscatter collectives for the cases where $n$
divides $c$. Even for the small $c=1152$ count, the mock-up
\texttt{Bcast\_lane} implementations is better than the native
\mpibcast operations, and as $c$ grows, becomes so by a factor of more
than 3. A particularly drastic result is for $c=115200$ where the
native \mpibcast is more than a factor of 20 off from the full-lane
mock-up; this points to a severe defect in the MPI library broadcast
implementation. Also the hierarchical broadcast of
Listing~\ref{lst:hierbcast} is better than the native \mpibcast, but
consistently worse than the full-lane implementation. It is
interesting to notice that the full-lane version that relies only on irregular
scatter and allgather operations seems to perform slightly better that
the version that applies specialized collectives for the divisible
case.  This contradicts our immediate expectation, and becomes more
prominent for the reduction collectives.

On the VSC-3 system, we have used $N=100$ nodes and counts $c$ ranging
from 16 to 1600000 and divisible by the $n=16$ processes per node.
From $c=1600$, the mock-up performs better than the native \mpibcast, for
$n=160000$ by a large factor of more than 7, indicating again a problem
with the broadcast implementation for the \intelmpiversion library.

\subsection{Gather and Scatter}

\begin{table}
  \caption{Results for the native \mpiscatter compared against the
    mock-up guideline implementations on the ``Hydra'' system.  The MPI library
    used is \openmpiversion.}
  \label{tab:scatteropenmpi}
  \begin{center}
    \begin{tabular}{rrrrrr}
      \toprule
      \multicolumn{6}{c}{ScatterLane} \\
      $n$ & $N$ & $p$ & $c$ & avg ($\mu s$) & min ($\mu s$) \\
      \midrule
32 & 36 & 1152 & 1 & 29.92 & 26.88 \\
32 & 36 & 1152 & 10 & 51.92 & 44.83 \\
32 & 36 & 1152 & 100 & 181.03 & 171.51 \\
32 & 36 & 1152 & 1000 & 1585.70 & 1556.09 \\
32 & 36 & 1152 & 10000 & 18520.08 & 18436.32 \\
      \midrule
      \multicolumn{6}{c}{ScatterHier} \\
      $n$ & $N$ & $p$ & $c$ & avg ($\mu s$) & min ($\mu s$) \\
      \midrule
32 & 36 & 1152 & 1 & 15.07 & 11.51 \\
32 & 36 & 1152 & 10 & 23.74 & 19.41 \\
32 & 36 & 1152 & 100 & 229.16 & 215.52 \\
32 & 36 & 1152 & 1000 & 949.85 & 943.94 \\
32 & 36 & 1152 & 10000 & 4589.65 & 4576.82 \\
      \midrule
      \multicolumn{6}{c}{\mpiscatter} \\
      $n$ & $N$ & $p$ & $c$ & avg ($\mu s$) & min ($\mu s$) \\
      \midrule
32 & 36 & 1152 & 1 & 28.01 & 25.23 \\
32 & 36 & 1152 & 10 & 51.77 & 45.18 \\
32 & 36 & 1152 & 100 & 499.95 & 491.64 \\
32 & 36 & 1152 & 1000 & 1067.63 & 1060.21 \\
32 & 36 & 1152 & 10000 & 12930.14 & 12874.43 \\
      \bottomrule
    \end{tabular}
  \end{center}
\end{table}

\begin{table}
  \caption{Results for the native \mpigather compared against the
    mock-up guideline implementations on the VSC-3 system.  The MPI library
    used is \intelmpiversion.}
  \label{tab:gatherintelmpi}
  \begin{center}
    \begin{tabular}{rrrrrr}
      \toprule
      \multicolumn{6}{c}{GatherLane} \\
      $n$ & $N$ & $p$ & $c$ & avg ($\mu s$) & min ($\mu s$) \\
      \midrule
16 & 100 & 1600 & 1 & 27.17 & 21.93 \\
16 & 100 & 1600 & 10 & 44.05 & 33.86 \\
16 & 100 & 1600 & 100 & 202.58 & 176.91 \\
16 & 100 & 1600 & 1000 & 2792.12 & 1883.03 \\
      \midrule
      \multicolumn{6}{c}{GatherHier} \\
      $n$ & $N$ & $p$ & $c$ & avg ($\mu s$) & min ($\mu s$) \\
      \midrule
16 & 100 & 1600 & 1 & 27.86 & 21.22 \\
16 & 100 & 1600 & 10 & 69.22 & 51.02 \\
16 & 100 & 1600 & 100 & 245.51 & 236.99 \\
16 & 100 & 1600 & 1000 & 14977.19 & 1910.21 \\
      \midrule
      \multicolumn{6}{c}{\mpigather} \\
      $n$ & $N$ & $p$ & $c$ & avg ($\mu s$) & min ($\mu s$) \\
      \midrule
16 & 100 & 1600 & 1 & 175.62 & 112.06 \\
16 & 100 & 1600 & 10 & 188.83 & 144.00 \\
16 & 100 & 1600 & 100 & 6930.01 & 4875.90 \\
16 & 100 & 1600 & 1000 & 11536.76 & 5889.89 \\
      \bottomrule
    \end{tabular}
  \end{center}
\end{table}

\begin{table}
  \caption{Results for the native \mpiscatter compared against the
    mock-up guideline implementations on the VSC-3 system.  The MPI library
    used is \intelmpiversion.}
  \label{tab:scatterintelmpi}
  \begin{center}
    \begin{tabular}{rrrrrr}
      \toprule
      \multicolumn{6}{c}{ScatterLane} \\
      $n$ & $N$ & $p$ & $c$ & avg ($\mu s$) & min ($\mu s$) \\
      \midrule
16 & 100 & 1600 & 1 & 37.21 & 29.09 \\
16 & 100 & 1600 & 10 & 72.12 & 47.92 \\
16 & 100 & 1600 & 100 & 315.86 & 241.04 \\
16 & 100 & 1600 & 1000 & 2197.12 & 2171.99 \\
      \midrule
      \multicolumn{6}{c}{ScatterHier} \\
      $n$ & $N$ & $p$ & $c$ & avg ($\mu s$) & min ($\mu s$) \\
      \midrule
16 & 100 & 1600 & 1 & 32.83 & 26.94 \\
16 & 100 & 1600 & 10 & 80.03 & 73.19 \\
16 & 100 & 1600 & 100 & 332.90 & 325.92 \\
16 & 100 & 1600 & 1000 & 2154.81 & 2144.81 \\
      \midrule
      \multicolumn{6}{c}{\mpiscatter} \\
      $n$ & $N$ & $p$ & $c$ & avg ($\mu s$) & min ($\mu s$) \\
      \midrule
16 & 100 & 1600 & 1 & 27.06 & 21.93 \\
16 & 100 & 1600 & 10 & 73.24 & 67.95 \\
16 & 100 & 1600 & 100 & 1023.24 & 987.77 \\
16 & 100 & 1600 & 1000 & 3129.94 & 3085.85 \\
      \bottomrule
    \end{tabular}
  \end{center}
\end{table}

Results for \mpiscatter on the in ``Hydra'' system are shown in
Table~\ref{tab:scatteropenmpi} for the \openmpiversion library; in
this library there seems to be a bug in \mpigather with derived
datatypes that prevent us to produce results for \openmpiversion. The
results show that the mock-up \texttt{Scatter\_lane} might be
preferable only for small counts. The reason could be that the
bottleneck \mpiscatter on the \texttt{nodecomm} takes almost as much
time as a full scatter on the whole system. In effect, the
decomposition into node and lane communicators does not lead to a good
multi-lane algorithm for gather and scatter like operations. See the
discussion in Section~\ref{sec:modelalgorithms}. The hierarchical
scatter erforms much better, outperforming the native \mpiscatter by a
factor of three for large counts.

On the VSC-3, with the \intelmpiversion library, we give the results
for both gather and scatter in Table~\ref{tab:gatherintelmpi} and
Table~\ref{tab:scatterintelmpi}. For scatter, the full-lane
implementation seems significantly better than the native
\mpiscatter. For the gather operations, it seems that \mpigather is
severely broken for small counts $c=1,10,100$.

\subsection{Allgather}

\begin{table}
  \caption{Results for native \mpiallgather compared against the mock-up
    guideline implementations on the ``Hydra'' system.  The MPI library used
    is \openmpiversion.}
\label{tab:allgatheropenmpi}
  \begin{center}
    \begin{tabular}{crrrrrrr}
      \toprule
    \multicolumn{6}{c}{AllgatherLane} \\
      $n$ & $N$ & $p$ & $c$ & avg ($\mu s$) & min ($\mu s$) \\
      \midrule
      32 & 36 & 1152 & 1 & 53.88 & 39.55 \\
      32 & 36 & 1152 & 10 & 86.18 & 76.97 \\
      32 & 36 & 1152 & 100 & 636.36 & 560.67 \\
      32 & 36 & 1152 & 1000 & 10777.78 & 10023.95 \\
      32 & 36 & 1152 & 10000 & 135983.90 & 134931.78 \\
      \midrule
      \multicolumn{6}{c}{AllgatherLane without derived datatypes} \\
      $n$ & $N$ & $p$ & $c$ & avg ($\mu s$) & min ($\mu s$) \\
      \midrule
      32 & 36 & 1152 & 1 & 38.54 & 32.77 \\
      32 & 36 & 1152 & 10 & 70.85 & 62.18 \\
      32 & 36 & 1152 & 100 & 1101.75 & 1077.21 \\
      32 & 36 & 1152 & 1000 & 15638.91 & 15479.23 \\
      32 & 36 & 1152 & 10000 & 119918.19 & 118698.01 \\
      \midrule
      \multicolumn{6}{c}{AllgatherHier} \\
      $n$ & $N$ & $p$ & $c$ & avg ($\mu s$) & min ($\mu s$) \\
      \midrule
      32 & 36 & 1152 & 1 & 44.01 & 36.30 \\
      32 & 36 & 1152 & 10 & 140.52 & 123.17 \\
      32 & 36 & 1152 & 100 & 1232.35 & 1189.94 \\
      32 & 36 & 1152 & 1000 & 10553.83 & 10280.65 \\
      32 & 36 & 1152 & 10000 & 48009.40 & 44854.14 \\
      \midrule
      \multicolumn{6}{c}{\mpiallgather} \\
      $n$ & $N$ & $p$ & $c$ & avg ($\mu s$) & min ($\mu s$) \\
      \midrule
      32 & 36 & 1152 & 1 & 36.50 & 29.18 \\
      32 & 36 & 1152 & 10 & 119.15 & 108.54 \\
      32 & 36 & 1152 & 100 & 2087.12 & 1959.38 \\
      32 & 36 & 1152 & 1000 & 7282.44 & 6963.72 \\
      32 & 36 & 1152 & 10000 & 42437.28 & 41982.33 \\
      \bottomrule
    \end{tabular}
  \end{center}
\end{table}

\begin{table}
  \caption{Results for native \mpiallgather compared against the mock-up
    guideline implementations on the VSC-3 system.  The MPI library used
    is \intelmpiversion.}
\label{tab:allgatherintelmpi}
  \begin{center}
    \begin{tabular}{crrrrrrr}
      \toprule
    \multicolumn{6}{c}{AllgatherLane} \\
      $n$ & $N$ & $p$ & $c$ & avg ($\mu s$) & min ($\mu s$) \\
      \midrule
16 & 100 & 1600 & 1 & 77.62 & 65.09 \\
16 & 100 & 1600 & 10 & 169.42 & 163.08 \\
16 & 100 & 1600 & 100 & 1105.40 & 1050.95 \\
16 & 100 & 1600 & 1000 & 11963.54 & 11735.92 \\
    \midrule
    \multicolumn{6}{c}{AllgatherHier} \\
      $n$ & $N$ & $p$ & $c$ & avg ($\mu s$) & min ($\mu s$) \\
    \midrule
16 & 100 & 1600 & 1 & 59.69 & 50.07 \\
16 & 100 & 1600 & 10 & 346.97 & 319.00 \\
16 & 100 & 1600 & 100 & 1439.46 & 1361.85 \\
16 & 100 & 1600 & 1000 & 15435.92 & 14804.12 \\    
      \midrule
      \multicolumn{6}{c}{\mpiallgather} \\
    $n$ & $N$ & $p$ & $c$ & avg ($\mu s$) & min ($\mu s$) \\
    \midrule
16 & 100 & 1600 & 1 & 139.66 & 131.13 \\
16 & 100 & 1600 & 10 & 1452.82 & 1435.04 \\
16 & 100 & 1600 & 100 & 3080.33 & 2983.09 \\
16 & 100 & 1600 & 1000 & 13955.36 & 13719.08 \\
    \bottomrule
    \end{tabular}
    \end{center}
\end{table}

Results for the allgather implementations are shown in
Table~\ref{tab:allgatheropenmpi} for the ``Hydra'' system. Here, the
native \mpiallgather is compared against the mock-up, full-lane
\texttt{Allgather\_lane} implementation, both in the zero-copy version
given in Listing~\ref{lst:laneallgather} and in a version that does
the allgather operations on \texttt{lanecomm} and \texttt{nodecomm} in
consecutive buffers and therefore does not use derived datatypes with
the collective operations (a final, process local reordering is
necessary here, and done with an \mpisendrecv self-copy with a derived
datatype).

For small element block counts up to $c=100$ (meaning that a total of
$pc=115\,200$ elements per process are gathered), the full-lane
mock-up performs better than the native \mpiallgather, for $c=100$ by
a considerable factor of more than 3. For the large counts,
\mpiallgather is considerably better than the mock-up, for $c=10000$
by a factor of almost 3. Derived datatypes that were convenient for
the formulation of the mock-up performance guideline in
Listing~\ref{lst:laneallgather}, perform surprisingly well. For the
smaller counts, the version with derived datatypes is better than the
version doing communication in consecutive buffers, and only for the
very large $c=10000$ there is an advantage to not using datatypes.
The reason for the disappointing performance of the mock-up could be
the high, relative cost of the \mpiallgather operation on the
\texttt{nodecomm}, and will be discussed in more detail in
Section~\ref{sec:modelalgorithms}. The hierarchical implementation of
Listing~\ref{lst:hierallgather} performs considerably better than the
full-lane implementations, despite having a higher node-local overhead
of two collective operations (gather and broadcast). The performce seems
close to the \mpiallgather performance for large $c$, which
could indicate the the \openmpiversion library uses such an
implementation internally.

The results for $c=1$ to $c=1000$ on the VSC-3 system can be seen in
Table~\ref{tab:allgatherintelmpi}. Here, the mock-up is in all cases
better than the \mpiallgather operation of \intelmpiversion, for
$c=100$ by a factor of almost 3.

\subsection{Alltoall}

\begin{table}
  \caption{Results for native \mpialltoall compared against the mock-up
    guideline implementation on the ``Hydra'' system.  The MPI library used
    is \openmpiversion.}
  \label{tab:alltoallopenmpi}
  \begin{center}
    \begin{tabular}{crrrrrrr}
      \toprule
    \multicolumn{6}{c}{AlltoallLane} \\
      $n$ & $N$ & $p$ & $c$ & avg ($\mu s$) & min ($\mu s$) \\
    \midrule
32 & 36 & 1152 & 1 & 126.54 & 109.42 \\
32 & 36 & 1152 & 10 & 286.16 & 224.32 \\
32 & 36 & 1152 & 100 & 1664.43 & 1575.91 \\
32 & 36 & 1152 & 1000 & 12820.35 & 12408.89 \\
32 & 36 & 1152 & 10000 & 149306.46 & 146599.05 \\
    \midrule
      \multicolumn{6}{c}{\mpialltoall} \\
    $n$ & $N$ & $p$ & $c$ & avg ($\mu s$) & min ($\mu s$) \\
    \midrule
32 & 36 & 1152 & 1 & 427.25 & 189.96 \\
32 & 36 & 1152 & 10 & 931.15 & 455.64 \\
32 & 36 & 1152 & 100 & 153715.13 & 4054.19 \\
32 & 36 & 1152 & 1000 & 12699.41 & 11991.91 \\
32 & 36 & 1152 & 10000 & 118932.43 & 116798.09 \\
    \bottomrule
    \end{tabular}
    \end{center}
\end{table}

\begin{table}
  \caption{Results for native \mpialltoall compared against the mock-up
    guideline implementation on the VSC-3 system.  The MPI library used
    is \intelmpiversion.}
  \label{tab:alltoallintelmpi}
  \begin{center}
    \begin{tabular}{crrrrrrr}
      \toprule
    \multicolumn{6}{c}{AlltoallLane} \\
      $n$ & $N$ & $p$ & $c$ & avg ($\mu s$) & min ($\mu s$) \\
    \midrule
16 & 100 & 1600 & 1 & 189.39 & 178.10 \\
16 & 100 & 1600 & 10 & 683.74 & 639.92 \\
16 & 100 & 1600 & 100 & 6644.50 & 6286.14 \\
16 & 100 & 1600 & 1000 & 80892.50 & 71113.11 \\
    \midrule
      \multicolumn{6}{c}{\mpialltoall} \\
    $n$ & $N$ & $p$ & $c$ & avg ($\mu s$) & min ($\mu s$) \\
    \midrule
16 & 100 & 1600 & 1 & 422.02 & 406.03 \\
16 & 100 & 1600 & 10 & 1787.76 & 1653.19 \\
16 & 100 & 1600 & 100 & 9290.99 & 7869.96 \\
16 & 100 & 1600 & 1000 & 67436.21 & 64065.93 \\
    \bottomrule
    \end{tabular}
    \end{center}
\end{table}

Despite being less attractive as a multi-lane algorithm, we have also
compared the mock-up, full-lane alltoall implementation against the
native \mpialltoall operation. Results from the full ``Hydra'' system
are shown in Table~\ref{tab:alltoallopenmpi}. For smaller block sizes
with $c$ up to 100, the full-lane implementation has a clear advantage
of more than a factor of 2 over the native \mpialltoall. The results
on VSC-3, shown in Table~\ref{tab:alltoallintelmpi}, are similar, but
the advantage of the mock-up implementation somewhat lesser.

\subsection{Reduce, Allreduce and Reduce-scatter}

\begin{table}
  \caption{Results for native \mpiallreduce compared against the mock-up
    guideline implementations on the ``Hydra'' system.  The MPI library used
    is \openmpiversion.}
  \label{tab:allreduceopenmpi}
  \begin{center}
    \begin{tabular}{crrrrrrr}
      \toprule
    \multicolumn{6}{c}{AllreduceLane} \\
      $n$ & $N$ & $p$ & $c$ & avg ($\mu s$) & min ($\mu s$) \\
    \midrule
32 & 36 & 1152 & 1152 & 73.14 & 62.24 \\
32 & 36 & 1152 & 11520 & 346.12 & 334.99 \\
32 & 36 & 1152 & 115200 & 2697.34 & 2550.35 \\
32 & 36 & 1152 & 1152000 & 11526.37 & 10829.64 \\
32 & 36 & 1152 & 11520000 & 207529.85 & 206294.47 \\
    \midrule
    \multicolumn{6}{c}{AllreduceLane (irregular \texttt{nodecomm} collectives)} \\
      $n$ & $N$ & $p$ & $c$ & avg ($\mu s$) & min ($\mu s$) \\
    \midrule
    32 & 36 & 1152 & 1152 & 50.36 & 38.89 \\
    32 & 36 & 1152 & 11520 & 137.58 & 120.25 \\
    32 & 36 & 1152 & 115200 & 1475.70 & 1296.44 \\
    32 & 36 & 1152 & 1152000 & 10409.37 & 10133.58 \\
    32 & 36 & 1152 & 11520000 & 139247.27 & 135422.53 \\
    \midrule
    \multicolumn{6}{c}{AllreduceHier} \\
      $n$ & $N$ & $p$ & $c$ & avg ($\mu s$) & min ($\mu s$) \\
    \midrule
    32 & 36 & 1152 & 1152 & 83.97 & 68.81 \\
    32 & 36 & 1152 & 11520 & 500.59 & 477.73 \\
    32 & 36 & 1152 & 115200 & 3935.85 & 3615.86 \\
    32 & 36 & 1152 & 1152000 & 20164.34 & 19732.57 \\
    32 & 36 & 1152 & 11520000 & 239983.19 & 235367.89 \\
    \midrule
    \multicolumn{6}{c}{\mpiallreduce} \\
    $n$ & $N$ & $p$ & $c$ & avg ($\mu s$) & min ($\mu s$) \\
    \midrule
    32 & 36 & 1152 & 1152 & 99.79 & 87.67 \\
    32 & 36 & 1152 & 11520 & 4383.72 & 3885.67 \\
    32 & 36 & 1152 & 115200 & 5657.04 & 5161.85 \\
    32 & 36 & 1152 & 1152000 & 19652.38 & 18973.40 \\
    32 & 36 & 1152 & 11520000 & 132091.10 & 130392.48 \\
    \bottomrule
    \end{tabular}
  \end{center}
\end{table}

\begin{table}
  \caption{Results for native \mpireduce compared against the mock-up
    guideline implementations on the ``Hydra'' system.  The MPI library used
    is \openmpiversion.}
  \label{tab:reduceopenmpi}
  \begin{center}
    \begin{tabular}{crrrrrrr}
      \toprule
    \multicolumn{6}{c}{ReduceLane} \\
      $n$ & $N$ & $p$ & $c$ & avg ($\mu s$) & min ($\mu s$) \\
    \midrule
    32 & 36 & 1152 & 1152 & 63.77 & 53.86 \\
    32 & 36 & 1152 & 11520 & 316.14 & 295.28 \\
    32 & 36 & 1152 & 115200 & 2453.98 & 2366.82 \\
    32 & 36 & 1152 & 1152000 & 10621.21 & 10266.04 \\
    32 & 36 & 1152 & 11520000 & 192453.14 & 191687.36 \\
    \midrule
    \multicolumn{6}{c}{ReduceLane (irregular \texttt{nodecomm} collectives)} \\
      $n$ & $N$ & $p$ & $c$ & avg ($\mu s$) & min ($\mu s$) \\
    \midrule
    32 & 36 & 1152 & 1152 & 65.62 & 54.57 \\
    32 & 36 & 1152 & 11520 & 112.72 & 98.22 \\
    32 & 36 & 1152 & 115200 & 1008.79 & 882.32 \\
    32 & 36 & 1152 & 1152000 & 8261.34 & 7856.84 \\
    32 & 36 & 1152 & 11520000 & 111196.88 & 110272.81 \\
    \midrule
    \multicolumn{6}{c}{ReduceHier} \\
      $n$ & $N$ & $p$ & $c$ & avg ($\mu s$) & min ($\mu s$) \\
    \midrule
    32 & 36 & 1152 & 1152 & 36.12 & 29.43 \\
    32 & 36 & 1152 & 11520 & 285.04 & 274.30 \\
    32 & 36 & 1152 & 115200 & 1992.18 & 1931.74 \\
    32 & 36 & 1152 & 1152000 & 7375.23 & 7279.59 \\
    32 & 36 & 1152 & 11520000 & 173289.99 & 172883.97 \\
    \midrule
    \multicolumn{6}{c}{\mpireduce} \\
    $n$ & $N$ & $p$ & $c$ & avg ($\mu s$) & min ($\mu s$) \\
    \midrule
    32 & 36 & 1152 & 1152 & 61.55 & 56.32 \\
    32 & 36 & 1152 & 11520 & 496.95 & 484.74 \\
    32 & 36 & 1152 & 115200 & 6011.26 & 4611.81 \\
    32 & 36 & 1152 & 1152000 & 23208.37 & 22400.12 \\
    32 & 36 & 1152 & 11520000 & 598133.52 & 597492.11 \\
    \bottomrule
    \end{tabular}
  \end{center}
\end{table}

\begin{table}
  \caption{Results for native \mpireducescatterblock compared against the mock-up
    guideline implementations on the ``Hydra'' system.  The MPI library used
    is \openmpiversion.}
    \label{tab:reducescatteropenmpi}
  \begin{center}
    \begin{tabular}{crrrrrrr}
      \toprule
    \multicolumn{6}{c}{ReduceScatterBlockLane} \\
      $n$ & $N$ & $p$ & $c$ & avg ($\mu s$) & min ($\mu s$) \\
    \midrule
    32 & 36 & 1152 & 1 & 74.41 & 65.33 \\
    32 & 36 & 1152 & 10 & 326.67 & 311.69 \\
    32 & 36 & 1152 & 100 & 2455.81 & 2399.48 \\
    32 & 36 & 1152 & 1000 & 17232.92 & 16694.82 \\
    32 & 36 & 1152 & 10000 & 237464.42 & 236207.82 \\
    \midrule
    \multicolumn{6}{c}{ReduceScatterBlockHier} \\
    $n$ & $N$ & $p$ & $c$ & avg ($\mu s$) & min ($\mu s$) \\
    \midrule
    32 & 36 & 1152 & 1 & 76.45 & 69.27 \\
    32 & 36 & 1152 & 10 & 526.51 & 514.50 \\
    32 & 36 & 1152 & 100 & 2901.77 & 2824.44 \\
    32 & 36 & 1152 & 1000 & 21851.54 & 21156.09 \\
    32 & 36 & 1152 & 10000 & 366696.44 & 359164.49 \\
    \midrule
    \multicolumn{6}{c}{\mpireducescatterblock} \\
    $n$ & $N$ & $p$ & $c$ & avg ($\mu s$) & min ($\mu s$) \\
    \midrule
    32 & 36 & 1152 & 1 & 92.01 & 84.68 \\
    32 & 36 & 1152 & 10 & 547.00 & 534.55 \\
    32 & 36 & 1152 & 100 & 7157.95 & 7094.17 \\
    32 & 36 & 1152 & 1000 & 25876.54 & 25215.88 \\
    32 & 36 & 1152 & 10000 & 619801.31 & 619055.31 \\
    \bottomrule
    \end{tabular}
    \end{center}
\end{table}

\begin{table}
  \caption{Results for native \mpiallreduce compared against the mock-up
    guideline implementations on the VSC-3 system.  The MPI library used
    is \intelmpiversion.}
  \label{tab:allreduceintelmpi}
  \begin{center}
    \begin{tabular}{crrrrrrr}
      \toprule
    \multicolumn{6}{c}{AllreduceLane} \\
      $n$ & $N$ & $p$ & $c$ & avg ($\mu s$) & min ($\mu s$) \\
    \midrule
16 & 100 & 1600 & 16 & 42.39 & 30.04 \\
16 & 100 & 1600 & 160 & 44.20 & 35.05 \\
16 & 100 & 1600 & 1600 & 634.45 & 612.97 \\
16 & 100 & 1600 & 16000 & 204.81 & 196.93 \\
16 & 100 & 1600 & 160000 & 1772.77 & 1724.96 \\
16 & 100 & 1600 & 1600000 & 22930.49 & 22562.03 \\
    \midrule
    \multicolumn{6}{c}{AllreduceHier} \\
      $n$ & $N$ & $p$ & $c$ & avg ($\mu s$) & min ($\mu s$) \\
    \midrule
16 & 100 & 1600 & 16 & 50.42 & 30.04 \\
16 & 100 & 1600 & 160 & 634.12 & 618.93 \\
16 & 100 & 1600 & 1600 & 96.28 & 83.92 \\
16 & 100 & 1600 & 16000 & 371.92 & 358.10 \\
16 & 100 & 1600 & 160000 & 3633.29 & 3542.90 \\
16 & 100 & 1600 & 1600000 & 47110.39 & 46608.92 \\
    \midrule
    \multicolumn{6}{c}{\mpiallreduce} \\
    $n$ & $N$ & $p$ & $c$ & avg ($\mu s$) & min ($\mu s$) \\
    \midrule
16 & 100 & 1600 & 16 & 40.89 & 26.94 \\
16 & 100 & 1600 & 160 & 5505.00 & 4719.02 \\
16 & 100 & 1600 & 1600 & 1287.48 & 1262.19 \\
16 & 100 & 1600 & 16000 & 269.14 & 234.13 \\
16 & 100 & 1600 & 160000 & 2438.08 & 2349.85 \\
16 & 100 & 1600 & 1600000 & 32067.79 & 31504.15 \\
    \bottomrule
    \end{tabular}
    \end{center}
\end{table}

\begin{table}
  \caption{Results for native \mpireduce compared against the mock-up
    guideline implementations on the VSC-3 system.  The MPI library used
    is \intelmpiversion.}
  \label{tab:reduceintelmpi}
  \begin{center}
    \begin{tabular}{crrrrrrr}
      \toprule
    \multicolumn{6}{c}{ReduceLane} \\
      $n$ & $N$ & $p$ & $c$ & avg ($\mu s$) & min ($\mu s$) \\
    \midrule
16 & 100 & 1600 & 16 & 31.80 & 20.03 \\
16 & 100 & 1600 & 160 & 32.45 & 23.13 \\
16 & 100 & 1600 & 1600 & 39.06 & 31.95 \\
16 & 100 & 1600 & 16000 & 133.36 & 124.93 \\
16 & 100 & 1600 & 160000 & 1045.61 & 992.06 \\
16 & 100 & 1600 & 1600000 & 12893.12 & 12050.87 \\
    \midrule
    \multicolumn{6}{c}{ReduceHier} \\
      $n$ & $N$ & $p$ & $c$ & avg ($\mu s$) & min ($\mu s$) \\
    \midrule
16 & 100 & 1600 & 16 & 12.35 & 5.01 \\
16 & 100 & 1600 & 160 & 10.10 & 5.01 \\
16 & 100 & 1600 & 1600 & 17.01 & 12.16 \\
16 & 100 & 1600 & 16000 & 79.54 & 63.18 \\
16 & 100 & 1600 & 160000 & 1061.79 & 1032.83 \\
16 & 100 & 1600 & 1600000 & 17230.79 & 16795.16 \\
    \midrule
    \multicolumn{6}{c}{\mpireduce} \\
    $n$ & $N$ & $p$ & $c$ & avg ($\mu s$) & min ($\mu s$) \\
    \midrule
16 & 100 & 1600 & 16 & 23.48 & 15.97 \\
16 & 100 & 1600 & 160 & 24.65 & 17.88 \\
16 & 100 & 1600 & 1600 & 54.16 & 45.06 \\
16 & 100 & 1600 & 16000 & 174.55 & 159.03 \\
16 & 100 & 1600 & 160000 & 1427.88 & 1389.03 \\
16 & 100 & 1600 & 1600000 & 26323.91 & 26043.89 \\
    \bottomrule
    \end{tabular}
    \end{center}
\end{table}

\begin{table}
  \caption{Results for native \mpireducescatterblock compared against the mock-up
    guideline implementations on the VSC-3 system.  The MPI library used
    is \intelmpiversion.}
    \label{tab:reducescatterintelmpi}
  \begin{center}
    \begin{tabular}{crrrrrrr}
      \toprule
    \multicolumn{6}{c}{ReduceScatterBlockLane} \\
      $n$ & $N$ & $p$ & $c$ & avg ($\mu s$) & min ($\mu s$) \\
    \midrule
16 & 100 & 1600 & 1 & 61.28 & 41.01 \\
16 & 100 & 1600 & 10 & 127.61 & 122.07 \\
16 & 100 & 1600 & 100 & 1146.39 & 1073.84 \\
16 & 100 & 1600 & 1000 & 17092.15 & 16785.14 \\
    \midrule
    \multicolumn{6}{c}{ReduceScatterBlockHier} \\
      $n$ & $N$ & $p$ & $c$ & avg ($\mu s$) & min ($\mu s$) \\
    \midrule
16 & 100 & 1600 & 1 & 63.66 & 54.84 \\
16 & 100 & 1600 & 10 & 228.72 & 218.15 \\
16 & 100 & 1600 & 100 & 1980.13 & 1878.98 \\
16 & 100 & 1600 & 1000 & 25677.48 & 25256.87 \\
    \midrule
    \multicolumn{6}{c}{\mpireducescatterblock} \\
    $n$ & $N$ & $p$ & $c$ & avg ($\mu s$) & min ($\mu s$) \\
    \midrule
16 & 100 & 1600 & 1 & 74.41 & 63.18 \\
16 & 100 & 1600 & 10 & 487.26 & 453.00 \\
16 & 100 & 1600 & 100 & 13680.96 & 10559.08 \\
16 & 100 & 1600 & 1000 & 73036.80 & 68859.10 \\
    \bottomrule
    \end{tabular}
    \end{center}
\end{table}

Results for the three reduction collectives \mpiallreduce, \mpireduce
and \mpireducescatterblock for the full ``Hydra'' system with
\openmpiversion are shown in Table~\ref{tab:allreduceopenmpi},
Table~\ref{tab:reduceopenmpi} and
Table~\ref{tab:reducescatteropenmpi}. We also benchmark a version that
does not use the regular \mpireducescatterblock and \mpiallgather
operations for the cases where $c$ is divisible by $n$, in order to
test the expectation that using a regular collective is usually better
than using the corresponding, irregular variant.

In most cases, the mock-up implementations perform significantly
better than the native MPI collectives, especially for the large
counts $c=11520000$, and for reduce and reduce-scatter by a large
factor. It should also be noticed that the versions that do not revert
to regular collectives perform surprisingly much better, especially
for the allreduce and reduce operations. For reduce, the hierarchical
implementation is the better, for allreduce and reduce-scatter the
full-lane implementation performs better. The performance of the
library native \mpireduce is particularly bad for large counts, by a
large factor worse than the performance of \mpiallreduce and thus
violating a straight-forward performance guideline.

For the VSC-3, results can be seen in
Table~\ref{tab:allreduceintelmpi}, Table~\ref{tab:reduceintelmpi}, and
Table~\ref{tab:reducescatterintelmpi}. Also here, the mock-ups are
consistently much better than the corresponding MPI library
implementations, which in many cases seem to perform erratically (see
for instance allreduce for $c=16000$ in
Table~\ref{tab:allreduceintelmpi}).

\subsection{Scan}

\begin{table}
  \caption{Results for native \mpiscan compared against the mock-up
    guideline implementations on the ``Hydra'' system.  The MPI library used
    is \openmpiversion.}
    \label{tab:scanopenmpi}
  \begin{center}
    \begin{tabular}{crrrrrrr}
      \toprule
    \multicolumn{6}{c}{ScanLane} \\
      $n$ & $N$ & $p$ & $c$ & avg ($\mu s$) & min ($\mu s$) \\
    \midrule
32 & 36 & 1152 & 1152 & 263.46 & 252.22 \\
32 & 36 & 1152 & 11520 & 1166.34 & 1140.59 \\
32 & 36 & 1152 & 115200 & 10422.08 & 10279.18 \\
32 & 36 & 1152 & 1152000 & 115656.08 & 114821.24 \\
32 & 36 & 1152 & 11520000 & 1222838.61 & 1219355.94 \\
    \midrule
    \multicolumn{6}{c}{ScanHier} \\
    $n$ & $N$ & $p$ & $c$ & avg ($\mu s$) & min ($\mu s$) \\
    \midrule
32 & 36 & 1152 & 1152 & 383.47 & 300.92 \\
32 & 36 & 1152 & 11520 & 2204.69 & 1602.06 \\
32 & 36 & 1152 & 115200 & 14037.24 & 13805.57 \\
32 & 36 & 1152 & 1152000 & 217251.30 & 215795.00 \\
32 & 36 & 1152 & 11520000 & 1796502.39 & 1790700.34 \\
    \midrule
    \multicolumn{6}{c}{\mpiscan} \\
    $n$ & $N$ & $p$ & $c$ & avg ($\mu s$) & min ($\mu s$) \\
    \midrule
32 & 36 & 1152 & 1152 & 5837.00 & 5771.48 \\
32 & 36 & 1152 & 11520 & 26613.09 & 26503.31 \\
32 & 36 & 1152 & 115200 & 240756.60 & 238336.21 \\
32 & 36 & 1152 & 1152000 & 2708744.33 & 2672371.42 \\
    \bottomrule
    \end{tabular}
    \end{center}
\end{table}

\begin{table}
  \caption{Results for native \mpiexscan compared against the mock-up
    guideline implementations on the ``Hydra'' system.  The MPI library used
    is \openmpiversion.}
    \label{tab:exscanopenmpi}
  \begin{center}
    \begin{tabular}{crrrrrrr}
      \toprule
    \multicolumn{6}{c}{ExscanLane} \\
      $n$ & $N$ & $p$ & $c$ & avg ($\mu s$) & min ($\mu s$) \\
    \midrule
32 & 36 & 1152 & 1152 & 268.43 & 252.55 \\
32 & 36 & 1152 & 11520 & 1178.12 & 1130.44 \\
32 & 36 & 1152 & 115200 & 10274.34 & 10069.46 \\
32 & 36 & 1152 & 1152000 & 109790.37 & 109028.02 \\
32 & 36 & 1152 & 11520000 & 1019522.97 & 1017324.47 \\
    \midrule
    \multicolumn{6}{c}{ExscanHier} \\
      $n$ & $N$ & $p$ & $c$ & avg ($\mu s$) & min ($\mu s$) \\
    \midrule
32 & 36 & 1152 & 1152 & 306.10 & 299.23 \\
32 & 36 & 1152 & 11520 & 1602.54 & 1587.75 \\
32 & 36 & 1152 & 115200 & 13978.31 & 13840.19 \\
32 & 36 & 1152 & 1152000 & 160991.80 & 158590.71 \\
32 & 36 & 1152 & 11520000 & 1444707.53 & 1442427.39 \\
    \midrule
    \multicolumn{6}{c}{\mpiexscan} \\
    $n$ & $N$ & $p$ & $c$ & avg ($\mu s$) & min ($\mu s$) \\
    \midrule
32 & 36 & 1152 & 1152 & 5752.68 & 5644.74 \\
32 & 36 & 1152 & 11520 & 26670.16 & 26464.64 \\
32 & 36 & 1152 & 115200 & 239297.18 & 237834.06 \\
32 & 36 & 1152 & 1152000 & 2796278.55 & 2750186.98 \\
    \bottomrule
    \end{tabular}
    \end{center}
\end{table}

\begin{table}
  \caption{Results for native \mpiscan compared against the mock-up
    guideline implementations on the VSC-3 system.  The MPI library used
    is \intelmpiversion.}
    \label{tab:scanintelmpi}
  \begin{center}
    \begin{tabular}{crrrrrrr}
      \toprule
    \multicolumn{6}{c}{ScanLane} \\
      $n$ & $N$ & $p$ & $c$ & avg ($\mu s$) & min ($\mu s$) \\
    \midrule
16 & 100 & 1600 & 16 & 44.81 & 32.19 \\
16 & 100 & 1600 & 160 & 51.57 & 37.19 \\
16 & 100 & 1600 & 1600 & 85.53 & 76.06 \\
16 & 100 & 1600 & 16000 & 577.95 & 551.94 \\
16 & 100 & 1600 & 160000 & 8413.32 & 7979.87 \\
16 & 100 & 1600 & 1600000 & 97133.36 & 95335.96 \\
    \midrule
    \multicolumn{6}{c}{ScanHier} \\
    $n$ & $N$ & $p$ & $c$ & avg ($\mu s$) & min ($\mu s$) \\
    \midrule
16 & 100 & 1600 & 16 & 37.51 & 28.85 \\
16 & 100 & 1600 & 160 & 43.47 & 35.05 \\
16 & 100 & 1600 & 1600 & 104.30 & 92.03 \\
16 & 100 & 1600 & 16000 & 657.97 & 638.01 \\
16 & 100 & 1600 & 160000 & 8936.49 & 8641.96 \\
16 & 100 & 1600 & 1600000 & 109818.02 & 108233.93 \\
    \midrule
    \multicolumn{6}{c}{\mpiscan} \\
    $n$ & $N$ & $p$ & $c$ & avg ($\mu s$) & min ($\mu s$) \\
    \midrule
16 & 100 & 1600 & 16 & 41.73 & 28.85 \\
16 & 100 & 1600 & 160 & 59.55 & 41.01 \\
16 & 100 & 1600 & 1600 & 309.67 & 292.06 \\
16 & 100 & 1600 & 16000 & 3225.26 & 2952.81 \\
16 & 100 & 1600 & 160000 & 33378.06 & 30730.01 \\
16 & 100 & 1600 & 1600000 & 351551.30 & 332144.98 \\
    \bottomrule
    \end{tabular}
    \end{center}
\end{table}

\begin{table}
  \caption{Results for native \mpiexscan compared against the mock-up
    guideline implementations on the VSC-3 system.  The MPI library used
    is \intelmpiversion.}
    \label{tab:exscanintelmpi}
  \begin{center}
    \begin{tabular}{crrrrrrr}
      \toprule
    \multicolumn{6}{c}{ExscanLane} \\
      $n$ & $N$ & $p$ & $c$ & avg ($\mu s$) & min ($\mu s$) \\
    \midrule
16 & 100 & 1600 & 16 & 45.31 & 32.19 \\
16 & 100 & 1600 & 160 & 50.32 & 38.86 \\
16 & 100 & 1600 & 1600 & 87.35 & 77.01 \\
16 & 100 & 1600 & 16000 & 573.19 & 548.84 \\
16 & 100 & 1600 & 160000 & 8291.58 & 7800.10 \\
16 & 100 & 1600 & 1600000 & 94399.87 & 92570.07 \\
    \midrule
    \multicolumn{6}{c}{ExscanHier} \\
      $n$ & $N$ & $p$ & $c$ & avg ($\mu s$) & min ($\mu s$) \\
    \midrule
16 & 100 & 1600 & 16 & 51.48 & 31.95 \\
16 & 100 & 1600 & 160 & 61.19 & 37.91 \\
16 & 100 & 1600 & 1600 & 108.81 & 96.08 \\
16 & 100 & 1600 & 16000 & 655.47 & 620.84 \\
16 & 100 & 1600 & 160000 & 8105.06 & 7892.85 \\
16 & 100 & 1600 & 1600000 & 99006.44 & 97747.80 \\
    \midrule
    \multicolumn{6}{c}{\mpiexscan} \\
    $n$ & $N$ & $p$ & $c$ & avg ($\mu s$) & min ($\mu s$) \\
    \midrule
16 & 100 & 1600 & 16 & 44.83 & 29.09 \\
16 & 100 & 1600 & 160 & 51.77 & 42.20 \\
16 & 100 & 1600 & 1600 & 308.95 & 290.16 \\
16 & 100 & 1600 & 16000 & 3211.54 & 2961.87 \\
16 & 100 & 1600 & 160000 & 32257.30 & 30708.07 \\
16 & 100 & 1600 & 1600000 & 341652.73 & 326359.99 \\
    \bottomrule
    \end{tabular}
    \end{center}
\end{table}

The scan results are shown in Table~\ref{tab:scanopenmpi} and
Table~\ref{tab:exscanopenmpi} for \openmpiversion, and in
Table~\ref{tab:scanintelmpi} and Table~\ref{tab:exscanintelmpi} for
\intelmpiversion. For \openmpiversion, both full-lane and hierarchical
implementations are better than \mpiscan, and by very a large factor;
in fact, the performance of \mpiscan was so poor that benchmarking for
the large count $c=11520000$ was not feasible. The full-lane
implementation is for all counts preferable. The results for the
exclusive scan operation are similar.

\subsection{Evaluation summary}

The results of comparing the mock-up guideline implementations against
the MPI library native collectives in many cases showed quite severe
violations of the performance guideline expectations, and often more than
can be accounted for by failing multi-lane utilization. Other MPI
libraries (mpich, Intel MPI) on the ``Hydra'' system show similar (but
quantitatively different) results, see the appendix.

\section{Designing multi-lane algorithms}
\label{sec:modelalgorithms}

\begin{table}
  \caption{Lane ($p=N$) versus node ($p=n$) case for \mpiallgather on the
  ``Hydra'' system with \openmpiversion.}
  \label{tab:lanevsnode}
\begin{center}
\begin{tabular}{rrrrrr}
  \toprule
  \multicolumn{6}{c}{Lane} \\
  $n$ & $N$ & $p$ & $c$ & avg ($\mu s$) & min ($\mu s$) \\
  \midrule
1 & 32 & 32 & 1 & 11.22 & 8.74 \\
1 & 32 & 32 & 10 & 11.90 & 9.05 \\
1 & 32 & 32 & 100 & 15.03 & 11.84 \\
1 & 32 & 32 & 1000 & 63.46 & 56.91 \\
1 & 32 & 32 & 10000 & 414.20 & 404.24 \\
1 & 32 & 32 & 100000 & 2070.53 & 1940.57 \\
  \midrule
  \multicolumn{6}{c}{Node} \\
  $n$ & $N$ & $p$ & $c$ & avg ($\mu s$) & min ($\mu s$) \\
  \midrule
32 & 1 & 32 & 1 & 6.10 & 4.75 \\
32 & 1 & 32 & 10 & 7.63 & 6.59 \\
32 & 1 & 32 & 100 & 23.88 & 21.26 \\
32 & 1 & 32 & 1000 & 244.26 & 233.32 \\
32 & 1 & 32 & 10000 & 1104.06 & 1067.18 \\
32 & 1 & 32 & 100000 & 9495.30 & 9377.90 \\
  \bottomrule
\hline
\end{tabular}
\end{center}
\end{table}

The full-lane implementations presented in Section~\ref{sec:mockups}
split the total data $c$ into $c/n$ elements handled via the lane
communicators, and $\frac{n-1}{n}c$ handled via the node
communicator. Thus, a $k$-fold speed-up is only possible for the part
of the collective handled over the lane communicators. This structure,
however, seems difficult to avoid, given that for the collectives
considered here, all processes have to send and/or receive
$\frac{p-1}{p}c$ data elements. For guaranteed speed-up proportional
to $k$, the part of the collective handled via the node communicators
must be proportionally faster. Experience shows that this can be very
far from the case on current systems. For example,
Table~\ref{tab:lanevsnode} compares the \mpiallgather operations on
$p=32$ processes, with either one process per node ($n=1,N=32$, lane
case) or 32 processes per node ($n=32,N=1$, node case) on the ``Hydra'' system
with the \openmpiversion MPI library, see Table~\ref{tab:systems}. Contrary to
expectation, the lane \mpiallgather over the network is, starting from
$c=1000$ three to four times faster than \mpiallgather on a single
node with shared-memory communication. Thus, the node communicator
part in the full-lane collectives can be a real performance
bottleneck. 

Our question is what is required from a multi-lane system in order to
make it possible to get a $k$ fold speed up with $k$ physical lanes
for the MPI collectives; and how algorithms must look, in particular
how lane and node communication can be traded off against each other
in the most efficient way?  We are looking for a performance and
design model for collective algorithms that can exploit multiple
communication lanes and would allow algorithms to exhibit a speed-up
of a factor of $k$ with $k$ physical lanes compared to the best known
bounds for algorithms designed under homogeneous, single-ported
communication capabilities.

The algorithm implicit in the \mpibcast mock-up was particularly
problematic, having an overhead of $\ceiling{\log n}$ communication
rounds, and $c-c/n$ too much data. This comes from the fixed
\mpiscatter and \mpiallgather overheads before and after the lane
communication by which the algorithm eventually communicates too much
data. But also the other algorithms had the problem that almost all
the data elements needed to be communicated on the compute nodes via
\texttt{nodecomm}.

\paragraph{Model:}
We propose using the following variant of a $k$-ported, bidirectional
communication model. Processors are organized into compute nodes, each
with the same number of processors $k$.  In one communication step,
each processor can send $c$ data elements to a processor on another
node and receive data from a processor on another, possibly different
node, \emph{and} at the same time send and receive (and perform a
computation on) $c$ data elements to $k-1$ processors on the same node
as the processor itself. We analyze algorithms by the number of such
communication steps taken, and by the amount of data communicated per
processor. We call this the \emph{$k$-lane model}.

We will define the model to be
\emph{self-simulating}~\cite{KruskalRudolphSnir90} in the sense that a
communication step of $n$ processes per compute node with $k$ lanes
can be run in $\ceiling{n/k}$ steps by serializing the accesses to the
$k$ physical communication lanes. If the $n$ processes are run on $n$
physical processors on the node, the number of steps will still be at
least $\ceiling{n/k}$ since the processors have to compete for the $k$
physical lanes (but communication inside the compute nodes can
possibly be sped up). This justifies our \emph{full-lane
  implementations} of Section~\ref{sec:mockups} that always utilize
$n$ virtual lanes.

The algorithms implied by the mock-ups all fail to exploit the model,
even in the favorable \mpiallgather and \mpiallreduce cases, by having
the node communication in separate steps from the lane
communication. Algorithms that are better in the $k$-lane model must
more closely intertwine node and lane communication to exploit the
overlap capability of the model.  Overlap can possibly be exploited by
pipelining, and we will examine pipelined algorithms further in the
following.

\paragraph{Tree algorithms:}
A communication tree algorithm for a collective operation like, \eg,
broadcast can be described by a communication graph (tree) with
processors assigned to the vertices, and the sequence of steps to be
performed by each vertex of the tree.  The algorithm is single-ported
if in each step at most one send and one receive operation is
performed with neighboring vertices. Cost of the algorithm is
accounted for by the number of steps taken by the processors assigned
to the vertices, assuming they work in synchronized rounds, and the
amount of data communicated between adjacent processors in each step.
We would have liked the following result: Any single-ported tree
algorithm running at cost $T_{\textrm{single}}(p,c)$ on $p$
processors (vertices) with input of size $c$ can be transformed into
an algorithm in the $k$-lane model running in time
$T_{\texttt{$k$-lane}}(p/k,c/k)$. We do not have such a result, but
something similar up to constant terms.

\paragraph{Construction:}
We now sketch a modular construction that takes any pipelined tree
(including a path: linear pipeline) algorithm running in the
single-ported model, and converts it into an algorithm in the $k$-lane
model that reduces the number of nodes in each tree by a factor of $k$
and increases the number of steps taken by only a small constant. The
construction is useful for pipelined tree algorithms (linear pipeline
and fixed-degree trees); in particular it applies also to algorithms
using two binary trees for broadcast and
reduction~\cite{Traff09:twotree} although this needs some care to work
out.

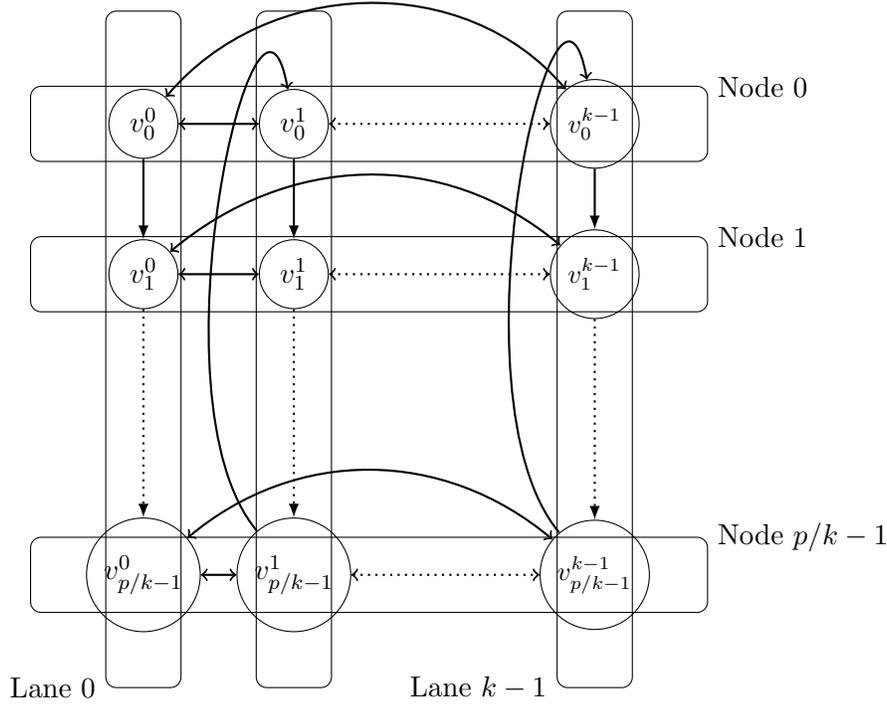
\begin{figure}
  \begin{center}
    \begin{tikzpicture}
  \draw[rounded corners] (0,1) rectangle (9,2) node[right] {Node $p/k-1$};
  \draw[rounded corners] (0,5) rectangle (9,6) node[right] {Node $1$};
  \draw[rounded corners] (0,7) rectangle (9,8) node[right] {Node $0$};

  \draw[rounded corners] (1,0) node[left] {Lane $0$} rectangle (2,9);
  \draw[rounded corners] (3,0) rectangle (4,9);
  \draw[rounded corners] (7,0) node[left] {Lane $k-1$} rectangle (8,9);

  \node[shape=circle,draw=black] (v00) at (1.5,7.5) {$v_0^0$};
  \node[shape=circle,draw=black] (v01) at (3.5,7.5) {$v_0^1$};
  \node[shape=circle,draw=black] (v0n-1) at (7.5,7.5) {\small $v_0^{k-1}$};

  \node[shape=circle,draw=black] (v10) at (1.5,5.5) {$v_1^0$};
  \node[shape=circle,draw=black] (v11) at (3.5,5.5) {$v_1^1$};
  \node[shape=circle,draw=black] (v1n-1) at (7.5,5.5) {\small $v_1^{k-1}$};

  \node[shape=circle,draw=black] (vN-10) at (1.5,1.5) {$v_{p/k-1}^0$};
  \node[shape=circle,draw=black] (vN-11) at (3.5,1.5) {$v_{p/k-1}^1$};
  \node[shape=circle,draw=black] (vN-1n-1) at (7.5,1.5) {\small $v_{p/k-1}^{k-1}$};

  \draw[-latex,thick] (v00) -- (v10);
  \draw[-latex,dotted,thick] (v10) -- (vN-10);

  \draw[-latex,thick] (v01) -- (v11);
  \draw[-latex,dotted,thick] (v11) -- (vN-11);

  \draw[-latex,thick] (v0n-1) -- (v1n-1);
  \draw[-latex,dotted,thick] (v1n-1) -- (vN-1n-1);

  \draw[<->,thick] (v00) -- (v01);
  \draw[<->,dotted,thick] (v01) -- (v0n-1);
  \draw[<->,thick] (v0n-1) to [out=130,in=50] (v00);

  \draw[<->,thick] (v10) -- (v11);
  \draw[<->,dotted,thick] (v11) -- (v1n-1);
  \draw[<->,thick] (v1n-1) to [out=140,in=40] (v10);

  \draw[<->,thick] (vN-10) -- (vN-11);
  \draw[<->,dotted,thick] (vN-11) -- (vN-1n-1);
  \draw[<->,thick] (vN-1n-1) to [out=140,in=40] (vN-10);

  \draw[->,thick] (vN-11) to [out=130,in=100] (v01);
  \draw[->,thick] (vN-1n-1) to [out=130,in=100] (v0n-1);
    \end{tikzpicture}
\end{center}
  \caption{The construction of the $k$-lane broadcast algorithm from a
    single-ported, linear pipeline. Each processor $v^i_j$ is connected to
    a successor on the next node $v^i_{j+1}$ (except for the last node), and
    is part of a $k$-clique inside the node. The last $k-1$ processors on
  the last node are connected to their counterparts on the first node.}
\label{fig:replica}
\end{figure}

Let $G=(V,E)$ be the communication tree of some tree algorithm for $p$
processors with $V=\{0\ldots,p-1\}$, and let $k$ be the number of
lanes. The communication graph for the $k$-lane algorithm is
constructed from a single-lane communication tree $G$ for $p/k$
processors, and replicates $G$ $k$ times into trees
$G^i,i=0,\ldots,k-1$. Each vertex $v$ of $G$ is replicated $k$ times,
and the $k$ replicates $v^0, v^1,\ldots,v^{k-1}$ are connected in a
$k$-clique. In the $k$-lane model, all replicas corresponding to $v$
are placed on the same compute node. Let $r\in V$ be the root of the
tree, and $r^0,r^1,\ldots,r^{k-1}$ be the replicas of $r$ with $r^0$
corresponding to $r$. In the replicas $G^i$ for $i=1,\ldots, k-1$,
there is an extra edge between the root $r^i$ and a leaf $v^i$. The
graph construction is illustrated in Figure~\ref{fig:replica}.

The $k$-lane algorithm for some vertex $v^i$ is derived
from the single-ported algorithm for vertex $v$, and is for all steps
almost identical. The idea is that each replica $G^i$ will handle only
$c/k$ of the data elements.  The root $r$ is special in either
initially being the only processor having data (broadcast) or finally
to receive a result (reduce).  We exemplify by transforming a
single-ported, linear pipeline for broadcast into a broadcast
algorithm in the $k$-lane model. As will become clear, this is a most
difficult case (binary tree broadcast for instance is easier).

Consider what the linear pipeline does in the steady state. In each
step in the single-ported algorithm, each non-root vertex $v$ receives
a new block of $C$ data elements from its predecessor, and sends the
block from the previous step to its successor. In the $k$-lane
algorithm, the replica vertices $v^i,i=0,\ldots,k-1$ receive a new
block of $C/k$ elements from their predecessor in the corresponding
$G^i$, and sends the previous block to their successor. In the same
step, each replica $v^i$ participates in an allgather operation in its
$k$-clique by receiving the previous blocks from all $v^j, j\neq i$
and sending its own previous block to all $v^j$. For the replicas of
the root node, a first special step is needed, since $r^i, i=1,\ldots
k-1$ do not have their $C/k$ data elements for their first round. In
this first, special step, the root $r^0$ which has all the $C$ data
elements, sends $C/k$ elements to each of the root replicas
$r^1,\ldots, r^{k-1}$. In the following, steady-state steps, the root
replicas receive a new block of data from $r^0$, and sends the
previous block to their successors. Also, all $r^i, i=1,\ldots,k-1$
exchange their previous block, as described above. Note that the root
cannot take part in this exchange, since it is already in the same
step engaged in sending the next block to each $r^i$. Therefore, the
vertices $r^i$ will miss the $C/k$ part of the data that are handled
by the replica $G^0$ of root $r^0$. But this missing part of the data
can be supplied by the leaves of $G^i$, which by the construction are
connected to $r^i$. When a leaf $v^i$ has received a block, it will in
the next step perform an exchange with the other leaves $v^j, j\neq
i$, and in particular receive the block from leaf $v^0$, the missing
block from $r^0$. This block can now be sent to $r^i$. Thus in the
steady state of the pipeline, vertices $v^i$ that correspond to the
leaf of $G$, are not really leaves, but both receive a previous block,
exchange, and send a block back to their root $r^i$.

All in all, this construction has the property that the linear
pipeline over $p$ processors with pipeline delay of $p-1$ steps,
is now transformed into $k$ pipelines with delay of $p/k-1+3=p/k+2$
steps. The extra steps come from the special first step for the root
vertices $r^i$ to get the pipelines started, the delay at the leaves
in waiting for the first exchange, and the extra communication step
back to the root. Each of the $k$ pipelines is responsible for only
$c/k$ of the total data elements.

In the $k$-lane model, the $k$ replicas of each vertex $v$ are placed
on the same compute node. Thus, for the linear pipeline broadcast
described above, in total $c$ data elements are sent from each node,
and $c$ data elements are received by each node, evenly distributed
over the $k$ lines (except for the root node, where only
$\frac{k-1}{k}c$ elements are received). Thus, the communication cost
can be improved by a factor of $k$ as desired using the $k$-lane
model.

Now, let $T_{\textrm{single}}(p,c) = (p-1)C+(c/C-1)C$ be the running
time (with additional constants and factors accounting more accurately
for communication latency and bandwidth) of the single-ported
pipelined broadcast algorithm with a pipeline block size of $C$
elements. The transformed, $k$-lane algorithm described will run in
$T_{\textrm{$k$-lane}}(p/k,c/k)=(p/k-1+3)C+(c/k/C-1)C$ steps which is
$T_{\textrm{$k$-lane}}(p/k,c/k)=T_{\textrm{single}}(p/k,c/k)+3=T_{\textrm{single}}(p/k,c/k)+O(1)$.

We claim the following general result from the construction.
\begin{proposition}
  Let $T_{\textrm{single}}(p,c)$ be the number of communication steps
  taken by a tree-based, pipelined, single-ported algorithm on $p$
  processors and $c$ data elements with a pipeline block size $C$. An
  algorithm in the $k$-lane model can be constructed which runs in
  $T_{\textrm{single}}(p/k,c/k)+O(1)$ communication steps.
\end{proposition}

A full proof of this claim follows by a more formal working out of the
description of the construction for the linear pipeline. For binary
trees, the construction is simpler, since the root processor will have
two steps in which to disseminate or collect its information to or
from the root replicas on the same node. The $O(1)$ constant will be
one smaller than for the linear pipeline case.

\paragraph{Non-pipelined tree algorithms:}
For non-pipelined tree algorithms as often used for broadcast (for
small problem sizes), gather and scatter, the construction does not
exploit the capability for the $k$ lanes to disseminate information to
$k$ new nodes efficiently. For such algorithms, we propose another
construction that makes use of more standard, $k$-ported
algorithms. We again exemplify with broadcast. Let $r$ be the root
processor. In the $k$-lane model, processor $r$ can in one step send
the data to a processor on another node and simultaneously to $k-1$
processors on the same node as $r$. We let $r$ send data first to the
$k-1$ non-root processors on the same node. Now, all $k$ processors on
the root node have the data, and can in the following steps send data
to $k$ other nodes. This is done using the pattern of the $k$-ported
algorithm, with each processor taking the role of one port, and
sending data to, say, the first processor of a new node. The first
time a processor on a non-root node receives data, it first sends the
data to all other processors on the node, which then continue as in
the $k$-ported algorithm. The number of steps of the $k$-lane
algorithm derived in this way from a $k$-ported algorithm is at most
twice that of the $k$-ported algorithm. It would be interesting to
implement this scheme for broadcast as well as the gather and scatter
operations. For gather and scatter, we believe that such an
implementation will perform better than the full-lane mock-up
implementation.

\section{Conclusion}

The two top ranked systems on the most recent Top500 list (June 2019,
see \url{www.top500.org}) both are dual-lane systems. We addressed the
question of whether MPI libraries for such systems efficiently exploit
multi-lane communication capabilities by proposing our full-lane
implementations for the MPI collectives, and more generally which
other capabilities and algorithm requirements must be posited in order
to get a $k$-fold speed-up with $k$ physical lanes for communication
operations like the MPI collectives. It would be fascinating to try
out the full-lane performance guidelines proposed in this report on
the large Top500 systems.

On small(er) HPC systems we demonstrated severe violations of the
full-lane performance guidelines for the regular MPI collectives,
indicating room for improvement that can possibly take better
advantage of multi-lane capabilities. Our implementations assumed
regular MPI communicators. It is an interesting question how
collective algorithms and implementations can look for the cases where
processes are not consecutively numbered and where compute nodes do
not carry the same number of MPI processes. Likewise, we did not
consider implementations for the irregular (vector) MPI
collectives. We suggested a model and presented a scheme for
constructing pipelined $k$-lane algorithms with $k$-fold speed-up, but
did not explore whether this will lead to practically better
implementations, nor whether there could be better trade-offs between
inter-node (lane) and intra-node (node) communication. All these
questions are highly interesting and relevant.

\bibliographystyle{plain}
\bibliography{traff,parallel}

\appendix
\section*{Appendix}

This appendix contains the benchmark results with the other MPI
libraries referred to in the main text, as well as experiments on an
older, single-rail cluster. The results confirm the findings of the
main text, but are quantitatively different, and sometimes reveal
effects so far not discussed. Interpretation is left for the reader
and for further research.

\section{Additional OpenMPI results}
\label{app:openmpi}

With the \openmpiversion library, we first show results with $c=1$
increasing to $c=10\,000\,000$ in multiples of $10$ and $n=32$. These
are listed in Tables~\ref{tab:lane.openmpi.n32.c1.A},
\ref{tab:lane.openmpi.n32.c1.B}, \ref{tab:multicoll.openmpi.n32.c1.A},
\ref{tab:multicoll.openmpi.n32.c1.B}, \ref{tab:bcast.openmpi.n32.c1},
\ref{tab:scatter.openmpi.n32.c1}, \ref{tab:allgather.openmpi.n32.c1},
\ref{tab:alltoall.openmpi.n32.c1}, \ref{tab:allreduce.openmpi.n32.c1},
\ref{tab:reduce.openmpi.n32.c1}, \ref{tab:reducescatter.openmpi.n32.c1},
\ref{tab:scan.openmpi.n32.c1} and~\ref{tab:exscan.openmpi.n32.c1}.

\begin{table}
\begin{center}
  \caption{Lane pattern benchmark results on ``Hydra'' for increasing
    number of virtual lanes $k$ used for communicating the data (count
    $c$ \mpiint) per compute node. The MPI library used is \openmpiversion.}
  \label{tab:lane.openmpi.n32.c1.A}
 \begin{tabular}{rrrrrrr}
      $k$ & $n$ & $N$ & $p$ & $c$ & avg ($\mu s$) & min ($\mu s$) \\
  \toprule
1 & 32 & 36 & 1152 & 1 & 65.92 & 62.66 \\
2 & 32 & 36 & 1152 & 1 & 65.65 & 62.81 \\
4 & 32 & 36 & 1152 & 1 & 66.37 & 62.84 \\
8 & 32 & 36 & 1152 & 1 & 66.36 & 62.94 \\
16 & 32 & 36 & 1152 & 1 & 67.37 & 62.80 \\
32 & 32 & 36 & 1152 & 1 & 73.27 & 65.68 \\
\midrule
1 & 32 & 36 & 1152 & 10 & 78.54 & 75.86 \\
2 & 32 & 36 & 1152 & 10 & 77.10 & 73.89 \\
4 & 32 & 36 & 1152 & 10 & 77.77 & 73.49 \\
8 & 32 & 36 & 1152 & 10 & 76.40 & 73.45 \\
16 & 32 & 36 & 1152 & 10 & 82.97 & 75.15 \\
32 & 32 & 36 & 1152 & 10 & 83.58 & 75.65 \\
\midrule
1 & 32 & 36 & 1152 & 100 & 83.56 & 78.96 \\
2 & 32 & 36 & 1152 & 100 & 81.11 & 76.31 \\
4 & 32 & 36 & 1152 & 100 & 81.08 & 76.85 \\
8 & 32 & 36 & 1152 & 100 & 82.42 & 76.64 \\
16 & 32 & 36 & 1152 & 100 & 80.50 & 75.85 \\
32 & 32 & 36 & 1152 & 100 & 85.67 & 80.14 \\
\midrule
1 & 32 & 36 & 1152 & 1000 & 131.30 & 125.67 \\
2 & 32 & 36 & 1152 & 1000 & 106.60 & 101.79 \\
4 & 32 & 36 & 1152 & 1000 & 94.43 & 89.43 \\
8 & 32 & 36 & 1152 & 1000 & 88.28 & 82.88 \\
16 & 32 & 36 & 1152 & 1000 & 89.55 & 80.07 \\
32 & 32 & 36 & 1152 & 1000 & 89.86 & 82.36 \\
\midrule
1 & 32 & 36 & 1152 & 10000 & 621.93 & 581.64 \\
2 & 32 & 36 & 1152 & 10000 & 356.62 & 318.08 \\
4 & 32 & 36 & 1152 & 10000 & 331.28 & 312.25 \\
8 & 32 & 36 & 1152 & 10000 & 176.13 & 162.81 \\
16 & 32 & 36 & 1152 & 10000 & 158.19 & 140.52 \\
32 & 32 & 36 & 1152 & 10000 & 146.60 & 130.26 \\
\midrule
1 & 32 & 36 & 1152 & 100000 & 3018.30 & 2845.26 \\
2 & 32 & 36 & 1152 & 100000 & 2021.81 & 1876.48 \\
4 & 32 & 36 & 1152 & 100000 & 1608.90 & 1516.95 \\
8 & 32 & 36 & 1152 & 100000 & 1277.15 & 1184.76 \\
16 & 32 & 36 & 1152 & 100000 & 1333.12 & 1307.83 \\
32 & 32 & 36 & 1152 & 100000 & 1290.69 & 1261.22 \\
\bottomrule
\end{tabular}
\end{center}
\end{table}

\begin{table}
\begin{center}
  \caption{Lane pattern benchmark results on ``Hydra'' for increasing
    number of virtual lanes $k$ used for communicating the data (count
    $c$ \mpiint) per compute node. The MPI library used is \openmpiversion.}
  \label{tab:lane.openmpi.n32.c1.B}
 \begin{tabular}{rrrrrrr}
      $k$ & $n$ & $N$ & $p$ & $c$ & avg ($\mu s$) & min ($\mu s$) \\
  \toprule
1 & 32 & 36 & 1152 & 1000000 & 23029.09 & 21726.68 \\
2 & 32 & 36 & 1152 & 1000000 & 12222.55 & 11589.04 \\
4 & 32 & 36 & 1152 & 1000000 & 9783.51 & 9630.14 \\
8 & 32 & 36 & 1152 & 1000000 & 9240.25 & 9138.92 \\
16 & 32 & 36 & 1152 & 1000000 & 8851.50 & 8785.14 \\
32 & 32 & 36 & 1152 & 1000000 & 9139.21 & 9004.22 \\
\midrule
1 & 32 & 36 & 1152 & 10000000 & 178161.50 & 175624.72 \\
2 & 32 & 36 & 1152 & 10000000 & 92741.40 & 91233.77 \\
4 & 32 & 36 & 1152 & 10000000 & 88634.77 & 87841.96 \\
8 & 32 & 36 & 1152 & 10000000 & 87965.27 & 87568.42 \\
16 & 32 & 36 & 1152 & 10000000 & 87497.30 & 87119.63 \\
32 & 32 & 36 & 1152 & 10000000 & 87978.13 & 87380.55 \\
\bottomrule
\end{tabular}
\end{center}
\end{table}

\begin{table}
  \caption{Multi-collective pattern benchmark results on ``Hydra'' for
    increasing number of virtual lanes $k$ used for communicating the
    data (count $c$ \mpiint) per lane. The collective function is
    \mpialltoall. The MPI library used is \openmpiversion.}
  \label{tab:multicoll.openmpi.n32.c1.A}
  \begin{center}
    \begin{tabular}{rrrrrrr}
      $k$ & $n$ & $N$ & $p$ & $c$ & avg ($\mu s$) & min ($\mu s$) \\
\toprule
1 & 32 & 36 & 1152 & 1 & 19.11 & 17.17 \\
2 & 32 & 36 & 1152 & 1 & 19.01 & 17.32 \\
4 & 32 & 36 & 1152 & 1 & 19.83 & 17.77 \\
8 & 32 & 36 & 1152 & 1 & 19.80 & 17.91 \\
16 & 32 & 36 & 1152 & 1 & 21.83 & 18.06 \\
32 & 32 & 36 & 1152 & 1 & 23.40 & 18.11 \\
\midrule
1 & 32 & 36 & 1152 & 10 & 18.71 & 17.22 \\
2 & 32 & 36 & 1152 & 10 & 18.71 & 17.28 \\
4 & 32 & 36 & 1152 & 10 & 19.11 & 17.68 \\
8 & 32 & 36 & 1152 & 10 & 19.18 & 17.70 \\
16 & 32 & 36 & 1152 & 10 & 23.48 & 17.91 \\
32 & 32 & 36 & 1152 & 10 & 25.83 & 18.01 \\
\midrule
1 & 32 & 36 & 1152 & 100 & 19.16 & 17.58 \\
2 & 32 & 36 & 1152 & 100 & 19.32 & 17.80 \\
4 & 32 & 36 & 1152 & 100 & 22.92 & 17.87 \\
8 & 32 & 36 & 1152 & 100 & 19.45 & 18.01 \\
16 & 32 & 36 & 1152 & 100 & 22.91 & 17.95 \\
32 & 32 & 36 & 1152 & 100 & 25.18 & 18.10 \\
\midrule
1 & 32 & 36 & 1152 & 1000 & 23.09 & 21.00 \\
2 & 32 & 36 & 1152 & 1000 & 23.54 & 21.23 \\
4 & 32 & 36 & 1152 & 1000 & 24.41 & 21.61 \\
8 & 32 & 36 & 1152 & 1000 & 29.01 & 22.25 \\
16 & 32 & 36 & 1152 & 1000 & 30.58 & 24.95 \\
32 & 32 & 36 & 1152 & 1000 & 36.59 & 32.15 \\
\midrule
1 & 32 & 36 & 1152 & 10000 & 39.61 & 36.83 \\
2 & 32 & 36 & 1152 & 10000 & 39.10 & 36.62 \\
4 & 32 & 36 & 1152 & 10000 & 41.95 & 37.86 \\
8 & 32 & 36 & 1152 & 10000 & 46.35 & 42.59 \\
16 & 32 & 36 & 1152 & 10000 & 67.94 & 60.46 \\
32 & 32 & 36 & 1152 & 10000 & 108.10 & 92.76 \\
\midrule
1 & 32 & 36 & 1152 & 100000 & 246.86 & 204.61 \\
2 & 32 & 36 & 1152 & 100000 & 251.29 & 226.00 \\
4 & 32 & 36 & 1152 & 100000 & 286.66 & 263.90 \\
8 & 32 & 36 & 1152 & 100000 & 358.35 & 342.99 \\
16 & 32 & 36 & 1152 & 100000 & 597.08 & 574.17 \\
32 & 32 & 36 & 1152 & 100000 & 934.63 & 837.98 \\
\bottomrule
    \end{tabular}
  \end{center}
\end{table}

\begin{table}
  \caption{Multi-collective pattern benchmark results on ``Hydra'' for
    increasing number of virtual lanes $k$ used for communicating the
    data (count $c$ \mpiint) per lane. The collective function is
    \mpialltoall. The MPI library used is \openmpiversion.}
  \label{tab:multicoll.openmpi.n32.c1.B}
  \begin{center}
    \begin{tabular}{rrrrrrr}
      $k$ & $n$ & $N$ & $p$ & $c$ & avg ($\mu s$) & min ($\mu s$) \\
\toprule
1 & 32 & 36 & 1152 & 1000000 & 1003.25 & 988.46 \\
2 & 32 & 36 & 1152 & 1000000 & 1020.33 & 995.54 \\
4 & 32 & 36 & 1152 & 1000000 & 1262.01 & 1229.66 \\
8 & 32 & 36 & 1152 & 1000000 & 1803.13 & 1764.90 \\
16 & 32 & 36 & 1152 & 1000000 & 3123.01 & 3066.27 \\
32 & 32 & 36 & 1152 & 1000000 & 6033.85 & 5982.96 \\
\midrule
1 & 32 & 36 & 1152 & 10000000 & 5628.15 & 5484.66 \\
2 & 32 & 36 & 1152 & 10000000 & 5670.64 & 5574.59 \\
4 & 32 & 36 & 1152 & 10000000 & 8396.63 & 8154.36 \\
8 & 32 & 36 & 1152 & 10000000 & 14815.66 & 14474.79 \\
16 & 32 & 36 & 1152 & 10000000 & 28365.53 & 27982.68 \\
32 & 32 & 36 & 1152 & 10000000 & 56336.55 & 55710.00 \\
\bottomrule
    \end{tabular}
  \end{center}
\end{table}

\begin{table}
  \caption{Results for native \mpibcast compared against the mock-up
    guideline implementations on the ``Hydra'' system.  The MPI library used
    is \openmpiversion.}
  \label{tab:bcast.openmpi.n32.c1}
  \begin{center}
    \begin{tabular}{crrrrrrr}
      \toprule
      \multicolumn{6}{c}{BcastLane} \\
      $n$ & $N$ & $p$ & $c$ & avg ($\mu s$) & min ($\mu s$) \\
      \midrule
32 & 36 & 1152 & 1 & 33.11 & 14.44 \\
32 & 36 & 1152 & 10 & 28.90 & 23.17 \\
32 & 36 & 1152 & 100 & 45.28 & 30.48 \\
32 & 36 & 1152 & 1000 & 78.04 & 46.03 \\
32 & 36 & 1152 & 10000 & 92.80 & 85.18 \\
32 & 36 & 1152 & 100000 & 313.21 & 279.70 \\
32 & 36 & 1152 & 1000000 & 3333.88 & 3255.67 \\
32 & 36 & 1152 & 10000000 & 34069.77 & 33854.16 \\
      \midrule
      \multicolumn{6}{c}{BcastHier} \\
      $n$ & $N$ & $p$ & $c$ & avg ($\mu s$) & min ($\mu s$) \\
      \midrule
32 & 36 & 1152 & 1 & 11.34 & 7.96 \\
32 & 36 & 1152 & 10 & 18.74 & 15.25 \\
32 & 36 & 1152 & 100 & 19.48 & 15.44 \\
32 & 36 & 1152 & 1000 & 44.71 & 27.22 \\
32 & 36 & 1152 & 10000 & 124.66 & 116.19 \\
32 & 36 & 1152 & 100000 & 1314.43 & 1188.10 \\
32 & 36 & 1152 & 1000000 & 11775.98 & 11430.09 \\
32 & 36 & 1152 & 10000000 & 43610.63 & 41302.93 \\
      \midrule
      \multicolumn{6}{c}{\mpibcast} \\
      $n$ & $N$ & $p$ & $c$ & avg ($\mu s$) & min ($\mu s$) \\
      \midrule
32 & 36 & 1152 & 1 & 14.61 & 8.17 \\
32 & 36 & 1152 & 10 & 18.02 & 13.33 \\
32 & 36 & 1152 & 100 & 20.87 & 15.38 \\
32 & 36 & 1152 & 1000 & 39.80 & 31.78 \\
32 & 36 & 1152 & 10000 & 123.98 & 116.90 \\
32 & 36 & 1152 & 100000 & 7736.69 & 7497.70 \\
32 & 36 & 1152 & 1000000 & 17625.11 & 17123.04 \\
32 & 36 & 1152 & 10000000 & 102445.25 & 91984.25 \\
      \bottomrule
    \end{tabular}
  \end{center}
\end{table}

\begin{table}
  \caption{Results for native \mpiscatter compared against the mock-up
    guideline implementations on the ``Hydra'' system.  The MPI library used
    is \openmpiversion.}
  \label{tab:scatter.openmpi.n32.c1}
  \begin{center}
    \begin{tabular}{crrrrrrr}
      \toprule
      \multicolumn{6}{c}{ScatterLane} \\
      $n$ & $N$ & $p$ & $c$ & avg ($\mu s$) & min ($\mu s$) \\
      \midrule
32 & 36 & 1152 & 1 & 28.79 & 25.89 \\
32 & 36 & 1152 & 9 & 49.50 & 43.83 \\
32 & 36 & 1152 & 87 & 162.52 & 155.45 \\
32 & 36 & 1152 & 869 & 1374.71 & 1349.95 \\
32 & 36 & 1152 & 8681 & 15813.94 & 15706.13 \\
      \midrule
      \multicolumn{6}{c}{ScatterHier} \\
      $n$ & $N$ & $p$ & $c$ & avg ($\mu s$) & min ($\mu s$) \\
      \midrule
32 & 36 & 1152 & 1 & 13.52 & 10.72 \\
32 & 36 & 1152 & 9 & 22.11 & 18.00 \\
32 & 36 & 1152 & 87 & 221.47 & 212.73 \\
32 & 36 & 1152 & 869 & 889.09 & 876.60 \\
32 & 36 & 1152 & 8681 & 4063.74 & 4049.34 \\
      \midrule
      \multicolumn{6}{c}{\mpiscatter} \\
      $n$ & $N$ & $p$ & $c$ & avg ($\mu s$) & min ($\mu s$) \\
      \midrule
32 & 36 & 1152 & 1 & 70.94 & 22.87 \\
32 & 36 & 1152 & 9 & 84.41 & 43.17 \\
32 & 36 & 1152 & 87 & 502.60 & 472.72 \\
32 & 36 & 1152 & 869 & 1017.95 & 967.59 \\
32 & 36 & 1152 & 8681 & 12645.31 & 12563.33 \\
      \bottomrule
    \end{tabular}
  \end{center}
\end{table}

\begin{table}
  \caption{Results for native \mpiallgather compared against the mock-up
    guideline implementations on the ``Hydra'' system.  The MPI library used
    is \openmpiversion.}
  \label{tab:allgather.openmpi.n32.c1}
  \begin{center}
    \begin{tabular}{crrrrrrr}
      \toprule
      \multicolumn{6}{c}{AllgatherLane} \\
      $n$ & $N$ & $p$ & $c$ & avg ($\mu s$) & min ($\mu s$) \\
      \midrule
32 & 36 & 1152 & 1 & 50.98 & 38.80 \\
32 & 36 & 1152 & 9 & 83.81 & 73.97 \\
32 & 36 & 1152 & 87 & 562.96 & 506.90 \\
32 & 36 & 1152 & 869 & 8579.20 & 8287.25 \\
32 & 36 & 1152 & 8681 & 120605.42 & 116649.86 \\
      \midrule
      \multicolumn{6}{c}{AllgatherHier} \\
      $n$ & $N$ & $p$ & $c$ & avg ($\mu s$) & min ($\mu s$) \\
      \midrule
32 & 36 & 1152 & 1 & 38.78 & 33.35 \\
32 & 36 & 1152 & 9 & 128.97 & 112.74 \\
32 & 36 & 1152 & 87 & 1070.70 & 1005.36 \\
32 & 36 & 1152 & 869 & 9106.83 & 8899.74 \\
32 & 36 & 1152 & 8681 & 41695.55 & 38108.44 \\
\midrule
      \multicolumn{6}{c}{\mpiallgather} \\
      $n$ & $N$ & $p$ & $c$ & avg ($\mu s$) & min ($\mu s$) \\
      \midrule
32 & 36 & 1152 & 1 & 39.21 & 26.82 \\
32 & 36 & 1152 & 9 & 105.43 & 97.97 \\
32 & 36 & 1152 & 87 & 1875.30 & 1808.54 \\
32 & 36 & 1152 & 869 & 6226.89 & 6109.79 \\
32 & 36 & 1152 & 8681 & 37704.78 & 37255.34 \\
      \bottomrule
    \end{tabular}
  \end{center}
\end{table}

\begin{table}
  \caption{Results for native \mpialltoall compared against the mock-up
    guideline implementation on the ``Hydra'' system.  The MPI library used
    is \openmpiversion.}
  \label{tab:alltoall.openmpi.n32.c1}
  \begin{center}
    \begin{tabular}{crrrrrrr}
      \toprule
      \multicolumn{6}{c}{AlltoallLane} \\
      $n$ & $N$ & $p$ & $c$ & avg ($\mu s$) & min ($\mu s$) \\
      \midrule
32 & 36 & 1152 & 1 & 118.51 & 109.80 \\
32 & 36 & 1152 & 9 & 269.35 & 214.85 \\
32 & 36 & 1152 & 87 & 1685.07 & 1403.78 \\
32 & 36 & 1152 & 869 & 11679.77 & 10782.07 \\
32 & 36 & 1152 & 8681 & 128363.61 & 125679.47 \\
      \midrule
      \multicolumn{6}{c}{\mpialltoall} \\
      $n$ & $N$ & $p$ & $c$ & avg ($\mu s$) & min ($\mu s$) \\
      \midrule
32 & 36 & 1152 & 1 & 209.31 & 199.08 \\
32 & 36 & 1152 & 9 & 472.29 & 436.10 \\
32 & 36 & 1152 & 87 & 35965.38 & 3607.93 \\
32 & 36 & 1152 & 869 & 10971.41 & 10517.80 \\
32 & 36 & 1152 & 8681 & 103127.94 & 101452.78 \\
      \bottomrule
    \end{tabular}
  \end{center}
\end{table}

\begin{table}
  \caption{Results for native \mpiallreduce compared against the mock-up
    guideline implementations on the ``Hydra'' system.  The MPI library used
    is \openmpiversion.}
  \label{tab:allreduce.openmpi.n32.c1}
  \begin{center}
    \begin{tabular}{crrrrrrr}
      \toprule
      \multicolumn{6}{c}{AllreduceLane} \\
      $n$ & $N$ & $p$ & $c$ & avg ($\mu s$) & min ($\mu s$) \\
      \midrule
32 & 36 & 1152 & 1 & 30.56 & 23.23 \\
32 & 36 & 1152 & 10 & 34.49 & 28.19 \\
32 & 36 & 1152 & 100 & 39.88 & 31.24 \\
32 & 36 & 1152 & 1000 & 48.63 & 39.82 \\
32 & 36 & 1152 & 10000 & 121.56 & 112.28 \\
32 & 36 & 1152 & 100000 & 2427.90 & 2244.68 \\
32 & 36 & 1152 & 1000000 & 9519.46 & 9252.62 \\
32 & 36 & 1152 & 10000000 & 190057.97 & 189271.82 \\
      \midrule
      \multicolumn{6}{c}{AllreduceHier} \\
      $n$ & $N$ & $p$ & $c$ & avg ($\mu s$) & min ($\mu s$) \\
      \midrule
      32 & 36 & 1152 & 1 & 33.32 & 17.73 \\
32 & 36 & 1152 & 10 & 25.90 & 22.37 \\
32 & 36 & 1152 & 100 & 42.68 & 24.74 \\
32 & 36 & 1152 & 1000 & 105.30 & 62.89 \\
32 & 36 & 1152 & 10000 & 456.63 & 433.97 \\
32 & 36 & 1152 & 100000 & 3482.99 & 3209.94 \\
32 & 36 & 1152 & 1000000 & 19292.62 & 17478.02 \\
32 & 36 & 1152 & 10000000 & 192041.39 & 186884.34 \\
      \midrule
      \multicolumn{6}{c}{\mpiallreduce} \\
      $n$ & $N$ & $p$ & $c$ & avg ($\mu s$) & min ($\mu s$) \\
      \midrule
32 & 36 & 1152 & 1 & 23.93 & 17.21 \\
32 & 36 & 1152 & 10 & 33.00 & 20.15 \\
32 & 36 & 1152 & 100 & 36.03 & 24.80 \\
32 & 36 & 1152 & 1000 & 85.04 & 76.39 \\
32 & 36 & 1152 & 10000 & 4477.71 & 3570.53 \\
32 & 36 & 1152 & 100000 & 5286.23 & 4591.92 \\
32 & 36 & 1152 & 1000000 & 17719.60 & 17029.32 \\
32 & 36 & 1152 & 10000000 & 109291.04 & 108124.69 \\
      \bottomrule
    \end{tabular}
  \end{center}
\end{table}

\begin{table}
  \caption{Results for native \mpireduce compared against the mock-up
    guideline implementations on the ``Hydra'' system.  The MPI library used
    is \openmpiversion.}
    \label{tab:reduce.openmpi.n32.c1}
  \begin{center}
    \begin{tabular}{crrrrrrr}
      \toprule
      \multicolumn{6}{c}{ReduceLane} \\
      $n$ & $N$ & $p$ & $c$ & avg ($\mu s$) & min ($\mu s$) \\
      \midrule
32 & 36 & 1152 & 1 & 16.58 & 12.20 \\
32 & 36 & 1152 & 10 & 19.08 & 15.89 \\
32 & 36 & 1152 & 100 & 57.58 & 50.20 \\
32 & 36 & 1152 & 1000 & 63.12 & 51.06 \\
32 & 36 & 1152 & 10000 & 104.72 & 90.98 \\
32 & 36 & 1152 & 100000 & 2462.59 & 2060.77 \\
32 & 36 & 1152 & 1000000 & 9484.57 & 8693.43 \\
32 & 36 & 1152 & 10000000 & 176732.63 & 175927.95 \\
\midrule
      \multicolumn{6}{c}{ReduceHier} \\
      $n$ & $N$ & $p$ & $c$ & avg ($\mu s$) & min ($\mu s$) \\
      \midrule
32 & 36 & 1152 & 1 & 9.42 & 4.26 \\
32 & 36 & 1152 & 10 & 10.77 & 5.33 \\
32 & 36 & 1152 & 100 & 9.83 & 6.11 \\
32 & 36 & 1152 & 1000 & 29.79 & 20.38 \\
32 & 36 & 1152 & 10000 & 247.29 & 205.23 \\
32 & 36 & 1152 & 100000 & 1798.93 & 1682.96 \\
32 & 36 & 1152 & 1000000 & 6888.60 & 6368.20 \\
32 & 36 & 1152 & 10000000 & 134329.80 & 133683.95 \\
\midrule
      \multicolumn{6}{c}{\mpireduce} \\
      $n$ & $N$ & $p$ & $c$ & avg ($\mu s$) & min ($\mu s$) \\
      \midrule
32 & 36 & 1152 & 1 & 13.95 & 10.48 \\
32 & 36 & 1152 & 10 & 16.01 & 12.50 \\
32 & 36 & 1152 & 100 & 18.72 & 14.07 \\
32 & 36 & 1152 & 1000 & 52.84 & 46.17 \\
32 & 36 & 1152 & 10000 & 437.61 & 424.87 \\
32 & 36 & 1152 & 100000 & 5112.91 & 4748.25 \\
32 & 36 & 1152 & 1000000 & 20729.30 & 19972.86 \\
32 & 36 & 1152 & 10000000 & 523609.17 & 522640.02 \\
      \bottomrule
    \end{tabular}
  \end{center}
\end{table}

\begin{table}
  \caption{Results for native \mpireducescatterblock compared against the mock-up
    guideline implementations on the ``Hydra'' system.  The MPI library used
    is \openmpiversion.}
    \label{tab:reducescatter.openmpi.n32.c1}
  \begin{center}
    \begin{tabular}{crrrrrrr}
      \toprule
      \multicolumn{6}{c}{ReduceScatterBlockLane} \\
      $n$ & $N$ & $p$ & $c$ & avg ($\mu s$) & min ($\mu s$) \\
      \midrule
32 & 36 & 1152 & 1 & 70.04 & 65.15 \\
32 & 36 & 1152 & 9 & 295.40 & 286.34 \\
32 & 36 & 1152 & 87 & 2129.55 & 2084.53 \\
32 & 36 & 1152 & 869 & 14863.69 & 14608.50 \\
32 & 36 & 1152 & 8681 & 201066.50 & 200022.63 \\
      \midrule
      \multicolumn{6}{c}{ReduceScatterBlockHier} \\
      $n$ & $N$ & $p$ & $c$ & avg ($\mu s$) & min ($\mu s$) \\
      \midrule
32 & 36 & 1152 & 1 & 73.92 & 69.62 \\
32 & 36 & 1152 & 9 & 480.17 & 468.69 \\
32 & 36 & 1152 & 87 & 2551.65 & 2494.95 \\
32 & 36 & 1152 & 869 & 18500.00 & 18056.31 \\
32 & 36 & 1152 & 8681 & 278281.59 & 272555.13 \\
      \midrule
      \multicolumn{6}{c}{\mpireducescatterblock} \\
      $n$ & $N$ & $p$ & $c$ & avg ($\mu s$) & min ($\mu s$) \\
      \midrule
32 & 36 & 1152 & 1 & 87.87 & 84.27 \\
32 & 36 & 1152 & 9 & 500.40 & 491.34 \\
32 & 36 & 1152 & 87 & 6088.07 & 6036.04 \\
32 & 36 & 1152 & 869 & 22971.92 & 22024.36 \\
32 & 36 & 1152 & 8681 & 545119.52 & 544187.12 \\
      \bottomrule
    \end{tabular}
  \end{center}
\end{table}

\begin{table}
  \caption{Results for native \mpiscan compared against the mock-up
    guideline implementations on the ``Hydra'' system.  The MPI library used
    is \openmpiversion.}
    \label{tab:scan.openmpi.n32.c1}
  \begin{center}
    \begin{tabular}{crrrrrrr}
      \toprule
    \multicolumn{6}{c}{ScanLane} \\
      $n$ & $N$ & $p$ & $c$ & avg ($\mu s$) & min ($\mu s$) \\
    \midrule
32 & 36 & 1152 & 1 & 59.24 & 54.91 \\
32 & 36 & 1152 & 10 & 72.05 & 68.22 \\
32 & 36 & 1152 & 100 & 133.09 & 126.18 \\
32 & 36 & 1152 & 1000 & 201.07 & 193.64 \\
32 & 36 & 1152 & 10000 & 850.61 & 833.16 \\
32 & 36 & 1152 & 100000 & 9206.98 & 8881.03 \\
32 & 36 & 1152 & 1000000 & 102285.15 & 101555.52 \\
32 & 36 & 1152 & 10000000 & 1051473.43 & 1045733.38 \\
    \midrule
    \multicolumn{6}{c}{ScanHier} \\
    $n$ & $N$ & $p$ & $c$ & avg ($\mu s$) & min ($\mu s$) \\
    \midrule
32 & 36 & 1152 & 1 & 85.02 & 79.29 \\
32 & 36 & 1152 & 10 & 113.85 & 110.49 \\
32 & 36 & 1152 & 100 & 130.69 & 126.27 \\
32 & 36 & 1152 & 1000 & 268.65 & 254.17 \\
32 & 36 & 1152 & 10000 & 1438.74 & 1426.12 \\
32 & 36 & 1152 & 100000 & 11997.64 & 11880.51 \\
32 & 36 & 1152 & 1000000 & 189560.58 & 187850.44 \\
32 & 36 & 1152 & 10000000 & 1540998.87 & 1538699.18 \\
    \midrule
    \multicolumn{6}{c}{\mpiscan} \\
    $n$ & $N$ & $p$ & $c$ & avg ($\mu s$) & min ($\mu s$) \\
    \midrule
32 & 36 & 1152 & 1 & 1480.67 & 1356.57 \\
32 & 36 & 1152 & 10 & 1989.61 & 1943.67 \\
32 & 36 & 1152 & 100 & 2292.26 & 2236.69 \\
32 & 36 & 1152 & 1000 & 4637.66 & 4543.73 \\
32 & 36 & 1152 & 10000 & 23502.69 & 23322.77 \\
32 & 36 & 1152 & 100000 & 207782.34 & 206138.31 \\
32 & 36 & 1152 & 1000000 & 2358807.74 & 2325487.40 \\
    \bottomrule
    \end{tabular}
    \end{center}
\end{table}

\begin{table}
  \caption{Results for native \mpiexscan compared against the mock-up
    guideline implementations on the ``Hydra'' system.  The MPI library used
    is \openmpiversion.}
    \label{tab:exscan.openmpi.n32.c1}
  \begin{center}
    \begin{tabular}{crrrrrrr}
      \toprule
    \multicolumn{6}{c}{ExscanLane} \\
      $n$ & $N$ & $p$ & $c$ & avg ($\mu s$) & min ($\mu s$) \\
    \midrule
32 & 36 & 1152 & 1 & 58.60 & 55.14 \\
32 & 36 & 1152 & 10 & 70.91 & 67.34 \\
32 & 36 & 1152 & 100 & 128.80 & 122.68 \\
32 & 36 & 1152 & 1000 & 200.46 & 192.73 \\
32 & 36 & 1152 & 10000 & 861.63 & 844.11 \\
32 & 36 & 1152 & 100000 & 8773.58 & 8648.98 \\
32 & 36 & 1152 & 1000000 & 97043.35 & 96400.16 \\
32 & 36 & 1152 & 10000000 & 878203.00 & 875551.52 \\
    \midrule
    \multicolumn{6}{c}{ExscanHier} \\
      $n$ & $N$ & $p$ & $c$ & avg ($\mu s$) & min ($\mu s$) \\
    \midrule
32 & 36 & 1152 & 1 & 84.10 & 78.69 \\
32 & 36 & 1152 & 10 & 114.98 & 109.71 \\
32 & 36 & 1152 & 100 & 133.17 & 126.87 \\
32 & 36 & 1152 & 1000 & 259.89 & 253.75 \\
32 & 36 & 1152 & 10000 & 1432.19 & 1417.94 \\
32 & 36 & 1152 & 100000 & 12003.40 & 11842.15 \\
32 & 36 & 1152 & 1000000 & 138995.78 & 135804.27 \\
32 & 36 & 1152 & 10000000 & 1247680.03 & 1244068.48 \\
    \midrule
    \multicolumn{6}{c}{\mpiexscan} \\
    $n$ & $N$ & $p$ & $c$ & avg ($\mu s$) & min ($\mu s$) \\
    \midrule
32 & 36 & 1152 & 1 & 1433.98 & 1331.95 \\
32 & 36 & 1152 & 10 & 2011.06 & 1930.80 \\
32 & 36 & 1152 & 100 & 2292.66 & 2222.87 \\
32 & 36 & 1152 & 1000 & 4629.42 & 4552.10 \\
32 & 36 & 1152 & 10000 & 23675.11 & 23471.04 \\
32 & 36 & 1152 & 100000 & 206577.39 & 205040.03 \\
32 & 36 & 1152 & 1000000 & 2387262.82 & 2357005.26 \\
    \bottomrule
    \end{tabular}
    \end{center}
\end{table}

\clearpage

The scalability of the collective full-lane mock-up implementations
with increasing number of MPI processes per node, that is results for
$c=1152$ up to $1152000$ for $n=2$ and $n=8$ processes per node is
shown in Tables~\ref{tab:lane.openmpi.n8.c1152},
\ref{tab:multicoll.openmpi.n8.c1152},
\ref{tab:bcast.openmpi.n8.c1152}, \ref{tab:scatter.openmpi.n8.c1152},
\ref{tab:allgather.openmpi.n8.c1152},
\ref{tab:alltoall.openmpi.n8.c1152},
\ref{tab:allreduce.openmpi.n8.c1152},
\ref{tab:reduce.openmpi.n8.c1152},
\ref{tab:reducescatter.openmpi.n8.c1152},
\ref{tab:scan.openmpi.n8.c1152}
and~\ref{tab:exscan.openmpi.n8.c1152}
for $n=8$, and in
Tables~\ref{tab:lane.openmpi.n2.c1152},
\ref{tab:multicoll.openmpi.n2.c1152},
\ref{tab:bcast.openmpi.n2.c1152}, \ref{tab:scatter.openmpi.n2.c1152},
\ref{tab:allgather.openmpi.n2.c1152},
\ref{tab:alltoall.openmpi.n2.c1152},
\ref{tab:allreduce.openmpi.n2.c1152},
\ref{tab:reduce.openmpi.n2.c1152},
\ref{tab:reducescatter.openmpi.n2.c1152}
\ref{tab:scan.openmpi.n2.c1152}
and~\ref{tab:exscan.openmpi.n2.c1152}
for $n=2$.

\begin{table}
  \caption{Lane pattern benchmark results on ``Hydra'' for increasing
    number of virtual lanes $k$ used for communicating the data (count
    $c$ \mpiint) per compute node. The MPI library used is \openmpiversion.}
  \label{tab:lane.openmpi.n8.c1152}
  \begin{center}
    \begin{tabular}{rrrrrrr}
      $k$ & $n$ & $N$ & $p$ & $c$ & avg ($\mu s$) & min ($\mu s$) \\
\toprule
1 & 8 & 36 & 288 & 1152 & 141.16 & 133.37 \\
2 & 8 & 36 & 288 & 1152 & 108.09 & 103.82 \\
4 & 8 & 36 & 288 & 1152 & 94.87 & 90.13 \\
8 & 8 & 36 & 288 & 1152 & 89.47 & 82.96 \\
\midrule
1 & 8 & 36 & 288 & 11520 & 726.08 & 676.10 \\
2 & 8 & 36 & 288 & 11520 & 370.73 & 316.04 \\
4 & 8 & 36 & 288 & 11520 & 323.92 & 303.22 \\
8 & 8 & 36 & 288 & 11520 & 187.24 & 175.44 \\
\midrule
1 & 8 & 36 & 288 & 115200 & 3354.20 & 3142.61 \\
2 & 8 & 36 & 288 & 115200 & 2031.78 & 1965.05 \\
4 & 8 & 36 & 288 & 115200 & 1705.15 & 1634.92 \\
8 & 8 & 36 & 288 & 115200 & 1377.99 & 1296.82 \\
\midrule
1 & 8 & 36 & 288 & 1152000 & 25697.63 & 24480.74 \\
2 & 8 & 36 & 288 & 1152000 & 13859.98 & 13369.08 \\
4 & 8 & 36 & 288 & 1152000 & 11163.61 & 11022.93 \\
8 & 8 & 36 & 288 & 1152000 & 10527.36 & 10457.54 \\
\midrule
1 & 8 & 36 & 288 & 11520000 & 203455.77 & 201404.82 \\
2 & 8 & 36 & 288 & 11520000 & 105710.45 & 104342.01 \\
4 & 8 & 36 & 288 & 11520000 & 101412.86 & 100607.49 \\
8 & 8 & 36 & 288 & 11520000 & 101065.05 & 100614.58 \\
\bottomrule
    \end{tabular}
  \end{center}
\end{table}

\begin{table}
  \caption{Multi-collective pattern benchmark results on ``Hydra'' for
    increasing number of virtual lanes $k$ used for communicating the
    data (count $c$ \mpiint) per lane. The collective function is
    \mpialltoall. The MPI library used is \openmpiversion.}
  \label{tab:multicoll.openmpi.n8.c1152}
  \begin{center}
    \begin{tabular}{rrrrrrr}
      $k$ & $n$ & $N$ & $p$ & $c$ & avg ($\mu s$) & min ($\mu s$) \\
\toprule
1 & 8 & 36 & 288 & 1152 & 24.10 & 21.52 \\
2 & 8 & 36 & 288 & 1152 & 24.03 & 21.62 \\
4 & 8 & 36 & 288 & 1152 & 24.75 & 21.87 \\
8 & 8 & 36 & 288 & 1152 & 26.16 & 22.89 \\
\midrule
1 & 8 & 36 & 288 & 11520 & 38.79 & 36.27 \\
2 & 8 & 36 & 288 & 11520 & 38.30 & 35.83 \\
4 & 8 & 36 & 288 & 11520 & 40.58 & 37.42 \\
8 & 8 & 36 & 288 & 11520 & 45.59 & 40.54 \\
\midrule
1 & 8 & 36 & 288 & 115200 & 257.25 & 215.44 \\
2 & 8 & 36 & 288 & 115200 & 262.83 & 238.09 \\
4 & 8 & 36 & 288 & 115200 & 293.09 & 276.27 \\
8 & 8 & 36 & 288 & 115200 & 378.33 & 354.46 \\
\midrule
1 & 8 & 36 & 288 & 1152000 & 2438.17 & 2140.28 \\
2 & 8 & 36 & 288 & 1152000 & 2456.42 & 2165.67 \\
4 & 8 & 36 & 288 & 1152000 & 1359.21 & 1340.96 \\
8 & 8 & 36 & 288 & 1152000 & 2134.69 & 1980.00 \\
\midrule
1 & 8 & 36 & 288 & 11520000 & 6213.04 & 6091.31 \\
2 & 8 & 36 & 288 & 11520000 & 6325.08 & 6146.27 \\
4 & 8 & 36 & 288 & 11520000 & 9428.74 & 9206.51 \\
8 & 8 & 36 & 288 & 11520000 & 16930.04 & 16573.59 \\
\bottomrule
    \end{tabular}
  \end{center}
\end{table}

\begin{table}
  \caption{Results for native \mpibcast compared against the mock-up
    guideline implementations on the ``Hydra'' system.  The MPI library used
    is \openmpiversion.}
  \label{tab:bcast.openmpi.n8.c1152}
  \begin{center}
    \begin{tabular}{crrrrrrr}
      \toprule
      \multicolumn{6}{c}{BcastLane} \\
      $n$ & $N$ & $p$ & $c$ & avg ($\mu s$) & min ($\mu s$) \\
      \midrule
      8 & 36 & 288 & 1152 & 42.14 & 20.77 \\
8 & 36 & 288 & 11520 & 62.40 & 55.05 \\
8 & 36 & 288 & 115200 & 323.94 & 315.02 \\
8 & 36 & 288 & 1152000 & 2896.77 & 2810.48 \\
8 & 36 & 288 & 11520000 & 34966.91 & 34438.42 \\
\midrule
\multicolumn{6}{c}{BcastHier} \\
      $n$ & $N$ & $p$ & $c$ & avg ($\mu s$) & min ($\mu s$) \\
\midrule
8 & 36 & 288 & 1152 & 36.41 & 23.00 \\
8 & 36 & 288 & 11520 & 112.80 & 106.65 \\
8 & 36 & 288 & 115200 & 806.93 & 764.99 \\
8 & 36 & 288 & 1152000 & 7476.56 & 6859.10 \\
8 & 36 & 288 & 11520000 & 39340.92 & 39073.37 \\
      \midrule
      \multicolumn{6}{c}{\mpibcast} \\
      $n$ & $N$ & $p$ & $c$ & avg ($\mu s$) & min ($\mu s$) \\
      \midrule
8 & 36 & 288 & 1152 & 28.16 & 22.62 \\
8 & 36 & 288 & 11520 & 82.46 & 75.84 \\
8 & 36 & 288 & 115200 & 2272.32 & 2217.43 \\
8 & 36 & 288 & 1152000 & 8473.07 & 8157.56 \\
8 & 36 & 288 & 11520000 & 81125.13 & 68593.24 \\
      \bottomrule
    \end{tabular}
  \end{center}
\end{table}

\begin{table}
  \caption{Results for native \mpiscatter compared against the mock-up
    guideline implementations on the ``Hydra'' system.  The MPI library used
    is \openmpiversion.}
  \label{tab:scatter.openmpi.n8.c1152}
  \begin{center}
    \begin{tabular}{crrrrrrr}
      \toprule
      \multicolumn{6}{c}{ScatterLane} \\
      $n$ & $N$ & $p$ & $c$ & avg ($\mu s$) & min ($\mu s$) \\
      \midrule
8 & 36 & 288 & 4 & 19.79 & 16.46 \\
8 & 36 & 288 & 40 & 34.54 & 31.60 \\
8 & 36 & 288 & 400 & 195.52 & 166.86 \\
8 & 36 & 288 & 4000 & 1508.41 & 1487.00 \\
8 & 36 & 288 & 40000 & 17002.34 & 16802.41 \\
\midrule
      \multicolumn{6}{c}{ScatterHier} \\
      $n$ & $N$ & $p$ & $c$ & avg ($\mu s$) & min ($\mu s$) \\
      \midrule
8 & 36 & 288 & 4 & 13.52 & 10.56 \\
8 & 36 & 288 & 40 & 24.95 & 21.37 \\
8 & 36 & 288 & 400 & 184.02 & 178.26 \\
8 & 36 & 288 & 4000 & 949.84 & 940.58 \\
8 & 36 & 288 & 40000 & 4479.74 & 4463.26 \\
\midrule
      \multicolumn{6}{c}{\mpiscatter} \\
      $n$ & $N$ & $p$ & $c$ & avg ($\mu s$) & min ($\mu s$) \\
      \midrule
8 & 36 & 288 & 4 & 22.01 & 18.68 \\
8 & 36 & 288 & 40 & 49.64 & 35.90 \\
8 & 36 & 288 & 400 & 112.31 & 110.53 \\
8 & 36 & 288 & 4000 & 1631.07 & 1577.86 \\
8 & 36 & 288 & 40000 & 9411.69 & 8936.62 \\
      \bottomrule
    \end{tabular}
  \end{center}
\end{table}

\begin{table}
  \caption{Results for native \mpiallgather compared against the mock-up
    guideline implementations on the ``Hydra'' system.  The MPI library used
    is \openmpiversion.}
  \label{tab:allgather.openmpi.n8.c1152}
  \begin{center}
    \begin{tabular}{crrrrrrr}
      \toprule
      \multicolumn{6}{c}{AllgatherLane} \\
      $n$ & $N$ & $p$ & $c$ & avg ($\mu s$) & min ($\mu s$) \\
      \midrule
8 & 36 & 288 & 4 & 29.84 & 26.88 \\
8 & 36 & 288 & 40 & 61.53 & 51.25 \\
8 & 36 & 288 & 400 & 477.00 & 450.70 \\
8 & 36 & 288 & 4000 & 8306.88 & 7536.17 \\
8 & 36 & 288 & 40000 & 67536.82 & 65308.33 \\
      \midrule
      \multicolumn{6}{c}{AllgatherHier} \\
      $n$ & $N$ & $p$ & $c$ & avg ($\mu s$) & min ($\mu s$) \\
      \midrule
8 & 36 & 288 & 4 & 41.18 & 28.23 \\
8 & 36 & 288 & 40 & 169.47 & 100.84 \\
8 & 36 & 288 & 400 & 608.62 & 563.47 \\
8 & 36 & 288 & 4000 & 4037.87 & 3909.20 \\
8 & 36 & 288 & 40000 & 34769.18 & 34229.73 \\
      \midrule
      \multicolumn{6}{c}{\mpiallgather} \\
      $n$ & $N$ & $p$ & $c$ & avg ($\mu s$) & min ($\mu s$) \\
      \midrule
8 & 36 & 288 & 4 & 28.65 & 21.78 \\
8 & 36 & 288 & 40 & 58.84 & 47.95 \\
8 & 36 & 288 & 400 & 492.88 & 471.06 \\
8 & 36 & 288 & 4000 & 3631.79 & 3590.41 \\
8 & 36 & 288 & 40000 & 27126.61 & 27064.33 \\
      \bottomrule
    \end{tabular}
  \end{center}
\end{table}

\begin{table}
  \caption{Results for native \mpialltoall compared against the mock-up
    guideline implementation on the ``Hydra'' system.  The MPI library used
    is \openmpiversion.}
  \label{tab:alltoall.openmpi.n8.c1152}
  \begin{center}
    \begin{tabular}{crrrrrrr}
      \toprule
      \multicolumn{6}{c}{AlltoallLane} \\
      $n$ & $N$ & $p$ & $c$ & avg ($\mu s$) & min ($\mu s$) \\
      \midrule
8 & 36 & 288 & 4 & 58.44 & 46.34 \\
8 & 36 & 288 & 40 & 100.20 & 89.38 \\
8 & 36 & 288 & 400 & 758.82 & 728.29 \\
8 & 36 & 288 & 4000 & 5824.25 & 5727.73 \\
8 & 36 & 288 & 40000 & 82041.85 & 80360.95 \\
      \midrule
      \multicolumn{6}{c}{\mpialltoall} \\
      $n$ & $N$ & $p$ & $c$ & avg ($\mu s$) & min ($\mu s$) \\
      \midrule
8 & 36 & 288 & 4 & 71.37 & 59.81 \\
8 & 36 & 288 & 40 & 179.81 & 171.08 \\
8 & 36 & 288 & 400 & 854.16 & 477.77 \\
8 & 36 & 288 & 4000 & 4479.28 & 4392.48 \\
8 & 36 & 288 & 40000 & 43612.51 & 42423.55 \\
      \bottomrule
    \end{tabular}
  \end{center}
\end{table}

\begin{table}
  \caption{Results for native \mpiallreduce compared against the mock-up
    guideline implementations on the ``Hydra'' system.  The MPI library used
    is \openmpiversion.}
  \label{tab:allreduce.openmpi.n8.c1152}
  \begin{center}
    \begin{tabular}{crrrrrrr}
      \toprule
      \multicolumn{6}{c}{AllreduceLane} \\
      $n$ & $N$ & $p$ & $c$ & avg ($\mu s$) & min ($\mu s$) \\
      \midrule
8 & 36 & 288 & 1152 & 46.20 & 41.41 \\
8 & 36 & 288 & 11520 & 156.70 & 149.52 \\
8 & 36 & 288 & 115200 & 999.65 & 984.33 \\
8 & 36 & 288 & 1152000 & 8128.16 & 8082.43 \\
8 & 36 & 288 & 11520000 & 126301.28 & 125800.00 \\
      \midrule
      \multicolumn{6}{c}{AllreduceHier} \\
      $n$ & $N$ & $p$ & $c$ & avg ($\mu s$) & min ($\mu s$) \\
      \midrule
      8 & 36 & 288 & 1152 & 74.84 & 50.43 \\
8 & 36 & 288 & 11520 & 289.05 & 282.39 \\
8 & 36 & 288 & 115200 & 1417.24 & 1352.60 \\
8 & 36 & 288 & 1152000 & 11380.07 & 10943.03 \\
8 & 36 & 288 & 11520000 & 134179.21 & 133247.32 \\
      \midrule
      \multicolumn{6}{c}{\mpiallreduce} \\
      $n$ & $N$ & $p$ & $c$ & avg ($\mu s$) & min ($\mu s$) \\
      \midrule
8 & 36 & 288 & 1152 & 48.09 & 43.30 \\
8 & 36 & 288 & 11520 & 920.49 & 862.08 \\
8 & 36 & 288 & 115200 & 1517.65 & 1469.00 \\
8 & 36 & 288 & 1152000 & 10012.95 & 8017.87 \\
8 & 36 & 288 & 11520000 & 71062.14 & 70544.27 \\
      \bottomrule
    \end{tabular}
  \end{center}
\end{table}

\begin{table}
  \caption{Results for native \mpireduce compared against the mock-up
    guideline implementations on the ``Hydra'' system.  The MPI library used
    is \openmpiversion.}
    \label{tab:reduce.openmpi.n8.c1152}
  \begin{center}
    \begin{tabular}{crrrrrrr}
      \toprule
      \multicolumn{6}{c}{ReduceLane} \\
      $n$ & $N$ & $p$ & $c$ & avg ($\mu s$) & min ($\mu s$) \\
      \midrule
8 & 36 & 288 & 1152 & 37.74 & 33.82 \\
8 & 36 & 288 & 11520 & 147.38 & 143.05 \\
8 & 36 & 288 & 115200 & 1134.58 & 1089.39 \\
8 & 36 & 288 & 1152000 & 7581.07 & 7119.93 \\
8 & 36 & 288 & 11520000 & 118367.30 & 117785.49 \\
      \midrule
      \multicolumn{6}{c}{ReduceHier} \\
      $n$ & $N$ & $p$ & $c$ & avg ($\mu s$) & min ($\mu s$) \\
      \midrule
8 & 36 & 288 & 1152 & 22.03 & 18.54 \\
8 & 36 & 288 & 11520 & 100.26 & 94.41 \\
8 & 36 & 288 & 115200 & 728.84 & 705.64 \\
8 & 36 & 288 & 1152000 & 5711.07 & 5600.72 \\
8 & 36 & 288 & 11520000 & 88521.86 & 83230.71 \\
      \midrule
      \multicolumn{6}{c}{\mpireduce} \\
      $n$ & $N$ & $p$ & $c$ & avg ($\mu s$) & min ($\mu s$) \\
      \midrule
8 & 36 & 288 & 1152 & 47.90 & 42.53 \\
8 & 36 & 288 & 11520 & 386.89 & 347.89 \\
8 & 36 & 288 & 115200 & 4089.59 & 4052.21 \\
8 & 36 & 288 & 1152000 & 15617.08 & 15132.57 \\
8 & 36 & 288 & 11520000 & 185448.64 & 184743.46 \\
      \bottomrule
    \end{tabular}
  \end{center}
\end{table}

\begin{table}
  \caption{Results for native \mpireducescatterblock compared against the mock-up
    guideline implementations on the ``Hydra'' system.  The MPI library used
    is \openmpiversion.}
    \label{tab:reducescatter.openmpi.n8.c1152}
  \begin{center}
    \begin{tabular}{crrrrrrr}
      \toprule
      \multicolumn{6}{c}{ReduceScatterBlockLane} \\
      $n$ & $N$ & $p$ & $c$ & avg ($\mu s$) & min ($\mu s$) \\
      \midrule
8 & 36 & 288 & 4 & 48.10 & 44.25 \\
8 & 36 & 288 & 40 & 150.90 & 145.84 \\
8 & 36 & 288 & 400 & 1077.49 & 1066.64 \\
8 & 36 & 288 & 4000 & 8956.53 & 8706.54 \\
8 & 36 & 288 & 40000 & 141993.53 & 140919.32 \\
\midrule
      \multicolumn{6}{c}{ReduceScatterBlockHier} \\
      $n$ & $N$ & $p$ & $c$ & avg ($\mu s$) & min ($\mu s$) \\
      \midrule
8 & 36 & 288 & 4 & 60.01 & 56.56 \\
8 & 36 & 288 & 40 & 358.37 & 354.80 \\
8 & 36 & 288 & 400 & 1388.08 & 1358.21 \\
8 & 36 & 288 & 4000 & 9418.23 & 9288.73 \\
8 & 36 & 288 & 40000 & 300314.45 & 291506.22 \\
\midrule
      \multicolumn{6}{c}{\mpireducescatterblock} \\
      $n$ & $N$ & $p$ & $c$ & avg ($\mu s$) & min ($\mu s$) \\
      \midrule
8 & 36 & 288 & 4 & 71.61 & 66.69 \\
8 & 36 & 288 & 40 & 430.30 & 423.73 \\
8 & 36 & 288 & 400 & 4273.21 & 4234.29 \\
8 & 36 & 288 & 4000 & 18177.84 & 17776.48 \\
8 & 36 & 288 & 40000 & 199387.96 & 198197.10 \\
      \bottomrule
    \end{tabular}
  \end{center}
\end{table}

\begin{table}
  \caption{Results for native \mpiscan compared against the mock-up
    guideline implementations on the ``Hydra'' system.  The MPI library used
    is \openmpiversion.}
    \label{tab:scan.openmpi.n8.c1152}
  \begin{center}
    \begin{tabular}{crrrrrrr}
      \toprule
    \multicolumn{6}{c}{ScanLane} \\
      $n$ & $N$ & $p$ & $c$ & avg ($\mu s$) & min ($\mu s$) \\
    \midrule
8 & 36 & 288 & 1152 & 129.72 & 125.11 \\
8 & 36 & 288 & 11520 & 450.78 & 445.27 \\
8 & 36 & 288 & 115200 & 3690.59 & 3663.88 \\
8 & 36 & 288 & 1152000 & 34565.44 & 34486.01 \\
8 & 36 & 288 & 11520000 & 435401.15 & 433144.02 \\
    \midrule
    \multicolumn{6}{c}{ScanHier} \\
    $n$ & $N$ & $p$ & $c$ & avg ($\mu s$) & min ($\mu s$) \\
    \midrule
8 & 36 & 288 & 1152 & 178.22 & 173.87 \\
8 & 36 & 288 & 11520 & 1067.24 & 1045.47 \\
8 & 36 & 288 & 115200 & 7643.10 & 7617.25 \\
8 & 36 & 288 & 1152000 & 80387.19 & 80179.20 \\
8 & 36 & 288 & 11520000 & 1044741.11 & 1042199.77 \\
    \midrule
    \multicolumn{6}{c}{\mpiscan} \\
    $n$ & $N$ & $p$ & $c$ & avg ($\mu s$) & min ($\mu s$) \\
    \midrule
8 & 36 & 288 & 1152 & 1423.55 & 1375.11 \\
8 & 36 & 288 & 11520 & 6604.12 & 6584.42 \\
8 & 36 & 288 & 115200 & 57905.42 & 57753.54 \\
8 & 36 & 288 & 1152000 & 610055.45 & 607143.24 \\
    \bottomrule
    \end{tabular}
    \end{center}
\end{table}

\begin{table}
  \caption{Results for native \mpiexscan compared against the mock-up
    guideline implementations on the ``Hydra'' system.  The MPI library used
    is \openmpiversion.}
    \label{tab:exscan.openmpi.n8.c1152}
  \begin{center}
    \begin{tabular}{crrrrrrr}
      \toprule
    \multicolumn{6}{c}{ExscanLane} \\
      $n$ & $N$ & $p$ & $c$ & avg ($\mu s$) & min ($\mu s$) \\
    \midrule
8 & 36 & 288 & 1152 & 127.91 & 123.25 \\
8 & 36 & 288 & 11520 & 440.94 & 433.51 \\
8 & 36 & 288 & 115200 & 3326.14 & 3280.89 \\
8 & 36 & 288 & 1152000 & 31697.24 & 31607.45 \\
8 & 36 & 288 & 11520000 & 375387.09 & 373537.59 \\
    \midrule
    \multicolumn{6}{c}{ExscanHier} \\
      $n$ & $N$ & $p$ & $c$ & avg ($\mu s$) & min ($\mu s$) \\
    \midrule
8 & 36 & 288 & 1152 & 177.67 & 173.76 \\
8 & 36 & 288 & 11520 & 1045.69 & 1039.15 \\
8 & 36 & 288 & 115200 & 7609.48 & 7584.75 \\
8 & 36 & 288 & 1152000 & 78914.20 & 78758.51 \\
8 & 36 & 288 & 11520000 & 812633.95 & 810983.33 \\
    \midrule
    \multicolumn{6}{c}{\mpiexscan} \\
    $n$ & $N$ & $p$ & $c$ & avg ($\mu s$) & min ($\mu s$) \\
    \midrule
8 & 36 & 288 & 1152 & 1385.95 & 1354.21 \\
8 & 36 & 288 & 11520 & 6585.03 & 6569.05 \\
8 & 36 & 288 & 115200 & 57619.11 & 57502.78 \\
8 & 36 & 288 & 1152000 & 609224.05 & 606579.81 \\
    \bottomrule
    \end{tabular}
    \end{center}
\end{table}

\begin{table}
  \caption{Lane pattern benchmark results on ``Hydra'' for increasing
    number of virtual lanes $k$ used for communicating the data (count
    $c$ \mpiint) per compute node. The MPI library used is \openmpiversion.}
  \label{tab:lane.openmpi.n2.c1152}
  \begin{center}
    \begin{tabular}{rrrrrrr}
      $k$ & $n$ & $N$ & $p$ & $c$ & avg ($\mu s$) & min ($\mu s$) \\
\toprule
1 & 2 & 36 & 72 & 1152 & 140.96 & 133.31 \\
2 & 2 & 36 & 72 & 1152 & 108.92 & 103.69 \\
\midrule
1 & 2 & 36 & 72 & 11520 & 719.73 & 677.00 \\
2 & 2 & 36 & 72 & 11520 & 367.57 & 317.15 \\
\midrule
1 & 2 & 36 & 72 & 115200 & 3269.44 & 3119.29 \\
2 & 2 & 36 & 72 & 115200 & 2046.33 & 1994.05 \\
\midrule
1 & 2 & 36 & 72 & 1152000 & 26100.98 & 24367.15 \\
2 & 2 & 36 & 72 & 1152000 & 13847.98 & 13419.57 \\
\midrule
1 & 2 & 36 & 72 & 11520000 & 203283.27 & 201459.53 \\
2 & 2 & 36 & 72 & 11520000 & 105707.08 & 103961.30 \\
\bottomrule
    \end{tabular}
  \end{center}
\end{table}

\begin{table}
  \caption{Multi-collective pattern benchmark results on ``Hydra'' for
    increasing number of virtual lanes $k$ used for communicating the
    data (count $c$ \mpiint) per lane. The collective function is
    \mpialltoall. The MPI library used is \openmpiversion.}
  \label{tab:multicoll.openmpi.n2.c1152}
  \begin{center}
    \begin{tabular}{rrrrrrr}
      $k$ & $n$ & $N$ & $p$ & $c$ & avg ($\mu s$) & min ($\mu s$) \\
\toprule
1 & 2 & 36 & 72 & 1152 & 23.44 & 21.37 \\
2 & 2 & 36 & 72 & 1152 & 23.51 & 21.28 \\
\midrule
1 & 2 & 36 & 72 & 11520 & 37.72 & 35.08 \\
2 & 2 & 36 & 72 & 11520 & 37.55 & 34.37 \\
\midrule
1 & 2 & 36 & 72 & 115200 & 258.06 & 192.28 \\
2 & 2 & 36 & 72 & 115200 & 261.02 & 236.10 \\
\midrule
1 & 2 & 36 & 72 & 1152000 & 2497.48 & 2188.18 \\
2 & 2 & 36 & 72 & 1152000 & 2506.30 & 2134.05 \\
\midrule
1 & 2 & 36 & 72 & 11520000 & 6328.54 & 6049.30 \\
2 & 2 & 36 & 72 & 11520000 & 6233.15 & 6169.83 \\
\bottomrule
    \end{tabular}
  \end{center}
\end{table}

\begin{table}
  \caption{Results for native \mpibcast compared against the mock-up
    guideline implementations on the ``Hydra'' system.  The MPI library used
    is \openmpiversion.}
  \label{tab:bcast.openmpi.n2.c1152}
  \begin{center}
    \begin{tabular}{crrrrrrr}
      \toprule
      \multicolumn{6}{c}{BcastLane} \\
      $n$ & $N$ & $p$ & $c$ & avg ($\mu s$) & min ($\mu s$) \\
      \midrule
2 & 36 & 72 & 1152 & 21.23 & 17.47 \\
2 & 36 & 72 & 11520 & 61.71 & 56.45 \\
2 & 36 & 72 & 115200 & 360.14 & 349.16 \\
2 & 36 & 72 & 1152000 & 3748.12 & 3529.28 \\
2 & 36 & 72 & 11520000 & 22121.47 & 21603.23 \\
\midrule
\multicolumn{6}{c}{BcastHier} \\
      $n$ & $N$ & $p$ & $c$ & avg ($\mu s$) & min ($\mu s$) \\
      \midrule
2 & 36 & 72 & 1152 & 23.03 & 18.51 \\
2 & 36 & 72 & 11520 & 84.43 & 78.47 \\
2 & 36 & 72 & 115200 & 654.04 & 595.27 \\
2 & 36 & 72 & 1152000 & 5275.19 & 4957.24 \\
2 & 36 & 72 & 11520000 & 23918.17 & 23852.87 \\
      \midrule
      \multicolumn{6}{c}{\mpibcast} \\
      $n$ & $N$ & $p$ & $c$ & avg ($\mu s$) & min ($\mu s$) \\
      \midrule
2 & 36 & 72 & 1152 & 18.94 & 16.48 \\
2 & 36 & 72 & 11520 & 73.21 & 63.84 \\
2 & 36 & 72 & 115200 & 767.02 & 738.17 \\
2 & 36 & 72 & 1152000 & 5059.17 & 4746.39 \\
2 & 36 & 72 & 11520000 & 25855.28 & 25357.43 \\
      \bottomrule
    \end{tabular}
  \end{center}
\end{table}

\begin{table}
  \caption{Results for native \mpiscatter compared against the mock-up
    guideline implementations on the ``Hydra'' system.  The MPI library used
    is \openmpiversion.}
  \label{tab:scatter.openmpi.n2.c1152}
  \begin{center}
    \begin{tabular}{crrrrrrr}
      \toprule
      \multicolumn{6}{c}{ScatterLane} \\
      $n$ & $N$ & $p$ & $c$ & avg ($\mu s$) & min ($\mu s$) \\
      \midrule
2 & 36 & 72 & 16 & 14.49 & 11.25 \\
2 & 36 & 72 & 160 & 26.36 & 24.71 \\
2 & 36 & 72 & 1600 & 126.63 & 123.22 \\
2 & 36 & 72 & 16000 & 1945.10 & 1841.19 \\
2 & 36 & 72 & 160000 & 11960.21 & 11883.59 \\
\midrule
      \multicolumn{6}{c}{ScatterHier} \\
      $n$ & $N$ & $p$ & $c$ & avg ($\mu s$) & min ($\mu s$) \\
      \midrule
      2 & 36 & 72 & 16 & 9.92 & 7.45 \\
2 & 36 & 72 & 160 & 15.44 & 12.96 \\
2 & 36 & 72 & 1600 & 184.21 & 175.29 \\
2 & 36 & 72 & 16000 & 906.55 & 901.48 \\
2 & 36 & 72 & 160000 & 4466.54 & 4458.72 \\
      \midrule
      \multicolumn{6}{c}{\mpiscatter} \\
      $n$ & $N$ & $p$ & $c$ & avg ($\mu s$) & min ($\mu s$) \\
      \midrule
2 & 36 & 72 & 16 & 14.52 & 11.86 \\
2 & 36 & 72 & 160 & 22.21 & 20.34 \\
2 & 36 & 72 & 1600 & 63.98 & 60.90 \\
2 & 36 & 72 & 16000 & 1114.57 & 1105.75 \\
2 & 36 & 72 & 160000 & 5339.85 & 4966.20 \\
      \bottomrule
    \end{tabular}
  \end{center}
\end{table}

\begin{table}
  \caption{Results for native \mpiallgather compared against the mock-up
    guideline implementations on the ``Hydra'' system.  The MPI library used
    is \openmpiversion.}
  \label{tab:allgather.openmpi.n2.c1152}
  \begin{center}
    \begin{tabular}{crrrrrrr}
      \toprule
      \multicolumn{6}{c}{AllgatherLane} \\
      $n$ & $N$ & $p$ & $c$ & avg ($\mu s$) & min ($\mu s$) \\
      \midrule
2 & 36 & 72 & 16 & 24.08 & 20.47 \\
2 & 36 & 72 & 160 & 50.22 & 43.68 \\
2 & 36 & 72 & 1600 & 469.29 & 457.15 \\
2 & 36 & 72 & 16000 & 4782.64 & 4715.21 \\
2 & 36 & 72 & 160000 & 37937.89 & 37522.34 \\
      \midrule
      \multicolumn{6}{c}{AllgatherHier} \\
      $n$ & $N$ & $p$ & $c$ & avg ($\mu s$) & min ($\mu s$) \\
      \midrule
      2 & 36 & 72 & 16 & 24.08 & 21.94 \\
2 & 36 & 72 & 160 & 74.99 & 70.95 \\
2 & 36 & 72 & 1600 & 399.04 & 333.84 \\
2 & 36 & 72 & 16000 & 1932.52 & 1903.74 \\
2 & 36 & 72 & 160000 & 19182.24 & 18823.21 \\
      \midrule
      \multicolumn{6}{c}{\mpiallgather} \\
      $n$ & $N$ & $p$ & $c$ & avg ($\mu s$) & min ($\mu s$) \\
      \midrule
2 & 36 & 72 & 16 & 18.90 & 15.43 \\
2 & 36 & 72 & 160 & 36.29 & 29.72 \\
2 & 36 & 72 & 1600 & 299.79 & 278.97 \\
2 & 36 & 72 & 16000 & 1892.93 & 1866.98 \\
2 & 36 & 72 & 160000 & 17196.10 & 16924.20 \\
      \bottomrule
    \end{tabular}
  \end{center}
\end{table}

\begin{table}
  \caption{Results for native \mpialltoall compared against the mock-up
    guideline implementation on the ``Hydra'' system.  The MPI library used
    is \openmpiversion.}
  \label{tab:alltoall.openmpi.n2.c1152}
  \begin{center}
    \begin{tabular}{crrrrrrr}
      \toprule
      \multicolumn{6}{c}{AlltoallLane} \\
      $n$ & $N$ & $p$ & $c$ & avg ($\mu s$) & min ($\mu s$) \\
      \midrule
2 & 36 & 72 & 16 & 29.16 & 26.07 \\
2 & 36 & 72 & 160 & 58.56 & 54.81 \\
2 & 36 & 72 & 1600 & 426.29 & 363.73 \\
2 & 36 & 72 & 16000 & 4552.82 & 3795.89 \\
2 & 36 & 72 & 160000 & 65023.81 & 62349.84 \\
      \midrule
      \multicolumn{6}{c}{\mpialltoall} \\
      $n$ & $N$ & $p$ & $c$ & avg ($\mu s$) & min ($\mu s$) \\
      \midrule
2 & 36 & 72 & 16 & 46.06 & 28.59 \\
2 & 36 & 72 & 160 & 79.84 & 65.84 \\
2 & 36 & 72 & 1600 & 351.08 & 281.57 \\
2 & 36 & 72 & 16000 & 2922.28 & 2502.92 \\
2 & 36 & 72 & 160000 & 11336.92 & 11202.52 \\
      \bottomrule
    \end{tabular}
  \end{center}
\end{table}

\begin{table}
  \caption{Results for native \mpiallreduce compared against the mock-up
    guideline implementations on the ``Hydra'' system.  The MPI library used
    is \openmpiversion.}
  \label{tab:allreduce.openmpi.n2.c1152}
  \begin{center}
    \begin{tabular}{crrrrrrr}
      \toprule
      \multicolumn{6}{c}{AllreduceLane} \\
      $n$ & $N$ & $p$ & $c$ & avg ($\mu s$) & min ($\mu s$) \\
      \midrule
2 & 36 & 72 & 1152 & 29.30 & 26.48 \\
2 & 36 & 72 & 11520 & 222.14 & 174.05 \\
2 & 36 & 72 & 115200 & 747.53 & 705.25 \\
2 & 36 & 72 & 1152000 & 6355.40 & 6275.81 \\
2 & 36 & 72 & 11520000 & 66803.81 & 66236.27 \\
\midrule
\multicolumn{6}{c}{AllreduceHier} \\
      $n$ & $N$ & $p$ & $c$ & avg ($\mu s$) & min ($\mu s$) \\
\midrule
2 & 36 & 72 & 1152 & 36.94 & 33.85 \\
2 & 36 & 72 & 11520 & 198.58 & 192.12 \\
2 & 36 & 72 & 115200 & 940.72 & 830.31 \\
2 & 36 & 72 & 1152000 & 6954.33 & 6843.44 \\
2 & 36 & 72 & 11520000 & 61959.06 & 60894.95 \\
      \midrule
      \multicolumn{6}{c}{\mpiallreduce} \\
      $n$ & $N$ & $p$ & $c$ & avg ($\mu s$) & min ($\mu s$) \\
      \midrule
2 & 36 & 72 & 1152 & 33.12 & 29.96 \\
2 & 36 & 72 & 11520 & 253.29 & 243.36 \\
2 & 36 & 72 & 115200 & 738.33 & 674.89 \\
2 & 36 & 72 & 1152000 & 6289.10 & 5941.67 \\
2 & 36 & 72 & 11520000 & 57918.30 & 56882.14 \\
      \bottomrule
    \end{tabular}
  \end{center}
\end{table}

\begin{table}
  \caption{Results for native \mpireduce compared against the mock-up
    guideline implementations on the ``Hydra'' system.  The MPI library used
    is \openmpiversion.}
    \label{tab:reduce.openmpi.n2.c1152}
  \begin{center}
    \begin{tabular}{crrrrrrr}
      \toprule
      \multicolumn{6}{c}{ReduceLane} \\
      $n$ & $N$ & $p$ & $c$ & avg ($\mu s$) & min ($\mu s$) \\
      \midrule
2 & 36 & 72 & 1152 & 28.17 & 25.95 \\
2 & 36 & 72 & 11520 & 177.56 & 175.43 \\
2 & 36 & 72 & 115200 & 701.46 & 693.16 \\
2 & 36 & 72 & 1152000 & 5289.11 & 5235.79 \\
2 & 36 & 72 & 11520000 & 119399.93 & 113051.42 \\
      \midrule
      \multicolumn{6}{c}{ReduceLane} \\
      $n$ & $N$ & $p$ & $c$ & avg ($\mu s$) & min ($\mu s$) \\
      \midrule
2 & 36 & 72 & 1152 & 8.56 & 5.34 \\
2 & 36 & 72 & 11520 & 54.01 & 48.81 \\
2 & 36 & 72 & 115200 & 416.25 & 406.77 \\
2 & 36 & 72 & 1152000 & 3534.64 & 3443.94 \\
2 & 36 & 72 & 11520000 & 31725.80 & 31333.03 \\
      \midrule
      \multicolumn{6}{c}{\mpireduce} \\
      $n$ & $N$ & $p$ & $c$ & avg ($\mu s$) & min ($\mu s$) \\
      \midrule
2 & 36 & 72 & 1152 & 36.41 & 26.75 \\
2 & 36 & 72 & 11520 & 307.81 & 275.05 \\
2 & 36 & 72 & 115200 & 1146.39 & 1125.63 \\
2 & 36 & 72 & 1152000 & 4792.21 & 4706.12 \\
2 & 36 & 72 & 11520000 & 143034.93 & 136661.46 \\
      \bottomrule
    \end{tabular}
  \end{center}
\end{table}

\begin{table}
  \caption{Results for native \mpireducescatterblock compared against the mock-up
    guideline implementations on the ``Hydra'' system.  The MPI library used
    is \openmpiversion.}
    \label{tab:reducescatter.openmpi.n2.c1152}
  \begin{center}
    \begin{tabular}{crrrrrrr}
      \toprule
      \multicolumn{6}{c}{ReduceScatterBlockLane} \\
      $n$ & $N$ & $p$ & $c$ & avg ($\mu s$) & min ($\mu s$) \\
      \midrule
2 & 36 & 72 & 16 & 39.80 & 37.72 \\
2 & 36 & 72 & 160 & 187.77 & 184.89 \\
2 & 36 & 72 & 1600 & 819.45 & 806.21 \\
2 & 36 & 72 & 16000 & 7038.31 & 6913.67 \\
2 & 36 & 72 & 160000 & 149069.54 & 139383.67 \\
      \midrule
      \multicolumn{6}{c}{ReduceScatterBlockHier} \\
      $n$ & $N$ & $p$ & $c$ & avg ($\mu s$) & min ($\mu s$) \\
      \midrule
      2 & 36 & 72 & 16 & 43.56 & 41.73 \\
2 & 36 & 72 & 160 & 292.63 & 289.01 \\
2 & 36 & 72 & 1600 & 1012.63 & 981.17 \\
2 & 36 & 72 & 16000 & 6887.88 & 6743.28 \\
2 & 36 & 72 & 160000 & 227898.44 & 217439.09 \\
      \midrule
      \multicolumn{6}{c}{\mpireducescatterblock} \\
      $n$ & $N$ & $p$ & $c$ & avg ($\mu s$) & min ($\mu s$) \\
      \midrule
2 & 36 & 72 & 16 & 46.11 & 43.59 \\
2 & 36 & 72 & 160 & 332.34 & 324.80 \\
2 & 36 & 72 & 1600 & 1239.54 & 1200.81 \\
2 & 36 & 72 & 16000 & 6314.41 & 6184.79 \\
2 & 36 & 72 & 160000 & 160900.66 & 152184.06 \\
      \bottomrule
    \end{tabular}
  \end{center}
\end{table}

\begin{table}
  \caption{Results for native \mpiscan compared against the mock-up
    guideline implementations on the ``Hydra'' system.  The MPI library used
    is \openmpiversion.}
    \label{tab:scan.openmpi.n2.c1152}
  \begin{center}
    \begin{tabular}{crrrrrrr}
      \toprule
    \multicolumn{6}{c}{ScanLane} \\
      $n$ & $N$ & $p$ & $c$ & avg ($\mu s$) & min ($\mu s$) \\
    \midrule
2 & 36 & 72 & 1152 & 107.82 & 104.59 \\
2 & 36 & 72 & 11520 & 457.29 & 448.03 \\
2 & 36 & 72 & 115200 & 3830.63 & 3811.99 \\
2 & 36 & 72 & 1152000 & 35999.02 & 35927.91 \\
2 & 36 & 72 & 11520000 & 386772.31 & 383366.63 \\
    \midrule
    \multicolumn{6}{c}{ScanHier} \\
    $n$ & $N$ & $p$ & $c$ & avg ($\mu s$) & min ($\mu s$) \\
    \midrule
2 & 36 & 72 & 1152 & 141.26 & 137.55 \\
2 & 36 & 72 & 11520 & 857.83 & 852.09 \\
2 & 36 & 72 & 115200 & 6175.40 & 6162.67 \\
2 & 36 & 72 & 1152000 & 62988.78 & 62730.44 \\
2 & 36 & 72 & 11520000 & 842670.57 & 839779.57 \\
    \midrule
    \multicolumn{6}{c}{\mpiscan} \\
    $n$ & $N$ & $p$ & $c$ & avg ($\mu s$) & min ($\mu s$) \\
    \midrule
2 & 36 & 72 & 1152 & 327.42 & 311.43 \\
2 & 36 & 72 & 11520 & 1621.13 & 1602.68 \\
2 & 36 & 72 & 115200 & 13150.62 & 13133.40 \\
2 & 36 & 72 & 1152000 & 124136.48 & 123727.02 \\
    \bottomrule
    \end{tabular}
    \end{center}
\end{table}

\begin{table}
  \caption{Results for native \mpiexscan compared against the mock-up
    guideline implementations on the ``Hydra'' system.  The MPI library used
    is \openmpiversion.}
    \label{tab:exscan.openmpi.n2.c1152}
  \begin{center}
    \begin{tabular}{crrrrrrr}
      \toprule
    \multicolumn{6}{c}{ExscanLane} \\
      $n$ & $N$ & $p$ & $c$ & avg ($\mu s$) & min ($\mu s$) \\
    \midrule
2 & 36 & 72 & 1152 & 107.93 & 104.87 \\
2 & 36 & 72 & 11520 & 443.30 & 431.42 \\
2 & 36 & 72 & 115200 & 3576.82 & 3564.83 \\
2 & 36 & 72 & 1152000 & 33378.59 & 33327.63 \\
2 & 36 & 72 & 11520000 & 355990.61 & 355037.94 \\
    \midrule
    \multicolumn{6}{c}{ExscanHier} \\
      $n$ & $N$ & $p$ & $c$ & avg ($\mu s$) & min ($\mu s$) \\
    \midrule
2 & 36 & 72 & 1152 & 139.09 & 136.10 \\
2 & 36 & 72 & 11520 & 842.23 & 837.04 \\
2 & 36 & 72 & 115200 & 6080.95 & 6069.89 \\
2 & 36 & 72 & 1152000 & 60884.96 & 60666.57 \\
2 & 36 & 72 & 11520000 & 629020.75 & 627280.74 \\
    \midrule
    \multicolumn{6}{c}{\mpiexscan} \\
    $n$ & $N$ & $p$ & $c$ & avg ($\mu s$) & min ($\mu s$) \\
    \midrule
2 & 36 & 72 & 1152 & 313.81 & 307.59 \\
2 & 36 & 72 & 11520 & 1608.24 & 1583.86 \\
2 & 36 & 72 & 115200 & 12937.17 & 12914.09 \\
2 & 36 & 72 & 1152000 & 121714.78 & 121597.82 \\
    \bottomrule
    \end{tabular}
    \end{center}
\end{table}

\clearpage

\section{Results with mpich}
\label{app:mpich}

With the \mpichversion library, we show results with $c=1152$
increasing to $c=11\,520\,000$ in multiples of $10$ and $n=32$. These
are listed in Tables~\ref{tab:lane.mpich.n32.c1152},
\ref{tab:multicoll.mpich.n32.c1152}, \ref{tab:bcast.mpich.n32.c1152},
\ref{tab:gather.mpich.n32.c1152}, \ref{tab:scatter.mpich.n32.c1152},
\ref{tab:allgather.mpich.n32.c1152},
\ref{tab:alltoall.mpich.n32.c1152},
\ref{tab:allreduce.mpich.n32.c1152}, \ref{tab:reduce.mpich.n32.c1152},
\ref{tab:reducescatter.mpich.n32.c1152},
\ref{tab:scan.mpich.n32.c1152},
and ~\ref{tab:exscan.mpich.n32.c1152}.

\begin{table}
  \caption{Lane pattern benchmark results on ``Hydra'' for increasing
    number of virtual lanes $k$ used for communicating the data (count
    $c$ \mpiint) per compute node. The MPI library used is \mpichversion.}
  \label{tab:lane.mpich.n32.c1152}
  \begin{center}
    \begin{tabular}{rrrrrrr}
      $k$ & $n$ & $N$ & $p$ & $c$ & avg ($\mu s$) & min ($\mu s$) \\
\toprule
1 & 32 & 36 & 1152 & 1152 & 232.78 & 208.14 \\
2 & 32 & 36 & 1152 & 1152 & 229.97 & 181.44 \\
4 & 32 & 36 & 1152 & 1152 & 224.15 & 161.17 \\
8 & 32 & 36 & 1152 & 1152 & 212.08 & 155.93 \\
16 & 32 & 36 & 1152 & 1152 & 212.28 & 152.35 \\
32 & 32 & 36 & 1152 & 1152 & 276.73 & 155.21 \\
\midrule
1 & 32 & 36 & 1152 & 11520 & 776.13 & 703.10 \\
2 & 32 & 36 & 1152 & 11520 & 624.24 & 584.13 \\
4 & 32 & 36 & 1152 & 11520 & 350.29 & 312.09 \\
8 & 32 & 36 & 1152 & 11520 & 299.11 & 276.57 \\
16 & 32 & 36 & 1152 & 11520 & 275.73 & 237.46 \\
32 & 32 & 36 & 1152 & 11520 & 288.86 & 209.09 \\
\midrule
1 & 32 & 36 & 1152 & 115200 & 3246.62 & 3135.44 \\
2 & 32 & 36 & 1152 & 115200 & 1891.03 & 1841.31 \\
4 & 32 & 36 & 1152 & 115200 & 1707.55 & 1665.59 \\
8 & 32 & 36 & 1152 & 115200 & 1377.56 & 1306.53 \\
16 & 32 & 36 & 1152 & 115200 & 1214.40 & 1152.52 \\
32 & 32 & 36 & 1152 & 115200 & 1494.65 & 1357.56 \\
\midrule
1 & 32 & 36 & 1152 & 1152000 & 25690.46 & 24283.89 \\
2 & 32 & 36 & 1152 & 1152000 & 13894.80 & 13383.87 \\
4 & 32 & 36 & 1152 & 1152000 & 11155.57 & 10936.26 \\
8 & 32 & 36 & 1152 & 1152000 & 10569.33 & 10462.28 \\
16 & 32 & 36 & 1152 & 1152000 & 10209.06 & 10108.71 \\
32 & 32 & 36 & 1152 & 1152000 & 10466.76 & 10299.44 \\
\midrule
1 & 32 & 36 & 1152 & 11520000 & 203822.55 & 201493.98 \\
2 & 32 & 36 & 1152 & 11520000 & 105915.67 & 104921.58 \\
4 & 32 & 36 & 1152 & 11520000 & 101360.24 & 100575.21 \\
8 & 32 & 36 & 1152 & 11520000 & 100847.23 & 100315.57 \\
16 & 32 & 36 & 1152 & 11520000 & 100338.68 & 99900.01 \\
32 & 32 & 36 & 1152 & 11520000 & 100540.88 & 100175.86 \\
\bottomrule
    \end{tabular}
  \end{center}
\end{table}

\begin{table}
  \caption{Multi-collective pattern benchmark results on ``Hydra'' for
    increasing number of virtual lanes $k$ used for communicating the
    data (count $c$ \mpiint) per lane. The collective function is
    \mpialltoall. The MPI library used is \mpichversion.}
  \label{tab:multicoll.mpich.n32.c1152}
  \begin{center}
    \begin{tabular}{rrrrrrr}
      $k$ & $n$ & $N$ & $p$ & $c$ & avg ($\mu s$) & min ($\mu s$) \\
\toprule
1 & 32 & 36 & 1152 & 1152 & 43.33 & 35.05 \\
2 & 32 & 36 & 1152 & 1152 & 44.45 & 34.81 \\
4 & 32 & 36 & 1152 & 1152 & 49.66 & 34.81 \\
8 & 32 & 36 & 1152 & 1152 & 44.31 & 35.05 \\
16 & 32 & 36 & 1152 & 1152 & 49.09 & 38.39 \\
32 & 32 & 36 & 1152 & 1152 & 60.56 & 50.54 \\
\midrule
1 & 32 & 36 & 1152 & 11520 & 57.23 & 51.26 \\
2 & 32 & 36 & 1152 & 11520 & 60.23 & 49.83 \\
4 & 32 & 36 & 1152 & 11520 & 59.94 & 47.92 \\
8 & 32 & 36 & 1152 & 11520 & 61.38 & 54.84 \\
16 & 32 & 36 & 1152 & 11520 & 78.95 & 71.05 \\
32 & 32 & 36 & 1152 & 11520 & 117.63 & 105.14 \\
\midrule
1 & 32 & 36 & 1152 & 115200 & 170.67 & 147.82 \\
2 & 32 & 36 & 1152 & 115200 & 159.11 & 145.20 \\
4 & 32 & 36 & 1152 & 115200 & 185.11 & 173.33 \\
8 & 32 & 36 & 1152 & 115200 & 257.33 & 241.99 \\
16 & 32 & 36 & 1152 & 115200 & 438.57 & 416.52 \\
32 & 32 & 36 & 1152 & 115200 & 874.87 & 848.29 \\
\midrule
1 & 32 & 36 & 1152 & 1152000 & 1089.80 & 1066.45 \\
2 & 32 & 36 & 1152 & 1152000 & 1095.15 & 1063.59 \\
4 & 32 & 36 & 1152 & 1152000 & 1376.23 & 1340.63 \\
8 & 32 & 36 & 1152 & 1152000 & 2008.60 & 1950.98 \\
16 & 32 & 36 & 1152 & 1152000 & 3585.19 & 3460.65 \\
32 & 32 & 36 & 1152 & 1152000 & 6782.77 & 6628.51 \\
\midrule
1 & 32 & 36 & 1152 & 11520000 & 6285.61 & 6155.25 \\
2 & 32 & 36 & 1152 & 11520000 & 6279.12 & 6192.92 \\
4 & 32 & 36 & 1152 & 11520000 & 9423.31 & 9221.08 \\
8 & 32 & 36 & 1152 & 11520000 & 16636.27 & 16500.95 \\
16 & 32 & 36 & 1152 & 11520000 & 32473.36 & 31785.49 \\
32 & 32 & 36 & 1152 & 11520000 & 69398.77 & 68089.72 \\
\bottomrule
    \end{tabular}
  \end{center}
\end{table}

\begin{table}
  \caption{Results for native \mpibcast compared against the mock-up
    guideline implementations on the ``Hydra'' system.  The MPI library used
    is \mpichversion.}
  \label{tab:bcast.mpich.n32.c1152}
  \begin{center}
    \begin{tabular}{crrrrrrr}
      \toprule
      \multicolumn{6}{c}{BcastLane} \\
      $n$ & $N$ & $p$ & $c$ & avg ($\mu s$) & min ($\mu s$) \\
      \midrule
32 & 36 & 1152 & 1152 & 49.37 & 41.48 \\
32 & 36 & 1152 & 11520 & 198.89 & 87.50 \\
32 & 36 & 1152 & 115200 & 813.49 & 405.07 \\
32 & 36 & 1152 & 1152000 & 6270.97 & 5522.01 \\
32 & 36 & 1152 & 11520000 & 57303.51 & 56267.74 \\
      \midrule
      \multicolumn{6}{c}{BcastHier} \\
      $n$ & $N$ & $p$ & $c$ & avg ($\mu s$) & min ($\mu s$) \\
      \midrule
32 & 36 & 1152 & 1152 & 44.08 & 36.48 \\
32 & 36 & 1152 & 11520 & 297.51 & 201.70 \\
32 & 36 & 1152 & 115200 & 1017.08 & 603.20 \\
32 & 36 & 1152 & 1152000 & 6705.37 & 6529.09 \\
32 & 36 & 1152 & 11520000 & 63319.17 & 62561.04 \\
      \midrule
      \multicolumn{6}{c}{\mpibcast} \\
      $n$ & $N$ & $p$ & $c$ & avg ($\mu s$) & min ($\mu s$) \\
      \midrule
32 & 36 & 1152 & 1152 & 44.01 & 31.23 \\
32 & 36 & 1152 & 11520 & 196.65 & 186.20 \\
32 & 36 & 1152 & 115200 & 660.03 & 597.24 \\
32 & 36 & 1152 & 1152000 & 6705.24 & 6564.62 \\
32 & 36 & 1152 & 11520000 & 64502.77 & 63453.20 \\
      \bottomrule
    \end{tabular}
  \end{center}
\end{table}

\begin{table}
  \caption{Results for native \mpigather compared against the mock-up
    guideline implementations on the ``Hydra'' system.  The MPI library used
    is \mpichversion.}
  \label{tab:gather.mpich.n32.c1152}
  \begin{center}
    \begin{tabular}{crrrrrrr}
      \toprule
      \multicolumn{6}{c}{GatherLane} \\
      $n$ & $N$ & $p$ & $c$ & avg ($\mu s$) & min ($\mu s$) \\
      \midrule
32 & 36 & 1152 & 1 & 29.14 & 22.89 \\
32 & 36 & 1152 & 10 & 41.69 & 36.00 \\
32 & 36 & 1152 & 100 & 131.29 & 126.60 \\
32 & 36 & 1152 & 1000 & 1063.11 & 1044.51 \\
32 & 36 & 1152 & 10000 & 11862.39 & 11593.34 \\
      \midrule
      \multicolumn{6}{c}{GatherHier} \\
      $n$ & $N$ & $p$ & $c$ & avg ($\mu s$) & min ($\mu s$) \\
      \midrule
      32 & 36 & 1152 & 1 & 29.02 & 23.60 \\
32 & 36 & 1152 & 10 & 43.05 & 39.10 \\
32 & 36 & 1152 & 100 & 105.24 & 99.18 \\
32 & 36 & 1152 & 1000 & 506.79 & 498.77 \\
32 & 36 & 1152 & 10000 & 5072.92 & 4982.23 \\
      \midrule
      \multicolumn{6}{c}{\mpigather} \\
      $n$ & $N$ & $p$ & $c$ & avg ($\mu s$) & min ($\mu s$) \\
      \midrule
32 & 36 & 1152 & 1 & 27.11 & 23.84 \\
32 & 36 & 1152 & 10 & 38.57 & 34.33 \\
32 & 36 & 1152 & 100 & 100.41 & 97.04 \\
32 & 36 & 1152 & 1000 & 511.66 & 505.21 \\
32 & 36 & 1152 & 10000 & 4161.52 & 4138.71 \\
      \bottomrule
    \end{tabular}
  \end{center}
\end{table}

\begin{table}
  \caption{Results for native \mpiscatter compared against the mock-up
    guideline implementations on the ``Hydra'' system.  The MPI library used
    is \mpichversion.}
  \label{tab:scatter.mpich.n32.c1152}
  \begin{center}
    \begin{tabular}{crrrrrrr}
      \toprule
      \multicolumn{6}{c}{ScatterLane} \\
      $n$ & $N$ & $p$ & $c$ & avg ($\mu s$) & min ($\mu s$) \\
      \midrule
32 & 36 & 1152 & 1 & 24.64 & 21.46 \\
32 & 36 & 1152 & 10 & 43.00 & 40.29 \\
32 & 36 & 1152 & 100 & 121.18 & 117.06 \\
32 & 36 & 1152 & 1000 & 1022.21 & 979.90 \\
32 & 36 & 1152 & 10000 & 12023.90 & 11529.45 \\
      \midrule
      \multicolumn{6}{c}{ScatterHier} \\
      $n$ & $N$ & $p$ & $c$ & avg ($\mu s$) & min ($\mu s$) \\
      \midrule
32 & 36 & 1152 & 1 & 20.92 & 18.12 \\
32 & 36 & 1152 & 10 & 33.58 & 29.80 \\
32 & 36 & 1152 & 100 & 104.54 & 101.57 \\
32 & 36 & 1152 & 1000 & 507.18 & 495.20 \\
32 & 36 & 1152 & 10000 & 4248.73 & 4232.65 \\
      \midrule
      \multicolumn{6}{c}{\mpiscatter} \\
      $n$ & $N$ & $p$ & $c$ & avg ($\mu s$) & min ($\mu s$) \\
      \midrule
32 & 36 & 1152 & 1 & 20.75 & 18.36 \\
32 & 36 & 1152 & 10 & 32.07 & 30.04 \\
32 & 36 & 1152 & 100 & 103.31 & 101.09 \\
32 & 36 & 1152 & 1000 & 497.36 & 493.05 \\
32 & 36 & 1152 & 10000 & 4154.26 & 4101.28 \\
      \bottomrule
    \end{tabular}
  \end{center}
\end{table}

\begin{table}
  \caption{Results for native \mpiallgather compared against the mock-up
    guideline implementations on the ``Hydra'' system.  The MPI library used
    is \mpichversion.}
  \label{tab:allgather.mpich.n32.c1152}
  \begin{center}
    \begin{tabular}{crrrrrrr}
      \toprule
      \multicolumn{6}{c}{AllgatherLane} \\
      $n$ & $N$ & $p$ & $c$ & avg ($\mu s$) & min ($\mu s$) \\
      \midrule
32 & 36 & 1152 & 1 & 62.80 & 53.41 \\
32 & 36 & 1152 & 10 & 123.94 & 95.37 \\
32 & 36 & 1152 & 100 & 584.01 & 314.24 \\
32 & 36 & 1152 & 1000 & 6064.94 & 5856.99 \\
32 & 36 & 1152 & 10000 & 58873.78 & 58331.73 \\
      \midrule
      \multicolumn{6}{c}{AllgatherHier} \\
      $n$ & $N$ & $p$ & $c$ & avg ($\mu s$) & min ($\mu s$) \\
      \midrule
32 & 36 & 1152 & 1 & 44.57 & 39.58 \\
32 & 36 & 1152 & 10 & 106.78 & 92.74 \\
32 & 36 & 1152 & 100 & 529.55 & 471.35 \\
32 & 36 & 1152 & 1000 & 6162.46 & 6058.22 \\
32 & 36 & 1152 & 10000 & 59901.55 & 59194.56 \\
\midrule
      \multicolumn{6}{c}{\mpiallgather} \\
      $n$ & $N$ & $p$ & $c$ & avg ($\mu s$) & min ($\mu s$) \\
      \midrule
32 & 36 & 1152 & 1 & 51.47 & 40.53 \\
32 & 36 & 1152 & 10 & 141.02 & 113.73 \\
32 & 36 & 1152 & 100 & 5036.14 & 4267.93 \\
32 & 36 & 1152 & 1000 & 7572.00 & 6687.40 \\
32 & 36 & 1152 & 10000 & 42900.54 & 40843.73 \\
      \bottomrule
    \end{tabular}
  \end{center}
\end{table}

\begin{table}
  \caption{Results for native \mpialltoall compared against the mock-up
    guideline implementation on the ``Hydra'' system.  The MPI library used
    is \mpichversion.}
  \label{tab:alltoall.mpich.n32.c1152}
  \begin{center}
    \begin{tabular}{crrrrrrr}
      \toprule
      \multicolumn{6}{c}{AlltoallLane} \\
      $n$ & $N$ & $p$ & $c$ & avg ($\mu s$) & min ($\mu s$) \\
      \midrule
32 & 36 & 1152 & 1 & 187.12 & 113.96 \\
32 & 36 & 1152 & 10 & 248.91 & 200.99 \\
32 & 36 & 1152 & 100 & 1514.48 & 1384.02 \\
32 & 36 & 1152 & 1000 & 12765.28 & 12515.31 \\
32 & 36 & 1152 & 10000 & 128732.26 & 126021.62 \\
      \midrule
      \multicolumn{6}{c}{\mpialltoall} \\
      $n$ & $N$ & $p$ & $c$ & avg ($\mu s$) & min ($\mu s$) \\
      \midrule
32 & 36 & 1152 & 1 & 431.21 & 408.65 \\
32 & 36 & 1152 & 10 & 791.83 & 625.85 \\
32 & 36 & 1152 & 100 & 2425.90 & 2072.10 \\
32 & 36 & 1152 & 1000 & 12882.62 & 12226.34 \\
32 & 36 & 1152 & 10000 & 119425.71 & 116754.53 \\
      \bottomrule
    \end{tabular}
  \end{center}
\end{table}

\begin{table}
  \caption{Results for native \mpiallreduce compared against the mock-up
    guideline implementations on the ``Hydra'' system.  The MPI library used
    is \mpichversion.}
  \label{tab:allreduce.mpich.n32.c1152}
  \begin{center}
    \begin{tabular}{crrrrrrr}
      \toprule
      \multicolumn{6}{c}{AllreduceLane} \\
      $n$ & $N$ & $p$ & $c$ & avg ($\mu s$) & min ($\mu s$) \\
      \midrule
32 & 36 & 1152 & 1152 & 60.02 & 52.69 \\
32 & 36 & 1152 & 11520 & 114.48 & 100.61 \\
32 & 36 & 1152 & 115200 & 866.86 & 796.32 \\
32 & 36 & 1152 & 1152000 & 8707.33 & 8510.59 \\
32 & 36 & 1152 & 11520000 & 113937.38 & 112956.76 \\
      \midrule
      \multicolumn{6}{c}{AllreduceHier} \\
      $n$ & $N$ & $p$ & $c$ & avg ($\mu s$) & min ($\mu s$) \\
      \midrule
32 & 36 & 1152 & 1152 & 80.14 & 73.43 \\
32 & 36 & 1152 & 11520 & 264.99 & 218.39 \\
32 & 36 & 1152 & 115200 & 1582.83 & 1488.69 \\
32 & 36 & 1152 & 1152000 & 21246.47 & 20580.05 \\
32 & 36 & 1152 & 11520000 & 227814.92 & 224124.43 \\
      \midrule
      \multicolumn{6}{c}{\mpiallreduce} \\
      $n$ & $N$ & $p$ & $c$ & avg ($\mu s$) & min ($\mu s$) \\
      \midrule
32 & 36 & 1152 & 1152 & 98.56 & 74.63 \\
32 & 36 & 1152 & 11520 & 244.12 & 217.68 \\
32 & 36 & 1152 & 115200 & 1558.95 & 1478.67 \\
32 & 36 & 1152 & 1152000 & 21133.15 & 20495.41 \\
32 & 36 & 1152 & 11520000 & 223564.34 & 221547.60 \\
      \bottomrule
    \end{tabular}
  \end{center}
\end{table}

\begin{table}
  \caption{Results for native \mpireduce compared against the mock-up
    guideline implementations on the ``Hydra'' system.  The MPI library used
    is \mpichversion.}
  \label{tab:reduce.mpich.n32.c1152}
  \begin{center}
    \begin{tabular}{crrrrrrr}
      \toprule
      \multicolumn{6}{c}{ReduceLane} \\
      $n$ & $N$ & $p$ & $c$ & avg ($\mu s$) & min ($\mu s$) \\
      \midrule
32 & 36 & 1152 & 1152 & 42.74 & 34.57 \\
32 & 36 & 1152 & 11520 & 83.96 & 75.10 \\
32 & 36 & 1152 & 115200 & 533.64 & 507.35 \\
32 & 36 & 1152 & 1152000 & 5666.00 & 5358.46 \\
32 & 36 & 1152 & 11520000 & 79701.51 & 78959.94 \\
      \midrule
      \multicolumn{6}{c}{ReduceHier} \\
      $n$ & $N$ & $p$ & $c$ & avg ($\mu s$) & min ($\mu s$) \\
      \midrule
32 & 36 & 1152 & 1152 & 33.90 & 27.89 \\
32 & 36 & 1152 & 11520 & 127.07 & 92.03 \\
32 & 36 & 1152 & 115200 & 864.01 & 692.61 \\
32 & 36 & 1152 & 1152000 & 12584.36 & 11826.52 \\
32 & 36 & 1152 & 11520000 & 125895.31 & 123850.11 \\
      \midrule
      \multicolumn{6}{c}{\mpireduce} \\
      $n$ & $N$ & $p$ & $c$ & avg ($\mu s$) & min ($\mu s$) \\
      \midrule
32 & 36 & 1152 & 1152 & 84.94 & 76.29 \\
32 & 36 & 1152 & 11520 & 217.03 & 207.66 \\
32 & 36 & 1152 & 115200 & 1409.92 & 1377.82 \\
32 & 36 & 1152 & 1152000 & 22391.99 & 21353.72 \\
32 & 36 & 1152 & 11520000 & 240272.61 & 236975.67 \\
      \bottomrule
    \end{tabular}
  \end{center}
\end{table}

\begin{table}
  \caption{Results for native \mpireducescatterblock compared against the mock-up
    guideline implementations on the ``Hydra'' system.  The MPI library used
    is \mpichversion.}
  \label{tab:reducescatter.mpich.n32.c1152}
  \begin{center}
    \begin{tabular}{crrrrrrr}
      \toprule
      \multicolumn{6}{c}{ReduceScatterBlockLane} \\
      $n$ & $N$ & $p$ & $c$ & avg ($\mu s$) & min ($\mu s$) \\
      \midrule
32 & 36 & 1152 & 1 & 57.28 & 39.82 \\
32 & 36 & 1152 & 10 & 122.61 & 80.82 \\
32 & 36 & 1152 & 100 & 912.61 & 763.89 \\
32 & 36 & 1152 & 1000 & 9224.34 & 8980.75 \\
32 & 36 & 1152 & 10000 & 125734.05 & 124442.82 \\
      \midrule
      \multicolumn{6}{c}{ReduceScatterBlockLane} \\
      $n$ & $N$ & $p$ & $c$ & avg ($\mu s$) & min ($\mu s$) \\
      \midrule
32 & 36 & 1152 & 1 & 88.07 & 51.50 \\
32 & 36 & 1152 & 10 & 219.05 & 144.96 \\
32 & 36 & 1152 & 100 & 1253.69 & 1029.97 \\
32 & 36 & 1152 & 1000 & 15415.39 & 14077.66 \\
32 & 36 & 1152 & 10000 & 145719.78 & 142099.86 \\
      \midrule
      \multicolumn{6}{c}{\mpireducescatterblock} \\
      $n$ & $N$ & $p$ & $c$ & avg ($\mu s$) & min ($\mu s$) \\
      \midrule
32 & 36 & 1152 & 1 & 53.39 & 46.73 \\
32 & 36 & 1152 & 10 & 178.34 & 165.70 \\
32 & 36 & 1152 & 100 & 1579.15 & 1448.15 \\
32 & 36 & 1152 & 1000 & 13263.14 & 12807.37 \\
32 & 36 & 1152 & 10000 & 118800.54 & 116559.27 \\
      \bottomrule
    \end{tabular}
  \end{center}
\end{table}

\begin{table}
  \caption{Results for native \mpiscan compared against the mock-up
    guideline implementations on the ``Hydra'' system.  The MPI library used
    is \openmpiversion.}
    \label{tab:scan.mpich.n32.c1152}
  \begin{center}
    \begin{tabular}{crrrrrrr}
      \toprule
    \multicolumn{6}{c}{ScanLane} \\
      $n$ & $N$ & $p$ & $c$ & avg ($\mu s$) & min ($\mu s$) \\
    \midrule
32 & 36 & 1152 & 1152 & 104.34 & 91.79 \\
32 & 36 & 1152 & 11520 & 386.50 & 348.81 \\
32 & 36 & 1152 & 115200 & 4001.00 & 3946.07 \\
32 & 36 & 1152 & 1152000 & 66432.70 & 65996.17 \\
32 & 36 & 1152 & 11520000 & 727650.36 & 724220.28 \\
    \midrule
    \multicolumn{6}{c}{ScanHier} \\
    $n$ & $N$ & $p$ & $c$ & avg ($\mu s$) & min ($\mu s$) \\
    \midrule
32 & 36 & 1152 & 1152 & 100.40 & 85.35 \\
32 & 36 & 1152 & 11520 & 519.30 & 489.00 \\
32 & 36 & 1152 & 115200 & 4981.30 & 4903.32 \\
32 & 36 & 1152 & 1152000 & 87140.01 & 86146.59 \\
32 & 36 & 1152 & 11520000 & 857327.76 & 853897.57 \\
    \midrule
    \multicolumn{6}{c}{\mpiscan} \\
    $n$ & $N$ & $p$ & $c$ & avg ($\mu s$) & min ($\mu s$) \\
    \midrule
32 & 36 & 1152 & 1152 & 130.27 & 85.59 \\
32 & 36 & 1152 & 11520 & 558.80 & 492.81 \\
32 & 36 & 1152 & 115200 & 6730.49 & 6654.50 \\
32 & 36 & 1152 & 1152000 & 93993.21 & 93306.30 \\
32 & 36 & 1152 & 11520000 & 1122829.96 & 1116591.69 \\
    \bottomrule
    \end{tabular}
    \end{center}
\end{table}

\begin{table}
  \caption{Results for native \mpiexscan compared against the mock-up
    guideline implementations on the ``Hydra'' system.  The MPI library used
    is \openmpiversion.}
    \label{tab:exscan.mpich.n32.c1152}
  \begin{center}
    \begin{tabular}{crrrrrrr}
      \toprule
    \multicolumn{6}{c}{ExscanLane} \\
      $n$ & $N$ & $p$ & $c$ & avg ($\mu s$) & min ($\mu s$) \\
    \midrule
32 & 36 & 1152 & 1152 & 126.86 & 82.97 \\
32 & 36 & 1152 & 11520 & 524.73 & 308.99 \\
32 & 36 & 1152 & 115200 & 4186.33 & 3329.99 \\
32 & 36 & 1152 & 1152000 & 57939.11 & 57158.47 \\
32 & 36 & 1152 & 11520000 & 633589.22 & 631589.41 \\
    \midrule
    \multicolumn{6}{c}{ExscanHier} \\
      $n$ & $N$ & $p$ & $c$ & avg ($\mu s$) & min ($\mu s$) \\
    \midrule
32 & 36 & 1152 & 1152 & 96.80 & 77.96 \\
32 & 36 & 1152 & 11520 & 510.28 & 461.58 \\
32 & 36 & 1152 & 115200 & 4994.10 & 4896.88 \\
32 & 36 & 1152 & 1152000 & 76358.60 & 74630.74 \\
32 & 36 & 1152 & 11520000 & 824986.27 & 819083.45 \\
    \midrule
    \multicolumn{6}{c}{\mpiexscan} \\
    $n$ & $N$ & $p$ & $c$ & avg ($\mu s$) & min ($\mu s$) \\
    \midrule
32 & 36 & 1152 & 1152 & 132.09 & 113.73 \\
32 & 36 & 1152 & 11520 & 785.83 & 648.50 \\
32 & 36 & 1152 & 115200 & 6210.79 & 6103.28 \\
32 & 36 & 1152 & 1152000 & 91143.54 & 89541.44 \\
32 & 36 & 1152 & 11520000 & 915955.27 & 896526.34 \\
    \bottomrule
    \end{tabular}
    \end{center}
\end{table}

\clearpage

\section{Results with Intel MPI}
\label{app:intelmpi}

With the \intelmpiversion library, we show results with $c=1152$
increasing to $c=11\,520\,000$ in multiples of $10$ and $n=32$. These
are listed in Tables~\ref{tab:lane.intel.n32.c1152},
\ref{tab:multicoll.intel.n32.c1152}, \ref{tab:bcast.intel.n32.c1152},
\ref{tab:gather.intel.n32.c1152}, \ref{tab:scatter.intel.n32.c1152},
\ref{tab:allgather.intel.n32.c1152},
\ref{tab:alltoall.intel.n32.c1152},
\ref{tab:allreduce.intel.n32.c1152}, \ref{tab:reduce.intel.n32.c1152},
\ref{tab:reducescatter.intel.n32.c1152},
\ref{tab:scan.intel.n32.c1152}
and~\ref{tab:exscan.intel.n32.c1152}.

\begin{table}
  \caption{Lane pattern benchmark results on ``Hydra'' for increasing
    number of virtual lanes $k$ used for communicating the data (count
    $c$ \mpiint) per compute node. The MPI library used is \intelmpiversion.}
  \label{tab:lane.intel.n32.c1152}
  \begin{center}
    \begin{tabular}{rrrrrrr}
      $k$ & $n$ & $N$ & $p$ & $c$ & avg ($\mu s$) & min ($\mu s$) \\
\toprule
1 & 32 & 36 & 1152 & 1152 & 204.22 & 188.11 \\
2 & 32 & 36 & 1152 & 1152 & 171.62 & 159.98 \\
4 & 32 & 36 & 1152 & 1152 & 165.24 & 154.97 \\
8 & 32 & 36 & 1152 & 1152 & 162.68 & 153.78 \\
16 & 32 & 36 & 1152 & 1152 & 160.97 & 149.97 \\
32 & 32 & 36 & 1152 & 1152 & 159.45 & 149.01 \\
\midrule
1 & 32 & 36 & 1152 & 11520 & 689.79 & 653.03 \\
2 & 32 & 36 & 1152 & 11520 & 535.96 & 508.07 \\
4 & 32 & 36 & 1152 & 11520 & 353.42 & 331.88 \\
8 & 32 & 36 & 1152 & 11520 & 293.10 & 283.00 \\
16 & 32 & 36 & 1152 & 11520 & 253.03 & 241.04 \\
32 & 32 & 36 & 1152 & 11520 & 228.48 & 215.05 \\
\midrule
1 & 32 & 36 & 1152 & 115200 & 3350.98 & 3163.81 \\
2 & 32 & 36 & 1152 & 115200 & 1961.88 & 1895.90 \\
4 & 32 & 36 & 1152 & 115200 & 1724.61 & 1679.90 \\
8 & 32 & 36 & 1152 & 115200 & 1528.14 & 1399.99 \\
16 & 32 & 36 & 1152 & 115200 & 1234.70 & 1168.97 \\
32 & 32 & 36 & 1152 & 115200 & 1404.77 & 1367.09 \\
\midrule
1 & 32 & 36 & 1152 & 1152000 & 25827.80 & 24452.21 \\
2 & 32 & 36 & 1152 & 1152000 & 14015.52 & 13465.88 \\
4 & 32 & 36 & 1152 & 1152000 & 11126.28 & 10977.98 \\
8 & 32 & 36 & 1152 & 1152000 & 10546.33 & 10473.01 \\
16 & 32 & 36 & 1152 & 1152000 & 10227.28 & 10149.96 \\
32 & 32 & 36 & 1152 & 1152000 & 10515.76 & 10395.05 \\
\midrule
1 & 32 & 36 & 1152 & 11520000 & 203769.09 & 202217.10 \\
2 & 32 & 36 & 1152 & 11520000 & 105724.21 & 104517.94 \\
4 & 32 & 36 & 1152 & 11520000 & 101395.20 & 100646.02 \\
8 & 32 & 36 & 1152 & 11520000 & 101092.90 & 100641.97 \\
16 & 32 & 36 & 1152 & 11520000 & 100803.36 & 100234.03 \\
32 & 32 & 36 & 1152 & 11520000 & 102319.14 & 101988.08 \\
\bottomrule
    \end{tabular}
  \end{center}
\end{table}

\begin{table}
  \caption{Multi-collective pattern benchmark results on ``Hydra'' for
    increasing number of virtual lanes $k$ used for communicating the
    data (count $c$ \mpiint) per lane. The collective function is
    \mpialltoall. The MPI library used is \intelmpiversion.}
  \label{tab:multicoll.intel.n32.c1152}
  \begin{center}
    \begin{tabular}{rrrrrrr}
      $k$ & $n$ & $N$ & $p$ & $c$ & avg ($\mu s$) & min ($\mu s$) \\
\toprule
1 & 32 & 36 & 1152 & 1152 & 32.36 & 28.13 \\
2 & 32 & 36 & 1152 & 1152 & 33.56 & 30.04 \\
4 & 32 & 36 & 1152 & 1152 & 34.06 & 30.99 \\
8 & 32 & 36 & 1152 & 1152 & 35.55 & 30.99 \\
16 & 32 & 36 & 1152 & 1152 & 42.00 & 35.05 \\
32 & 32 & 36 & 1152 & 1152 & 54.31 & 50.07 \\
\midrule
1 & 32 & 36 & 1152 & 11520 & 44.10 & 41.01 \\
2 & 32 & 36 & 1152 & 11520 & 45.11 & 41.01 \\
4 & 32 & 36 & 1152 & 11520 & 50.83 & 42.20 \\
8 & 32 & 36 & 1152 & 11520 & 53.04 & 47.92 \\
16 & 32 & 36 & 1152 & 11520 & 72.36 & 66.04 \\
32 & 32 & 36 & 1152 & 11520 & 111.61 & 102.04 \\
\midrule
1 & 32 & 36 & 1152 & 115200 & 264.05 & 200.99 \\
2 & 32 & 36 & 1152 & 115200 & 275.46 & 226.97 \\
4 & 32 & 36 & 1152 & 115200 & 296.57 & 280.14 \\
8 & 32 & 36 & 1152 & 115200 & 386.16 & 364.78 \\
16 & 32 & 36 & 1152 & 115200 & 651.77 & 612.02 \\
32 & 32 & 36 & 1152 & 115200 & 1028.79 & 926.02 \\
\midrule
1 & 32 & 36 & 1152 & 1152000 & 1118.67 & 1049.04 \\
2 & 32 & 36 & 1152 & 1152000 & 1080.26 & 1058.82 \\
4 & 32 & 36 & 1152 & 1152000 & 1365.93 & 1343.97 \\
8 & 32 & 36 & 1152 & 1152000 & 1997.04 & 1956.94 \\
16 & 32 & 36 & 1152 & 1152000 & 3527.34 & 3479.96 \\
32 & 32 & 36 & 1152 & 1152000 & 6867.92 & 6818.06 \\
\midrule
1 & 32 & 36 & 1152 & 11520000 & 6449.66 & 6165.03 \\
2 & 32 & 36 & 1152 & 11520000 & 6328.89 & 6166.93 \\
4 & 32 & 36 & 1152 & 11520000 & 9396.34 & 9255.89 \\
8 & 32 & 36 & 1152 & 11520000 & 16681.35 & 16541.96 \\
16 & 32 & 36 & 1152 & 11520000 & 32052.58 & 31856.06 \\
32 & 32 & 36 & 1152 & 11520000 & 63829.26 & 63449.86 \\
\bottomrule
    \end{tabular}
  \end{center}
\end{table}

\begin{table}
  \caption{Results for native \mpibcast compared against the mock-up
    guideline implementations on the ``Hydra'' system.  The MPI library used
    is \intelmpiversion.}
  \label{tab:bcast.intel.n32.c1152}
  \begin{center}
    \begin{tabular}{crrrrrrr}
      \toprule
      \multicolumn{6}{c}{BcastLane} \\
      $n$ & $N$ & $p$ & $c$ & avg ($\mu s$) & min ($\mu s$) \\
      \midrule
32 & 36 & 1152 & 1152 & 150.64 & 144.00 \\
32 & 36 & 1152 & 11520 & 196.56 & 183.82 \\
32 & 36 & 1152 & 115200 & 786.34 & 701.90 \\
32 & 36 & 1152 & 1152000 & 5006.78 & 4487.04 \\
32 & 36 & 1152 & 11520000 & 47107.13 & 44135.09 \\
      \midrule
      \multicolumn{6}{c}{BcastHier} \\
      $n$ & $N$ & $p$ & $c$ & avg ($\mu s$) & min ($\mu s$) \\
      \midrule
32 & 36 & 1152 & 1152 & 174.13 & 170.95 \\
32 & 36 & 1152 & 11520 & 378.98 & 360.97 \\
32 & 36 & 1152 & 115200 & 1060.61 & 962.02 \\
32 & 36 & 1152 & 1152000 & 8810.80 & 8598.09 \\
32 & 36 & 1152 & 11520000 & 95585.24 & 92481.85 \\
      \midrule
      \multicolumn{6}{c}{\mpibcast} \\
      $n$ & $N$ & $p$ & $c$ & avg ($\mu s$) & min ($\mu s$) \\
      \midrule
32 & 36 & 1152 & 1152 & 2838.98 & 2816.92 \\
32 & 36 & 1152 & 11520 & 5935.11 & 5890.13 \\
32 & 36 & 1152 & 115200 & 7702.32 & 7539.03 \\
32 & 36 & 1152 & 1152000 & 16376.67 & 15994.07 \\
32 & 36 & 1152 & 11520000 & 127934.29 & 113408.09 \\
      \bottomrule
    \end{tabular}
  \end{center}
\end{table}

\begin{table}
  \caption{Results for native \mpigather compared against the mock-up
    guideline implementations on the ``Hydra'' system.  The MPI library used
    is \intelmpiversion.}
  \label{tab:gather.intel.n32.c1152}
  \begin{center}
    \begin{tabular}{crrrrrrr}
      \toprule
      \multicolumn{6}{c}{GatherLane} \\
      $n$ & $N$ & $p$ & $c$ & avg ($\mu s$) & min ($\mu s$) \\
      \midrule
32 & 36 & 1152 & 1 & 27.86 & 22.89 \\
32 & 36 & 1152 & 10 & 73.60 & 58.89 \\
32 & 36 & 1152 & 100 & 102.22 & 97.99 \\
32 & 36 & 1152 & 1000 & 1496.20 & 1362.09 \\
32 & 36 & 1152 & 10000 & 14730.07 & 14376.88 \\
      \midrule
      \multicolumn{6}{c}{GatherHier} \\
      $n$ & $N$ & $p$ & $c$ & avg ($\mu s$) & min ($\mu s$) \\
      \midrule
32 & 36 & 1152 & 1 & 25.23 & 20.03 \\
32 & 36 & 1152 & 10 & 37.28 & 30.99 \\
32 & 36 & 1152 & 100 & 71.76 & 68.19 \\
32 & 36 & 1152 & 1000 & 414.29 & 411.99 \\
32 & 36 & 1152 & 10000 & 3876.30 & 3865.00 \\
      \midrule
      \multicolumn{6}{c}{\mpigather} \\
      $n$ & $N$ & $p$ & $c$ & avg ($\mu s$) & min ($\mu s$) \\
      \midrule
32 & 36 & 1152 & 1 & 73.04 & 69.86 \\
32 & 36 & 1152 & 10 & 78.11 & 75.10 \\
32 & 36 & 1152 & 100 & 3200.29 & 2674.10 \\
32 & 36 & 1152 & 1000 & 3752.01 & 3566.98 \\
32 & 36 & 1152 & 10000 & 1005838.56 & 649173.02 \\
      \bottomrule
    \end{tabular}
  \end{center}
\end{table}

\begin{table}
  \caption{Results for native \mpiscatter compared against the mock-up
    guideline implementations on the ``Hydra'' system.  The MPI library used
    is \intelmpiversion.}
  \label{tab:scatter.intel.n32.c1152}
  \begin{center}
    \begin{tabular}{crrrrrrr}
      \toprule
      \multicolumn{6}{c}{ScatterLane} \\
      $n$ & $N$ & $p$ & $c$ & avg ($\mu s$) & min ($\mu s$) \\
      \midrule
32 & 36 & 1152 & 1 & 18.84 & 14.07 \\
32 & 36 & 1152 & 10 & 53.24 & 51.02 \\
32 & 36 & 1152 & 100 & 150.88 & 149.01 \\
32 & 36 & 1152 & 1000 & 815.14 & 802.04 \\
32 & 36 & 1152 & 10000 & 6417.43 & 6255.87 \\
      \midrule
      \multicolumn{6}{c}{ScatterHier} \\
      $n$ & $N$ & $p$ & $c$ & avg ($\mu s$) & min ($\mu s$) \\
      \midrule
32 & 36 & 1152 & 1 & 20.01 & 17.88 \\
32 & 36 & 1152 & 10 & 19.68 & 19.07 \\
32 & 36 & 1152 & 100 & 67.24 & 66.04 \\
32 & 36 & 1152 & 1000 & 419.74 & 416.99 \\
32 & 36 & 1152 & 10000 & 4036.87 & 4009.01 \\
      \midrule
      \multicolumn{6}{c}{\mpiscatter} \\
      $n$ & $N$ & $p$ & $c$ & avg ($\mu s$) & min ($\mu s$) \\
      \midrule
32 & 36 & 1152 & 1 & 20.15 & 16.93 \\
32 & 36 & 1152 & 10 & 31.81 & 29.09 \\
32 & 36 & 1152 & 100 & 559.12 & 550.03 \\
32 & 36 & 1152 & 1000 & 986.76 & 978.95 \\
32 & 36 & 1152 & 10000 & 4109.13 & 4098.89 \\
      \bottomrule
    \end{tabular}
  \end{center}
\end{table}

\begin{table}
  \caption{Results for native \mpiallgather compared against the mock-up
    guideline implementations on the ``Hydra'' system.  The MPI library used
    is \intelmpiversion.}
  \label{tab:allgather.intel.n32.c1152}
  \begin{center}
    \begin{tabular}{crrrrrrr}
      \toprule
      \multicolumn{6}{c}{AllgatherLane} \\
      $n$ & $N$ & $p$ & $c$ & avg ($\mu s$) & min ($\mu s$) \\
      \midrule
32 & 36 & 1152 & 1 & 125.80 & 118.97 \\
32 & 36 & 1152 & 10 & 149.84 & 142.81 \\
32 & 36 & 1152 & 100 & 600.08 & 591.99 \\
32 & 36 & 1152 & 1000 & 5539.53 & 5331.04 \\
32 & 36 & 1152 & 10000 & 38672.78 & 38158.89 \\
      \midrule
      \multicolumn{6}{c}{AllgatherHier} \\
      $n$ & $N$ & $p$ & $c$ & avg ($\mu s$) & min ($\mu s$) \\
      \midrule
32 & 36 & 1152 & 1 & 112.84 & 108.00 \\
32 & 36 & 1152 & 10 & 298.53 & 288.96 \\
32 & 36 & 1152 & 100 & 827.54 & 803.95 \\
32 & 36 & 1152 & 1000 & 6506.40 & 6441.12 \\
32 & 36 & 1152 & 10000 & 62407.69 & 58815.96 \\
      \midrule
      \multicolumn{6}{c}{\mpiallgather} \\
      $n$ & $N$ & $p$ & $c$ & avg ($\mu s$) & min ($\mu s$) \\
      \midrule
32 & 36 & 1152 & 1 & 54.92 & 45.06 \\
32 & 36 & 1152 & 10 & 450.48 & 442.03 \\
32 & 36 & 1152 & 100 & 1193.44 & 1168.01 \\
32 & 36 & 1152 & 1000 & 8319.15 & 7808.21 \\
32 & 36 & 1152 & 10000 & 78019.10 & 51890.13 \\
      \bottomrule
    \end{tabular}
  \end{center}
\end{table}

\begin{table}
  \caption{Results for native \mpialltoall compared against the mock-up
    guideline implementation on the ``Hydra'' system.  The MPI library used
    is \intelmpiversion.}
  \label{tab:alltoall.intel.n32.c1152}
  \begin{center}
    \begin{tabular}{crrrrrrr}
      \toprule
      \multicolumn{6}{c}{AlltoallLane} \\
      $n$ & $N$ & $p$ & $c$ & avg ($\mu s$) & min ($\mu s$) \\
      \midrule
32 & 36 & 1152 & 1 & 112.23 & 103.95 \\
32 & 36 & 1152 & 10 & 193.19 & 185.97 \\
32 & 36 & 1152 & 100 & 1500.13 & 1458.88 \\
32 & 36 & 1152 & 1000 & 12183.86 & 12088.06 \\
32 & 36 & 1152 & 10000 & 129355.99 & 120849.85 \\
      \midrule
      \multicolumn{6}{c}{\mpialltoall} \\
      $n$ & $N$ & $p$ & $c$ & avg ($\mu s$) & min ($\mu s$) \\
      \midrule
32 & 36 & 1152 & 1 & 390.54 & 190.02 \\
32 & 36 & 1152 & 10 & 644.00 & 420.09 \\
32 & 36 & 1152 & 100 & 5081.86 & 4900.93 \\
32 & 36 & 1152 & 1000 & 14234.51 & 14036.18 \\
32 & 36 & 1152 & 10000 & 118267.09 & 116883.04 \\
      \bottomrule
    \end{tabular}
  \end{center}
\end{table}

\begin{table}
  \caption{Results for native \mpiallreduce compared against the mock-up
    guideline implementations on the ``Hydra'' system.  The MPI library used
    is \intelmpiversion.}
  \label{tab:allreduce.intel.n32.c1152}
  \begin{center}
    \begin{tabular}{crrrrrrr}
      \toprule
      \multicolumn{6}{c}{AllreduceLane} \\
      $n$ & $N$ & $p$ & $c$ & avg ($\mu s$) & min ($\mu s$) \\
      \midrule
32 & 36 & 1152 & 1152 & 123.86 & 108.96 \\
32 & 36 & 1152 & 11520 & 177.15 & 158.07 \\
32 & 36 & 1152 & 115200 & 894.14 & 867.13 \\
32 & 36 & 1152 & 1152000 & 7516.50 & 7138.97 \\
32 & 36 & 1152 & 11520000 & 84169.85 & 81623.08 \\
      \midrule
      \multicolumn{6}{c}{AllreduceHier} \\
      $n$ & $N$ & $p$ & $c$ & avg ($\mu s$) & min ($\mu s$) \\
      \midrule
32 & 36 & 1152 & 1152 & 249.06 & 125.89 \\
32 & 36 & 1152 & 11520 & 391.29 & 319.00 \\
32 & 36 & 1152 & 115200 & 1535.27 & 1477.00 \\
32 & 36 & 1152 & 1152000 & 22787.18 & 21923.07 \\
32 & 36 & 1152 & 11520000 & 241627.20 & 236961.84 \\
      \midrule
      \multicolumn{6}{c}{\mpiallreduce} \\
      $n$ & $N$ & $p$ & $c$ & avg ($\mu s$) & min ($\mu s$) \\
      \midrule
32 & 36 & 1152 & 1152 & 56.40 & 51.02 \\
32 & 36 & 1152 & 11520 & 179.23 & 174.05 \\
32 & 36 & 1152 & 115200 & 1232.02 & 1213.07 \\
32 & 36 & 1152 & 1152000 & 16907.19 & 16779.90 \\
32 & 36 & 1152 & 11520000 & 181196.77 & 179758.07 \\
      \bottomrule
    \end{tabular}
  \end{center}
\end{table}

\begin{table}
  \caption{Results for native \mpireduce compared against the mock-up
    guideline implementations on the ``Hydra'' system.  The MPI library used
    is \intelmpiversion.}
  \label{tab:reduce.intel.n32.c1152}
  \begin{center}
    \begin{tabular}{crrrrrrr}
      \toprule
      \multicolumn{6}{c}{ReduceLane} \\
      $n$ & $N$ & $p$ & $c$ & avg ($\mu s$) & min ($\mu s$) \\
      \midrule
32 & 36 & 1152 & 1152 & 28.26 & 20.03 \\
32 & 36 & 1152 & 11520 & 65.95 & 58.17 \\
32 & 36 & 1152 & 115200 & 349.92 & 339.03 \\
32 & 36 & 1152 & 1152000 & 4031.86 & 3878.12 \\
32 & 36 & 1152 & 11520000 & 58459.48 & 57580.95 \\
      \midrule
      \multicolumn{6}{c}{ReduceHier} \\
      $n$ & $N$ & $p$ & $c$ & avg ($\mu s$) & min ($\mu s$) \\
      \midrule
32 & 36 & 1152 & 1152 & 24.04 & 15.02 \\
32 & 36 & 1152 & 11520 & 84.63 & 61.04 \\
32 & 36 & 1152 & 115200 & 678.27 & 640.15 \\
32 & 36 & 1152 & 1152000 & 11027.48 & 10645.15 \\
32 & 36 & 1152 & 11520000 & 108368.30 & 106885.91 \\
      \midrule
      \multicolumn{6}{c}{\mpireduce} \\
      $n$ & $N$ & $p$ & $c$ & avg ($\mu s$) & min ($\mu s$) \\
      \midrule
32 & 36 & 1152 & 1152 & 48.92 & 41.96 \\
32 & 36 & 1152 & 11520 & 145.84 & 113.96 \\
32 & 36 & 1152 & 115200 & 794.06 & 780.11 \\
32 & 36 & 1152 & 1152000 & 14897.52 & 14566.90 \\
32 & 36 & 1152 & 11520000 & 153258.38 & 152194.02 \\
      \bottomrule
    \end{tabular}
  \end{center}
\end{table}

\begin{table}
  \caption{Results for native \mpireducescatterblock compared against the mock-up
    guideline implementations on the ``Hydra'' system.  The MPI library used
    is \intelmpiversion.}
  \label{tab:reducescatter.intel.n32.c1152}
  \begin{center}
    \begin{tabular}{crrrrrrr}
      \toprule
      \multicolumn{6}{c}{ReduceScatterBlockLane} \\
      $n$ & $N$ & $p$ & $c$ & avg ($\mu s$) & min ($\mu s$) \\
      \midrule
32 & 36 & 1152 & 1 & 40.53 & 25.99 \\
32 & 36 & 1152 & 10 & 68.59 & 56.98 \\
32 & 36 & 1152 & 100 & 401.31 & 380.04 \\
32 & 36 & 1152 & 1000 & 6409.99 & 6297.11 \\
32 & 36 & 1152 & 10000 & 85173.27 & 84170.10 \\
      \midrule
      \multicolumn{6}{c}{ReduceScatterBlockHier} \\
      $n$ & $N$ & $p$ & $c$ & avg ($\mu s$) & min ($\mu s$) \\
      \midrule
32 & 36 & 1152 & 1 & 49.73 & 39.10 \\
32 & 36 & 1152 & 10 & 127.11 & 113.01 \\
32 & 36 & 1152 & 100 & 805.08 & 783.21 \\
32 & 36 & 1152 & 1000 & 12735.31 & 12542.01 \\
32 & 36 & 1152 & 10000 & 120062.00 & 118673.80 \\
      \midrule
      \multicolumn{6}{c}{\mpireducescatterblock} \\
      $n$ & $N$ & $p$ & $c$ & avg ($\mu s$) & min ($\mu s$) \\
      \midrule
32 & 36 & 1152 & 1 & 43.15 & 36.95 \\
32 & 36 & 1152 & 10 & 158.59 & 152.83 \\
32 & 36 & 1152 & 100 & 1471.60 & 1365.90 \\
32 & 36 & 1152 & 1000 & 12450.03 & 12302.16 \\
32 & 36 & 1152 & 10000 & 117739.30 & 116584.06 \\
      \bottomrule
    \end{tabular}
  \end{center}
\end{table}

\begin{table}
  \caption{Results for native \mpiscan compared against the mock-up
    guideline implementations on the ``Hydra'' system.  The MPI library used
    is \openmpiversion.}
    \label{tab:scan.intel.n32.c1152}
  \begin{center}
    \begin{tabular}{crrrrrrr}
      \toprule
    \multicolumn{6}{c}{ScanLane} \\
      $n$ & $N$ & $p$ & $c$ & avg ($\mu s$) & min ($\mu s$) \\
    \midrule
32 & 36 & 1152 & 1152 & 134.23 & 122.79 \\
32 & 36 & 1152 & 11520 & 280.89 & 262.02 \\
32 & 36 & 1152 & 115200 & 2551.50 & 2464.06 \\
32 & 36 & 1152 & 1152000 & 55289.51 & 54286.00 \\
32 & 36 & 1152 & 11520000 & 589505.96 & 586055.99 \\
    \midrule
    \multicolumn{6}{c}{ScanHier} \\
    $n$ & $N$ & $p$ & $c$ & avg ($\mu s$) & min ($\mu s$) \\
    \midrule
32 & 36 & 1152 & 1152 & 127.61 & 118.97 \\
32 & 36 & 1152 & 11520 & 434.86 & 413.89 \\
32 & 36 & 1152 & 115200 & 2920.06 & 2882.00 \\
32 & 36 & 1152 & 1152000 & 64231.87 & 63405.04 \\
32 & 36 & 1152 & 11520000 & 663403.53 & 659806.97 \\
    \midrule
    \multicolumn{6}{c}{\mpiscan} \\
    $n$ & $N$ & $p$ & $c$ & avg ($\mu s$) & min ($\mu s$) \\
    \midrule
32 & 36 & 1152 & 1152 & 109.31 & 97.99 \\
32 & 36 & 1152 & 11520 & 708.85 & 586.03 \\
32 & 36 & 1152 & 115200 & 5394.87 & 5317.93 \\
32 & 36 & 1152 & 1152000 & 87939.59 & 86171.87 \\
32 & 36 & 1152 & 11520000 & 864992.75 & 852500.92 \\
    \bottomrule
    \end{tabular}
    \end{center}
\end{table}

\begin{table}
  \caption{Results for native \mpiexscan compared against the mock-up
    guideline implementations on the ``Hydra'' system.  The MPI library used
    is \openmpiversion.}
    \label{tab:exscan.intel.n32.c1152}
  \begin{center}
    \begin{tabular}{crrrrrrr}
      \toprule
    \multicolumn{6}{c}{ExscanLane} \\
      $n$ & $N$ & $p$ & $c$ & avg ($\mu s$) & min ($\mu s$) \\
    \midrule
32 & 36 & 1152 & 1152 & 690.78 & 122.07 \\
32 & 36 & 1152 & 11520 & 351.29 & 259.88 \\
32 & 36 & 1152 & 115200 & 3279.53 & 2403.97 \\
32 & 36 & 1152 & 1152000 & 52718.65 & 52138.09 \\
32 & 36 & 1152 & 11520000 & 560860.16 & 558268.07 \\
    \midrule
    \multicolumn{6}{c}{ExscanHier} \\
      $n$ & $N$ & $p$ & $c$ & avg ($\mu s$) & min ($\mu s$) \\
    \midrule
32 & 36 & 1152 & 1152 & 557.01 & 119.92 \\
32 & 36 & 1152 & 11520 & 516.37 & 410.08 \\
32 & 36 & 1152 & 115200 & 2920.24 & 2874.14 \\
32 & 36 & 1152 & 1152000 & 61787.60 & 60901.88 \\
32 & 36 & 1152 & 11520000 & 638908.31 & 634175.06 \\
    \midrule
    \multicolumn{6}{c}{\mpiexscan} \\
    $n$ & $N$ & $p$ & $c$ & avg ($\mu s$) & min ($\mu s$) \\
    \midrule
32 & 36 & 1152 & 1152 & 106.77 & 101.09 \\
32 & 36 & 1152 & 11520 & 722.93 & 586.03 \\
32 & 36 & 1152 & 115200 & 5410.01 & 5298.85 \\
32 & 36 & 1152 & 1152000 & 85420.58 & 83101.99 \\
32 & 36 & 1152 & 11520000 & 835976.28 & 825301.89 \\
    \bottomrule
    \end{tabular}
    \end{center}
\end{table}

\clearpage

\section{Results on a single-lane system}
\label{app:singlelane}

We have run all benchmarks on an older single-lane InfiniBand based
system with AMD Opteron processors. The ``Jupiter'' system consists of
36 dual-socket 8-core AMD Opteron 6134 processors at 2.3GHz with a
Mellanox MT4036 QDR InFiniband switch. We ran the benchmarks on 36
nodes with the \mvapichversion MPI library, compiled with \texttt{gcc}
4.8.5. The results can be found in
Tables~\ref{tab:lane.mvapich.n16.c56},
\ref{tab:multicoll.mvapich.n16.c56}, \ref{tab:bcast.mvapich.n16.c56},
\ref{tab:gather.mvapich.n16.c56}, \ref{tab:scatter.mvapich.n16.c56},
\ref{tab:allgather.mvapich.n16.c56},
\ref{tab:alltoall.mvapich.n16.c56},
\ref{tab:allreduce.mvapich.n16.c56}, \ref{tab:reduce.mvapich.n16.c56},
\ref{tab:reducescatter.mvapich.n16.c56},
\ref{tab:scan.mvapich.n16.c56}. and~\ref{tab:exscan.mvapich.n16.c56}.

\begin{table}
  \caption{Lane pattern benchmark results on ``Jupiter'' for increasing
    number of virtual lanes $k$ used for communicating the data (count
    $c$ \mpiint) per compute node. The MPI library used is \mvapichversion.}
  \label{tab:lane.mvapich.n16.c56}
  \begin{center}
    \begin{tabular}{rrrrrrr}
      $k$ & $n$ & $N$ & $p$ & $c$ & avg ($\mu s$) & min ($\mu s$) \\
\toprule
1 & 16 & 35 & 560 & 56 & 242.62 & 203.13 \\
2 & 16 & 35 & 560 & 56 & 288.71 & 164.27 \\
4 & 16 & 35 & 560 & 56 & 242.51 & 169.28 \\
8 & 16 & 35 & 560 & 56 & 273.14 & 196.93 \\
16 & 16 & 35 & 560 & 56 & 458.97 & 429.15 \\
\midrule
1 & 16 & 35 & 560 & 560 & 367.36 & 334.26 \\
2 & 16 & 35 & 560 & 560 & 336.74 & 294.45 \\
4 & 16 & 35 & 560 & 560 & 308.08 & 269.41 \\
8 & 16 & 35 & 560 & 560 & 325.65 & 286.82 \\
16 & 16 & 35 & 560 & 560 & 538.94 & 511.41 \\
\midrule
1 & 16 & 35 & 560 & 5600 & 1102.91 & 1058.58 \\
2 & 16 & 35 & 560 & 5600 & 1036.42 & 998.74 \\
4 & 16 & 35 & 560 & 5600 & 823.46 & 773.91 \\
8 & 16 & 35 & 560 & 5600 & 731.70 & 705.96 \\
16 & 16 & 35 & 560 & 5600 & 930.64 & 897.41 \\
\midrule
1 & 16 & 35 & 560 & 56000 & 5051.42 & 4987.48 \\
2 & 16 & 35 & 560 & 56000 & 4889.84 & 4576.44 \\
4 & 16 & 35 & 560 & 56000 & 5482.81 & 5418.06 \\
8 & 16 & 35 & 560 & 56000 & 5692.70 & 5537.51 \\
16 & 16 & 35 & 560 & 56000 & 6121.68 & 5691.77 \\
\midrule
1 & 16 & 35 & 560 & 560000 & 43333.85 & 43246.75 \\
2 & 16 & 35 & 560 & 560000 & 44350.01 & 42528.87 \\
4 & 16 & 35 & 560 & 560000 & 45020.53 & 44866.80 \\
8 & 16 & 35 & 560 & 560000 & 44864.57 & 44694.19 \\
16 & 16 & 35 & 560 & 560000 & 53223.24 & 52467.58 \\
\midrule
1 & 16 & 35 & 560 & 5600000 & 457877.05 & 457665.44 \\
2 & 16 & 35 & 560 & 5600000 & 452755.73 & 452425.48 \\
4 & 16 & 35 & 560 & 5600000 & 451086.13 & 450649.74 \\
8 & 16 & 35 & 560 & 5600000 & 457505.72 & 457067.25 \\
16 & 16 & 35 & 560 & 5600000 & 515301.02 & 514506.10 \\
\bottomrule
    \end{tabular}
  \end{center}
\end{table}

\begin{table}
  \caption{Multi-collective pattern benchmark results on ``Jupiter'' for
    increasing number of virtual lanes $k$ used for communicating the
    data (count $c$ \mpiint) per lane. The collective function is
    \mpialltoall. The MPI library used is \mvapichversion.}
  \label{tab:multicoll.mvapich.n16.c56}
  \begin{center}
    \begin{tabular}{rrrrrrr}
      $k$ & $n$ & $N$ & $p$ & $c$ & avg ($\mu s$) & min ($\mu s$) \\
\toprule
1 & 16 & 35 & 560 & 56 & 61.57 & 47.21 \\
2 & 16 & 35 & 560 & 56 & 97.72 & 59.13 \\
4 & 16 & 35 & 560 & 56 & 112.34 & 70.33 \\
8 & 16 & 35 & 560 & 56 & 171.84 & 101.80 \\
16 & 16 & 35 & 560 & 56 & 259.94 & 201.94 \\
\midrule
1 & 16 & 35 & 560 & 560 & 78.74 & 61.04 \\
2 & 16 & 35 & 560 & 560 & 112.09 & 82.49 \\
4 & 16 & 35 & 560 & 560 & 182.85 & 105.14 \\
8 & 16 & 35 & 560 & 560 & 384.52 & 261.78 \\
16 & 16 & 35 & 560 & 560 & 740.63 & 632.52 \\
\midrule
1 & 16 & 35 & 560 & 5600 & 148.22 & 123.50 \\
2 & 16 & 35 & 560 & 5600 & 188.59 & 149.49 \\
4 & 16 & 35 & 560 & 5600 & 218.25 & 169.99 \\
8 & 16 & 35 & 560 & 5600 & 308.81 & 242.95 \\
16 & 16 & 35 & 560 & 5600 & 542.58 & 492.10 \\
\midrule
1 & 16 & 35 & 560 & 56000 & 261.41 & 247.00 \\
2 & 16 & 35 & 560 & 56000 & 329.97 & 306.37 \\
4 & 16 & 35 & 560 & 56000 & 520.12 & 501.39 \\
8 & 16 & 35 & 560 & 56000 & 1121.05 & 1016.62 \\
16 & 16 & 35 & 560 & 56000 & 2472.19 & 2424.00 \\
\midrule
1 & 16 & 35 & 560 & 560000 & 1465.88 & 1353.74 \\
2 & 16 & 35 & 560 & 560000 & 3195.35 & 2979.28 \\
4 & 16 & 35 & 560 & 560000 & 6939.54 & 6412.03 \\
8 & 16 & 35 & 560 & 560000 & 14327.03 & 12862.21 \\
16 & 16 & 35 & 560 & 560000 & 33430.03 & 29813.77 \\
\midrule
1 & 16 & 35 & 560 & 5600000 & 10657.12 & 10425.09 \\
2 & 16 & 35 & 560 & 5600000 & 21702.23 & 21114.35 \\
4 & 16 & 35 & 560 & 5600000 & 46349.53 & 44035.20 \\
8 & 16 & 35 & 560 & 5600000 & 97650.14 & 94068.05 \\
16 & 16 & 35 & 560 & 5600000 & 218991.88 & 209491.49 \\
\bottomrule
    \end{tabular}
  \end{center}
\end{table}

\begin{table}
  \caption{Results for native \mpibcast compared against the mock-up
    guideline implementations on the ``Jupiter'' system.  The MPI library used
    is \mvapichversion.}
  \label{tab:bcast.mvapich.n16.c56}
  \begin{center}
    \begin{tabular}{crrrrrrr}
      \toprule
      \multicolumn{6}{c}{BcastLane} \\
      $n$ & $N$ & $p$ & $c$ & avg ($\mu s$) & min ($\mu s$) \\
      \midrule
16 & 35 & 560 & 56 & 249.71 & 33.62 \\
16 & 35 & 560 & 560 & 138.08 & 26.70 \\
16 & 35 & 560 & 5600 & 237.42 & 179.77 \\
16 & 35 & 560 & 56000 & 1760.19 & 1677.75 \\
16 & 35 & 560 & 560000 & 10197.44 & 10102.51 \\
16 & 35 & 560 & 5600000 & 95233.03 & 94971.90 \\
      \midrule
      \multicolumn{6}{c}{BcastHier} \\
      $n$ & $N$ & $p$ & $c$ & avg ($\mu s$) & min ($\mu s$) \\
      \midrule
16 & 35 & 560 & 56 & 347.47 & 43.15 \\
16 & 35 & 560 & 560 & 136.54 & 106.57 \\
16 & 35 & 560 & 5600 & 236.01 & 211.24 \\
16 & 35 & 560 & 56000 & 1233.39 & 1158.24 \\
16 & 35 & 560 & 560000 & 8042.91 & 7828.00 \\
16 & 35 & 560 & 5600000 & 63423.96 & 62008.38 \\
      \midrule
      \multicolumn{6}{c}{\mpibcast} \\
      $n$ & $N$ & $p$ & $c$ & avg ($\mu s$) & min ($\mu s$) \\
      \midrule
16 & 35 & 560 & 56 & 118.14 & 9.78 \\
16 & 35 & 560 & 560 & 114.36 & 30.04 \\
16 & 35 & 560 & 5600 & 232.76 & 170.95 \\
16 & 35 & 560 & 56000 & 1455.54 & 1189.23 \\
16 & 35 & 560 & 560000 & 18024.08 & 15295.27 \\
16 & 35 & 560 & 5600000 & 178340.84 & 172038.79 \\
      \bottomrule
    \end{tabular}
  \end{center}
\end{table}

\begin{table}
  \caption{Results for native \mpigather compared against the mock-up
    guideline implementations on the ``Jupiter'' system.  The MPI library used
    is \mvapichversion.}
  \label{tab:gather.mvapich.n16.c56}
  \begin{center}
    \begin{tabular}{crrrrrrr}
      \toprule
      \multicolumn{6}{c}{GatherLane} \\
      $n$ & $N$ & $p$ & $c$ & avg ($\mu s$) & min ($\mu s$) \\
      \midrule
16 & 35 & 560 & 1 & 64.92 & 50.78 \\
16 & 35 & 560 & 10 & 73.62 & 64.37 \\
16 & 35 & 560 & 100 & 150.62 & 140.67 \\
16 & 35 & 560 & 1000 & 1429.44 & 1397.61 \\
16 & 35 & 560 & 10000 & 7359.84 & 7340.91 \\
      \midrule
      \multicolumn{6}{c}{GatherHier} \\
      $n$ & $N$ & $p$ & $c$ & avg ($\mu s$) & min ($\mu s$) \\
      \midrule
16 & 35 & 560 & 1 & 185.35 & 163.56 \\
16 & 35 & 560 & 10 & 182.32 & 163.56 \\
16 & 35 & 560 & 100 & 443.43 & 358.10 \\
16 & 35 & 560 & 1000 & 2463.18 & 2295.97 \\
16 & 35 & 560 & 10000 & 25577.00 & 25001.53 \\
      \midrule
      \multicolumn{6}{c}{\mpigather} \\
      $n$ & $N$ & $p$ & $c$ & avg ($\mu s$) & min ($\mu s$) \\
      \midrule
16 & 35 & 560 & 1 & 80.90 & 61.75 \\
16 & 35 & 560 & 10 & 90.13 & 80.35 \\
16 & 35 & 560 & 100 & 163.56 & 137.57 \\
16 & 35 & 560 & 1000 & 2931.76 & 2769.71 \\
16 & 35 & 560 & 10000 & 25336.85 & 25165.08 \\
      \bottomrule
    \end{tabular}
  \end{center}
\end{table}

\begin{table}
  \caption{Results for native \mpiscatter compared against the mock-up
    guideline implementations on the ``Jupiter'' system.  The MPI library used
    is \mvapichversion.}
  \label{tab:scatter.mvapich.n16.c56}
  \begin{center}
    \begin{tabular}{crrrrrrr}
      \toprule
      \multicolumn{6}{c}{ScatterLane} \\
      $n$ & $N$ & $p$ & $c$ & avg ($\mu s$) & min ($\mu s$) \\
      \midrule
16 & 35 & 560 & 1 & 28.37 & 22.41 \\
16 & 35 & 560 & 10 & 32.37 & 24.56 \\
16 & 35 & 560 & 100 & 118.71 & 106.57 \\
16 & 35 & 560 & 1000 & 826.08 & 813.72 \\
16 & 35 & 560 & 10000 & 8175.01 & 8111.24 \\
      \midrule
      \multicolumn{6}{c}{ScatterHier} \\
      $n$ & $N$ & $p$ & $c$ & avg ($\mu s$) & min ($\mu s$) \\
      \midrule
16 & 35 & 560 & 1 & 131.45 & 112.53 \\
16 & 35 & 560 & 10 & 182.46 & 150.92 \\
16 & 35 & 560 & 100 & 295.86 & 271.80 \\
16 & 35 & 560 & 1000 & 1825.72 & 1690.39 \\
16 & 35 & 560 & 10000 & 18900.18 & 18716.81 \\
\midrule
      \multicolumn{6}{c}{\mpiscatter} \\
      $n$ & $N$ & $p$ & $c$ & avg ($\mu s$) & min ($\mu s$) \\
\midrule
16 & 35 & 560 & 1 & 28.45 & 21.46 \\
16 & 35 & 560 & 10 & 33.46 & 26.23 \\
16 & 35 & 560 & 100 & 112.03 & 102.28 \\
16 & 35 & 560 & 1000 & 834.30 & 818.73 \\
16 & 35 & 560 & 10000 & 8900.76 & 8809.33 \\
      \bottomrule
    \end{tabular}
  \end{center}
\end{table}

\begin{table}
  \caption{Results for native \mpiallgather compared against the mock-up
    guideline implementations on the ``Jupiter'' system.  The MPI library used
    is \mvapichversion.}
  \label{tab:allgather.mvapich.n16.c56}
  \begin{center}
    \begin{tabular}{crrrrrrr}
      \toprule
      \multicolumn{6}{c}{AllgatherLane} \\
      $n$ & $N$ & $p$ & $c$ & avg ($\mu s$) & min ($\mu s$) \\
      \midrule
16 & 35 & 560 & 1 & 71.34 & 54.60 \\
16 & 35 & 560 & 10 & 186.97 & 164.75 \\
16 & 35 & 560 & 100 & 1044.70 & 989.20 \\
16 & 35 & 560 & 1000 & 9591.04 & 9233.95 \\
16 & 35 & 560 & 10000 & 79466.16 & 79236.75 \\
      \midrule
      \multicolumn{6}{c}{AllgatherHier} \\
      $n$ & $N$ & $p$ & $c$ & avg ($\mu s$) & min ($\mu s$) \\
      \midrule
16 & 35 & 560 & 1 & 208.38 & 169.52 \\
16 & 35 & 560 & 10 & 281.16 & 249.62 \\
16 & 35 & 560 & 100 & 1017.88 & 963.93 \\
16 & 35 & 560 & 1000 & 8235.08 & 8000.85 \\
16 & 35 & 560 & 10000 & 55201.42 & 54624.32 \\
      \midrule
      \multicolumn{6}{c}{\mpiallgather} \\
      $n$ & $N$ & $p$ & $c$ & avg ($\mu s$) & min ($\mu s$) \\
      \midrule
16 & 35 & 560 & 1 & 192.48 & 168.32 \\
16 & 35 & 560 & 10 & 382.21 & 332.83 \\
16 & 35 & 560 & 100 & 2356.35 & 2252.82 \\
16 & 35 & 560 & 1000 & 12104.26 & 5220.89 \\
16 & 35 & 560 & 10000 & 55844.84 & 52297.35 \\
      \bottomrule
    \end{tabular}
  \end{center}
\end{table}

\begin{table}
  \caption{Results for native \mpialltoall compared against the mock-up
    guideline implementation on the ``Jupiter'' system.  The MPI library used
    is \mvapichversion.}
  \label{tab:alltoall.mvapich.n16.c56}
  \begin{center}
    \begin{tabular}{crrrrrrr}
      \toprule
      \multicolumn{6}{c}{AlltoallLane} \\
      $n$ & $N$ & $p$ & $c$ & avg ($\mu s$) & min ($\mu s$) \\
      \midrule
16 & 35 & 560 & 1 & 559.29 & 535.49 \\
16 & 35 & 560 & 10 & 615.50 & 582.70 \\
16 & 35 & 560 & 100 & 3371.99 & 3116.13 \\
16 & 35 & 560 & 1000 & 43665.94 & 42370.80 \\
16 & 35 & 560 & 10000 & 291839.28 & 281864.40 \\
      \midrule
      \multicolumn{6}{c}{\mpialltoall} \\
      $n$ & $N$ & $p$ & $c$ & avg ($\mu s$) & min ($\mu s$) \\
      \midrule
16 & 35 & 560 & 1 & 276.18 & 242.95 \\
16 & 35 & 560 & 10 & 783.65 & 681.16 \\
16 & 35 & 560 & 100 & 6957.57 & 6736.52 \\
16 & 35 & 560 & 1000 & 29591.54 & 27791.98 \\
16 & 35 & 560 & 10000 & 218818.27 & 212746.86 \\
      \bottomrule
    \end{tabular}
  \end{center}
\end{table}

\begin{table}
  \caption{Results for native \mpiallreduce compared against the mock-up
    guideline implementations on the ``Jupiter'' system.  The MPI library used
    is \mvapichversion.}
  \label{tab:allreduce.mvapich.n16.c56}
  \begin{center}
    \begin{tabular}{crrrrrrr}
      \toprule
      \multicolumn{6}{c}{AllreduceLane} \\
      $n$ & $N$ & $p$ & $c$ & avg ($\mu s$) & min ($\mu s$) \\
      \midrule
16 & 35 & 560 & 56 & 64.93 & 44.11 \\
16 & 35 & 560 & 560 & 103.65 & 86.07 \\
16 & 35 & 560 & 5600 & 392.94 & 356.44 \\
16 & 35 & 560 & 56000 & 2159.22 & 2066.85 \\
16 & 35 & 560 & 560000 & 21878.93 & 21579.27 \\
16 & 35 & 560 & 5600000 & 201373.88 & 199721.34 \\
      \midrule
      \multicolumn{6}{c}{AllreduceHier} \\
      $n$ & $N$ & $p$ & $c$ & avg ($\mu s$) & min ($\mu s$) \\
      \midrule
16 & 35 & 560 & 56 & 255.12 & 225.07 \\
16 & 35 & 560 & 560 & 277.80 & 240.56 \\
16 & 35 & 560 & 5600 & 435.70 & 405.55 \\
16 & 35 & 560 & 56000 & 1821.38 & 1658.68 \\
16 & 35 & 560 & 560000 & 12337.21 & 12032.75 \\
16 & 35 & 560 & 5600000 & 121412.55 & 118511.44 \\
      \midrule
      \multicolumn{6}{c}{\mpiallreduce} \\
      $n$ & $N$ & $p$ & $c$ & avg ($\mu s$) & min ($\mu s$) \\
      \midrule
16 & 35 & 560 & 56 & 61.74 & 41.01 \\
16 & 35 & 560 & 560 & 102.87 & 86.55 \\
16 & 35 & 560 & 5600 & 392.09 & 362.87 \\
16 & 35 & 560 & 56000 & 1491.41 & 1435.04 \\
16 & 35 & 560 & 560000 & 12012.53 & 11727.33 \\
16 & 35 & 560 & 5600000 & 110638.41 & 108487.13 \\
      \bottomrule
    \end{tabular}
  \end{center}
\end{table}

\begin{table}
  \caption{Results for native \mpireduce compared against the mock-up
    guideline implementations on the ``Jupiter'' system.  The MPI library used
    is \mvapichversion.}
  \label{tab:reduce.mvapich.n16.c56}
  \begin{center}
    \begin{tabular}{crrrrrrr}
      \toprule
      \multicolumn{6}{c}{ReduceLane} \\
      $n$ & $N$ & $p$ & $c$ & avg ($\mu s$) & min ($\mu s$) \\
      \midrule
16 & 35 & 560 & 56 & 16.39 & 7.63 \\
16 & 35 & 560 & 560 & 38.10 & 22.65 \\
16 & 35 & 560 & 5600 & 172.82 & 144.96 \\
16 & 35 & 560 & 56000 & 933.36 & 730.51 \\
16 & 35 & 560 & 560000 & 9814.77 & 9613.28 \\
16 & 35 & 560 & 5600000 & 98577.63 & 93741.66 \\
      \midrule
      \multicolumn{6}{c}{ReduceHier} \\
      $n$ & $N$ & $p$ & $c$ & avg ($\mu s$) & min ($\mu s$) \\
      \midrule
16 & 35 & 560 & 56 & 150.14 & 126.12 \\
16 & 35 & 560 & 560 & 157.99 & 137.81 \\
16 & 35 & 560 & 5600 & 221.80 & 200.75 \\
16 & 35 & 560 & 56000 & 1272.31 & 1231.19 \\
16 & 35 & 560 & 560000 & 9453.70 & 9216.55 \\
16 & 35 & 560 & 5600000 & 113722.58 & 111022.23 \\
      \midrule
      \multicolumn{6}{c}{\mpireduce} \\
      $n$ & $N$ & $p$ & $c$ & avg ($\mu s$) & min ($\mu s$) \\
      \midrule
16 & 35 & 560 & 56 & 34.21 & 17.64 \\
16 & 35 & 560 & 560 & 56.48 & 38.62 \\
16 & 35 & 560 & 5600 & 438.78 & 417.23 \\
16 & 35 & 560 & 56000 & 2038.72 & 1935.48 \\
16 & 35 & 560 & 560000 & 24265.43 & 24067.88 \\
16 & 35 & 560 & 5600000 & 255436.66 & 249367.00 \\
      \bottomrule
    \end{tabular}
  \end{center}
\end{table}

\begin{table}
  \caption{Results for native \mpireducescatterblock compared against the mock-up
    guideline implementations on the ``Jupiter'' system.  The MPI library used
    is \mvapichversion.}
  \label{tab:reducescatter.mvapich.n16.c56}
  \begin{center}
    \begin{tabular}{crrrrrrr}
      \toprule
      \multicolumn{6}{c}{ReduceScatterBlockLane} \\
      $n$ & $N$ & $p$ & $c$ & avg ($\mu s$) & min ($\mu s$) \\
      \midrule
16 & 35 & 560 & 1 & 84.69 & 66.04 \\
16 & 35 & 560 & 1 & 83.65 & 64.85 \\
16 & 35 & 560 & 10 & 249.35 & 228.88 \\
16 & 35 & 560 & 100 & 1295.34 & 1266.48 \\
16 & 35 & 560 & 1000 & 12192.04 & 11956.69 \\
16 & 35 & 560 & 10000 & 123022.56 & 121702.67 \\
      \midrule
      \multicolumn{6}{c}{ReduceScatterBlockHier} \\
      $n$ & $N$ & $p$ & $c$ & avg ($\mu s$) & min ($\mu s$) \\
      \midrule
16 & 35 & 560 & 1 & 103.75 & 81.78 \\
16 & 35 & 560 & 10 & 168.54 & 146.87 \\
16 & 35 & 560 & 100 & 923.81 & 854.02 \\
16 & 35 & 560 & 1000 & 8554.58 & 8355.62 \\
16 & 35 & 560 & 10000 & 103504.45 & 101781.13 \\
      \midrule
      \multicolumn{6}{c}{\mpireducescatterblock} \\
      $n$ & $N$ & $p$ & $c$ & avg ($\mu s$) & min ($\mu s$) \\
      \midrule
16 & 35 & 560 & 1 & 106.32 & 92.27 \\
16 & 35 & 560 & 10 & 266.66 & 245.09 \\
16 & 35 & 560 & 100 & 2369.18 & 2287.86 \\
16 & 35 & 560 & 1000 & 28420.42 & 27554.99 \\
16 & 35 & 560 & 10000 & 233104.79 & 224643.95 \\
      \bottomrule
    \end{tabular}
  \end{center}
\end{table}

\begin{table}
  \caption{Results for native \mpiscan compared against the mock-up
    guideline implementations on the ``Hydra'' system.  The MPI library used
    is \openmpiversion.}
    \label{tab:scan.mvapich.n16.c56}
  \begin{center}
    \begin{tabular}{crrrrrrr}
      \toprule
    \multicolumn{6}{c}{ScanLane} \\
      $n$ & $N$ & $p$ & $c$ & avg ($\mu s$) & min ($\mu s$) \\
    \midrule
16 & 35 & 560 & 56 & 95.02 & 66.76 \\
16 & 35 & 560 & 560 & 161.71 & 111.34 \\
16 & 35 & 560 & 5600 & 739.02 & 597.24 \\
16 & 35 & 560 & 56000 & 4983.79 & 4513.98 \\
16 & 35 & 560 & 560000 & 74520.20 & 74109.08 \\
16 & 35 & 560 & 5600000 & 698889.36 & 683559.18 \\
    \midrule
    \multicolumn{6}{c}{ScanHier} \\
    $n$ & $N$ & $p$ & $c$ & avg ($\mu s$) & min ($\mu s$) \\
    \midrule
16 & 35 & 560 & 56 & 115.29 & 92.51 \\
16 & 35 & 560 & 560 & 156.87 & 134.47 \\
16 & 35 & 560 & 5600 & 550.03 & 477.55 \\
16 & 35 & 560 & 56000 & 4157.84 & 3859.52 \\
16 & 35 & 560 & 560000 & 60258.77 & 57585.48 \\
16 & 35 & 560 & 5600000 & 571872.59 & 556946.52 \\
    \midrule
    \multicolumn{6}{c}{\mpiscan} \\
    $n$ & $N$ & $p$ & $c$ & avg ($\mu s$) & min ($\mu s$) \\
    \midrule
16 & 35 & 560 & 56 & 90.17 & 70.33 \\
16 & 35 & 560 & 560 & 131.65 & 116.11 \\
16 & 35 & 560 & 5600 & 1248.41 & 1189.47 \\
16 & 35 & 560 & 56000 & 7229.13 & 7132.77 \\
16 & 35 & 560 & 560000 & 96150.36 & 95041.99 \\
16 & 35 & 560 & 5600000 & 1125118.99 & 1102876.66 \\
    \bottomrule
    \end{tabular}
    \end{center}
\end{table}

\begin{table}
  \caption{Results for native \mpiexscan compared against the mock-up
    guideline implementations on the ``Hydra'' system.  The MPI library used
    is \openmpiversion.}
    \label{tab:exscan.mvapich.n16.c56}
  \begin{center}
    \begin{tabular}{crrrrrrr}
      \toprule
    \multicolumn{6}{c}{ExscanLane} \\
      $n$ & $N$ & $p$ & $c$ & avg ($\mu s$) & min ($\mu s$) \\
    \midrule
16 & 35 & 560 & 56 & 68.80 & 49.83 \\
16 & 35 & 560 & 560 & 114.21 & 92.27 \\
16 & 35 & 560 & 5600 & 604.34 & 548.36 \\
16 & 35 & 560 & 56000 & 4295.78 & 4259.59 \\
16 & 35 & 560 & 560000 & 69679.28 & 65251.35 \\
16 & 35 & 560 & 5600000 & 640817.78 & 612981.08 \\
    \midrule
    \multicolumn{6}{c}{ExscanHier} \\
      $n$ & $N$ & $p$ & $c$ & avg ($\mu s$) & min ($\mu s$) \\
    \midrule
16 & 35 & 560 & 56 & 111.11 & 89.65 \\
16 & 35 & 560 & 560 & 155.33 & 128.27 \\
16 & 35 & 560 & 5600 & 480.94 & 452.28 \\
16 & 35 & 560 & 56000 & 3677.62 & 3433.70 \\
16 & 35 & 560 & 560000 & 50137.12 & 47584.77 \\
16 & 35 & 560 & 5600000 & 493240.18 & 481934.55 \\
    \midrule
    \multicolumn{6}{c}{\mpiexscan} \\
    $n$ & $N$ & $p$ & $c$ & avg ($\mu s$) & min ($\mu s$) \\
    \midrule
16 & 35 & 560 & 56 & 104.78 & 85.35 \\
16 & 35 & 560 & 560 & 186.77 & 174.28 \\
16 & 35 & 560 & 5600 & 1454.64 & 1405.95 \\
16 & 35 & 560 & 56000 & 12973.22 & 12656.45 \\
16 & 35 & 560 & 560000 & 140288.79 & 136096.24 \\
16 & 35 & 560 & 5600000 & 1388062.30 & 1324449.06 \\
    \bottomrule
    \end{tabular}
    \end{center}
\end{table}

\clearpage

\end{document}